\documentclass[a4paper,11pt]{article}

\usepackage{amsmath,amsthm,amsbsy,amssymb,amsfonts,graphicx,mathrsfs,bm,cite,color}
\usepackage[utf8]{inputenc}
\usepackage[all]{xy}
\usepackage[scr=boondoxo]{mathalfa}

\makeatletter
\catcode`\@=11
\@addtoreset{equation}{section}

\makeatother

\makeatletter

\@addtoreset{table}{section}
\makeatother

\renewcommand{\thefootnote}{\arabic{footnote}}

\newcommand{\test}[1]{#1}

\newcommand{\Exp}[1]{\operatorname{e}^{#1}}
\newcommand{\diag}{\operatorname{diag}}
\newcommand{\abs}[1]{\lvert {#1} \rvert}
\newcommand{\bra}[1]{\langle {#1} \rvert}
\newcommand{\ket}[1]{\lvert {#1} \rangle}
\newcommand{\rmd}{{\mathrm{d}}}
\newcommand{\nn}{\nonumber}
\newcommand{\Lie}{\pounds}

\newcommand{\gs}{g_s}
\newcommand{\ls}{l_s}

\newcommand{\cA}{\mathcal A}\newcommand{\cB}{\mathcal B}

\newcommand{\cE}{\mathcal E}
\newcommand{\cG}{\mathcal G}\newcommand{\cH}{\mathcal H}

\newcommand{\cL}{\mathcal L}
\newcommand{\cM}{\mathcal M}\newcommand{\cN}{\mathcal N}

\newcommand{\sfa}{\mathsf{a}}
\newcommand{\sfb}{\mathsf{b}}
\newcommand{\sfI}{\mathsf{I}}
\newcommand{\sfK}{\mathsf{K}}
\newcommand{\sfR}{\mathsf{R}}

\newcommand{\rmE}{\mathrm{E}}

\makeatletter
\newcommand*{\rmT}{{\mathpalette\@transpose{}}}
\newcommand*{\@transpose}[2]{\raisebox{\depth}{$\m@th#1\intercal$}}
\makeatother

\newcommand{\SL}{\text{SL}}
\newcommand{\OO}{\text{O}}

\newcommand{\SLa}{\test{\alpha}}
\newcommand{\SLb}{\test{\beta}}
\newcommand{\SLc}{\test{\gamma}}
\newcommand{\SLd}{\test{\delta}}
\newcommand{\SLe}{\test{\epsilon}}
\newcommand{\SLf}{\test{\zeta}}
\newcommand{\SLg}{\test{\eta}}

\newcommand{\MF}{\test{\hat{F}}}
\newcommand{\MA}{\test{\hat{A}}}
\newcommand{\Mg}{\test{\hat{g}}}

\newcommand{\Mv}{\test{\hat{\lambda}}}
\newcommand{\MN}{\test{\hat{N}}}

\renewcommand{\AA}{\test{\mathscr{A}}}
\newcommand{\AB}{\test{\mathscr{B}}}
\newcommand{\AC}{\test{\mathscr{C}}}

\newcommand{\AH}{\test{\mathscr{H}}}
\newcommand{\APhi}{\test{\varphi}}
\newcommand{\Ag}{\test{\mathscr{g}}}

\newcommand{\BA}{\test{\mathsf{A}}}
\newcommand{\BB}{\test{\mathsf{B}}}
\newcommand{\BC}{\test{\mathsf{C}}}
\newcommand{\BD}{\test{\mathsf{D}}}
\newcommand{\BE}{\test{\mathsf{E}}}
\newcommand{\BF}{\test{\mathsf{F}}}

\newcommand{\BH}{\test{\mathsf{H}}}
\newcommand{\BN}{\test{\mathsf{N}}}
\newcommand{\BPhi}{\test{\Phi}}
\newcommand{\Bg}{\test{\mathsf{g}}}
\newcommand{\Bm}{\test{\mathsf{m}}}
\newcommand{\bBA}{\test{\bm{\mathsf{A}}}}

\newcommand{\bBC}{\test{\bm{\mathsf{C}}}}
\newcommand{\bBD}{\test{\bm{\mathsf{D}}}}
\newcommand{\bBE}{\test{\bm{\mathsf{E}}}}

\newcommand{\SLE}[1]{\bm{#1}}

\newcommand{\By}{\test{\mathsf{y}}}
\newcommand{\Ay}{\test{y}}
\newcommand{\Az}{\test{z}}

\newcommand{\TmC}{\test{A}}
\newcommand{\TmD}{\test{D}}
\newcommand{\TmE}{\test{E}}
\newcommand{\TmF}{\test{F}}

\def\sla#1{\setbox0=\hbox{$#1$} 
\dimen0=\wd0 
\setbox1=\hbox{/} \dimen1=\wd1 
\ifdim\dimen0>\dimen1 
\rlap{\hbox to \dimen0{\hfil/\hfil}} 
#1 
\else 
\rlap{\hbox to \dimen1{\hfil$#1$\hfil}} 
/ 
\fi}

\allowdisplaybreaks[3]

\begin{document}

\begin{titlepage}
\renewcommand{\thefootnote}{\fnsymbol{footnote}}

\vspace*{1cm}

\centerline{\Large\textbf{Duality rules for more mixed-symmetry potentials}}%

\vspace{1.5cm}

\centerline{\large Yuho Sakatani}

\vspace{0.2cm}

\begin{center}
{\it Department of Physics, Kyoto Prefectural University of Medicine,}\\
{\it Kyoto 606-0823, Japan}\\
{\small\texttt{yuho@koto.kpu-m.ac.jp}}
\end{center}

\vspace*{2mm}

\begin{abstract}
$T$- and $S$-duality rules among the gauge potentials in type II supergravities are studied. In particular, by following the approach of arXiv:1909.01335, we determine the $T$- and $S$-duality rules for certain mixed-symmetry potentials, which couple to supersymmetric branes with tension $T\propto g_s^{-n}$ ($n\leq 4$). Although the $T$-duality rules are rather intricate, we find a certain redefinition of potentials which considerably simplifies the duality rules. After the redefinition, potentials are identified with components of the $T$-duality-covariant potentials, which have been predicted by the $E_{11}$ conjecture. We also discuss the field strengths of the mixed-symmetry potentials.
\end{abstract}

\thispagestyle{empty}
\end{titlepage}

\setcounter{footnote}{0}

\newpage

\tableofcontents

\newpage

\section{Introduction}

Toroidally compactified 11D supergravity or type II supergravity has the $U$-duality symmetry but this is not manifest in the standard formulation. 
In order to exhibit the symmetry, the standard metric, scalar fields, and $p$-form gauge potentials are not enough \cite{hep-th/0104081}. 
In fact, we additionally need to introduce certain mixed-symmetry potentials, which are related to the standard potentials through a non-local relation, similar to the electric-magnetic duality. 
According to the $E_{11}$ conjecture \cite{hep-th/0104081,hep-th/0307098}, there are infinitely many mixed-symmetry potentials in each theory. 
By introducing an integer-valued parameter $\ell$, known as the level, the number of mixed-symmetry potentials with a fixed level $\ell$ is finite, and we can determine the full list of the mixed-symmetry potentials for each level $\ell$ (see \cite{0705.0752,1907.07177} and references therein). 

Although a list of mixed-symmetry potentials which constitutes the $U$-duality multiplets has been algebraically determined, their physical definitions are still obscure. 
In the case of the standard supergravity fields, their definitions can be fixed by the supergravity action, but the mixed-symmetry potentials do not appear in the standard supergravity action and it is not straightforward to specify their definitions. 
A possible way to specify their definitions is to construct the worldvolume actions for supersymmetric branes. 
As is well known, the Ramond--Ramond (R--R) fields couple to D-branes, and one can identify the definition of the R--R fields by looking at the Wess--Zumino (WZ) term. 
Similarly, mixed-symmetry potentials generally couple to certain exotic branes \cite{hep-th/9707217,hep-th/9712047,hep-th/9712075,hep-th/9712084,hep-th/9809039,hep-th/9908094,hep-th/0012051,0805.4451,1004.2521,1209.6056} and their definitions can be fixed by constructing the WZ term for exotic branes. 
For example, the WZ term of the Kaluza--Klein (KK) monopole has been constructed in \cite{hep-th/9802199,hep-th/9806169,hep-th/9812188} and a precise definition of the dual graviton has been given. 
However, at present, worldvolume actions have been constructed only for a few exotic branes. 

\begin{figure}[b]
\begin{center}
\includegraphics[width=\linewidth]{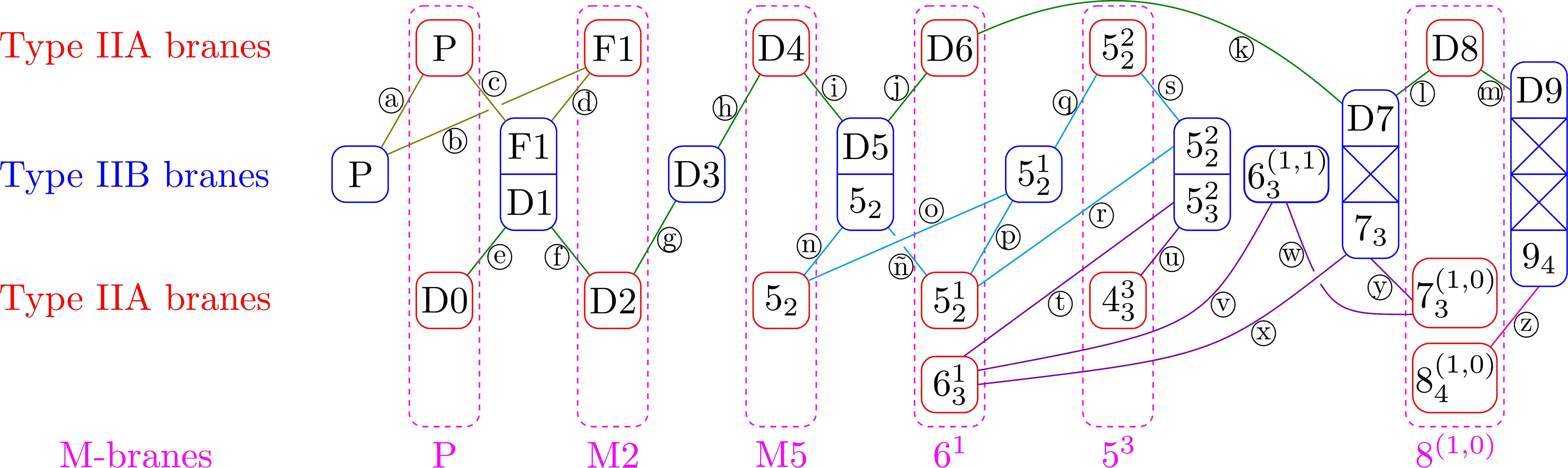}
\end{center}
\caption{Duality web studied in this paper. 
Type IIB branes are paired into $S$-duality multiplets, though only supersymmetric branes are shown explicitly. 
The M-theory uplifts of type IIA branes are displayed at the bottom. 
Several different names for branes are as follows:\newline
{\footnotesize\underline{M-theory:} $6^1=\!\Bigl\{{\text{\footnotesize MKK \cite{hep-th/9908094}}\atop\text{\footnotesize KK7 \cite{hep-th/9806120}}}$, $8^{(1,0)}=\!\Bigl\{{\text{\footnotesize M9 \cite{hep-th/9908094}}\atop\text{\footnotesize KK9 \cite{hep-th/9806120}}}$, \newline
\underline{Type IIA:} $5_2=\!\Bigl\{{\text{\footnotesize NS5A \cite{hep-th/9908094}}\atop\text{\footnotesize S5A \cite{hep-th/9806120}}}$, $5^1_2=\!\Bigl\{{\text{\footnotesize KK5A \cite{hep-th/9908094}} \atop \text{\footnotesize KK6A \cite{hep-th/9806120}}}$, $6^1_3\!=\Bigl\{{\text{\footnotesize KK6A \cite{hep-th/9908094}}\atop\text{\footnotesize KK7A \cite{hep-th/9806120}}}$, $7^{(1,0)}_3 =\!\Bigl\{{\text{\footnotesize KK8A \cite{hep-th/9908094}}\atop\text{\footnotesize KK8A \cite{hep-th/9806120}}}$, $8^{(1,0)}_4\!=\Bigl\{{\text{\footnotesize NS9A \cite{hep-th/9908094}}\atop\text{\footnotesize KK9A \cite{hep-th/9806120}}}$,\newline
\underline{Type IIB:} $5_2=\!\Bigl\{{\text{\footnotesize NS5B \cite{hep-th/9908094}}\atop\text{\footnotesize S5B \cite{hep-th/9806120}}}$, $5^1_2=\!\Bigl\{{\text{\footnotesize KK5B \cite{hep-th/9908094}}\atop\text{\footnotesize KK6B \cite{hep-th/9806120}}}$, $7_3=\!\Bigl\{{\text{\footnotesize NS7B \cite{hep-th/9908094}}\atop\text{\footnotesize Q7 \cite{hep-th/9806120}}}$, $6_3^{(1,1)}=\!\text{\footnotesize KK7B \cite{hep-th/9908094}}$, $9_4=\!\Bigl\{{\text{\footnotesize NS9B \cite{hep-th/9908094}}\atop\text{\footnotesize Q9 \cite{hep-th/9806120}}}$.}
\label{fig:duality-web}}
\end{figure}

To make precise definitions of mixed-symmetry potentials, it is more straightforward to determine the $T$- and $S$-duality transformation rule. 
The mixed-symmetry potentials which couple to supersymmetric branes are related to the standard $p$-form potentials under $T$-/$S$-duality transformations. 
Then, by determining the duality rule, we can fix the convention for the mixed-symmetry potentials. 
Recently, a systematic approach to determine the duality rules has been proposed in \cite{1909.01335}, and the $T$-/$S$-duality rules for the dual graviton have been determined. 
In this paper, we continue the analysis and obtain the $T$-/$S$-duality rules for more mixed-symmetry potentials. 
Concretely, we consider the duality web described in Figure \ref{fig:duality-web}. 
There, each line (with a circled alphabet appended) corresponds to a $T$-duality that connects a type IIA brane and a type IIB brane. 
For example, the $T$-duality \textcircled{q} connects the $5^2_2$-brane in type IIA theory and the $5^1_2$-brane in type IIB theory. 
Since the $5^2_2$-brane and the $5^1_2$-brane minimally couple to the potentials $\AA_{8,2}$ and $\BA_{7,1}$, respectively, $T$-duality \textcircled{q} corresponds to a $T$-duality rule for $\AA_{a_1\cdots a_7 \Ay, a\Ay} \leftrightarrow \BA_{a_1\cdots a_7, a}$\,, where $x^{\Ay}$ is the $T$-duality direction. 
We determine the $T$-duality rule for the 27 lines, \textcircled{a}--\textcircled{z}, including non-linear terms in the duality rules. 

In Figure \ref{fig:duality-web}, by following the notation of \cite{hep-th/9809039}, a $p$-brane in type II theory with tension 
\begin{align}
 T_p = \frac{\gs^{-n}}{\ls\,(2\pi l_s)^p} \Bigl(\frac{R_{n_1}\cdots R_{n_{c_2}}}{\ls^{c_2}}\Bigr)^2\cdots \Bigl(\frac{R_{q_1}\cdots R_{q_{c_s}}}{\ls^{c_s}}\Bigr)^s\qquad (R_n:\text{ toroidal radii})\,,
\end{align}
is denoted as a $p^{(c_s,\dotsc,c_2)}_{n}$-brane. 
In particular, the NS5-brane is denoted as $5_2\equiv 5^{(0,\dotsc,0)}_{2}$ and the fundamental string (F1) and the D$p$-brane may be denoted as $1_0\equiv 1^{(0,\dotsc,0)}_{0}$ and $p_1\equiv p^{(0,\dotsc,0)}_{1}$, respectively. 
The $T$-dualities \textcircled{a}--\textcircled{d} and \textcircled{e}--\textcircled{m} correspond to the standard $T$-dualities for the NS--NS fields and the R--R fields. 
The $T$-dualities \textcircled{n}--\textcircled{p} have been obtained in \cite{hep-th/9806169,0712.3235,0907.3614,1701.07819,1909.01335} and \textcircled{x} is obtained in \cite{hep-th/9908094}. 
To the author's knowledge, other $T$-dualities are new.\footnote{In \cite{0712.3235,0907.3614}, $T$-duality rules for some additional potentials, which are not studied here, have been studied.}
In our approach, type IIB fields are defined as $\SL(2)$ $S$-duality tensors, and the $S$-duality rules are simple. 

The structure of this paper is as follows. 
In section \ref{sec:sugra}, we fix our convention for the (bosonic) supergravity fields. 
The standard fields are defined through the action, and several higher $p$-form potentials are introduced through the electric-magnetic duality. 
Additional mixed-symmetry potentials are defined in section \ref{sec:1-form} by finding a consistent parameterization of the $U$-duality-covariant 1-form field $\cA_\mu^I$. 
In section \ref{sec:duality}, we determine the $T$- and $S$-duality rules by following the approach of \cite{1701.07819,1909.01335}. 
In particular, in sections \ref{sec:D-potential} and \ref{sec:EF-potential}, by considering certain field redefinitions, we show that our mixed-symmetry potentials can be packaged into the $\OO(10,10)$-covariant potentials $\TmD_{M_1\cdots M_4}$, $\TmE_{MN\dot{a}}$, and $\TmF^+_{M_1\cdots M_{10}}$. 
In section \ref{sec:gauge}, we discuss the gauge symmetries and field strengths in each theory. 
Section \ref{sec:conclusions} is devoted to conclusions and discussions. 
In Appendix \ref{app:notation}, we explain our conventions. 

\section{Supergravity fields}
\label{sec:sugra}

In this section, we fix our conventions for the bosonic supergravity fields. 

\subsection{11D supergravity}
\label{sec:11D-sugra}

In 11D supergravity, the bosonic fields are $\Mg_{ij}$ and $\MA_{\hat{3}}$, for which the Lagrangian is
\begin{align}
 \cL_{11} = \hat{*}\hat{R} -\tfrac{1}{2}\,\MF_{\hat{4}}\wedge \hat{*}\MF_{\hat{4}} - \tfrac{1}{3!}\,\MA_{\hat{3}}\wedge \MF_{\hat{4}}\wedge \MF_{\hat{4}} \,,
\end{align}
where we have defined $\MF_{\hat{4}}\equiv \rmd \MA_{\hat{3}}$\,. 
By introducing the dual field strength as
\begin{align}
 \MF_{\hat{7}} \equiv -*_{11} \MF_{\hat{4}} \,,
\label{eq:F4-F7}
\end{align}
the equation of motion for $\MA_{\hat{3}}$ is expressed as the Bianchi identity
\begin{align}
 \rmd \MF_{\hat{7}} - \tfrac{1}{2}\,\MF_{\hat{4}}\wedge \MF_{\hat{4}} = 0\,. 
\end{align}
This suggests us to introduce the 6-form potential $\MA_{\hat{6}}$ as
\begin{align}
 \MF_{\hat{7}} \equiv \rmd \MA_{\hat{6}} + \tfrac{1}{2}\,\MA_{\hat{3}}\wedge \MF_{\hat{4}}\,.
\end{align}

Although the potential $\MA_{\hat{6}}$ is not contained in the standard Lagrangian, it is necessary for manifesting the $U$-duality symmetry. 
For the manifest $U$-duality symmetry, in general, we need to introduce additional gauge potentials, which are generally mixed-symmetry tensors. 
Among these, we consider $\MA_{{\hat{8}},{\hat{1}}}$, $\MA_{{\hat{9}},{\hat{3}}}$, and $\MA_{{\hat{10}},{\hat{1}},{\hat{1}}}$ in this paper. 
The dual graviton $\MA_{{\hat{8}},{\hat{1}}}$ and the potential $\MA_{{\hat{10}},{\hat{1}},{\hat{1}}}$ respectively appear in the worldvolume action of the KK monopole \cite{hep-th/9802199} and the M9-brane \cite{hep-th/9806069,hep-th/9812225,hep-th/9912030,hep-th/0003240}, and their definitions are rather established. 
However, for the $5^3$-brane (which couples to $\MA_{{\hat{9}},{\hat{3}}}$), only the kinetic term has been constructed in \cite{1601.05589} and the WZ term including the potential $\MA_{{\hat{9}},{\hat{3}}}$ has not been known.
Thus the definition of $\MA_{{\hat{9}},{\hat{3}}}$ is still unclear. 

Here, instead of considering brane actions, we define the mixed-symmetry potentials by using the approach of \cite{1909.01335}. 
Namely, we consider the $E_{n(n)}$ $U$-duality-covariant 1-form $\cA_1^I$\,, which appears when the eleven-dimensional spacetime is compactified on an $n$-torus. 
It is uniquely defined (as the generalized graviphoton \cite{1909.01335}), and by parameterizing $\cA_1^I$ in terms of the mixed-symmetry potentials, we can fix the convention for the mixed-symmetry potentials. 
The concrete parameterizations are given in section \ref{sec:1-form}. 
After providing the parameterizations, we can straightforwardly obtain the $T$- and $S$-duality transformation rules as explained in \cite{1909.01335}. 

For convenience, below we summarize the correspondence between each gauge potential and the supersymmetric brane, which electrically couples to the potential:
\begin{align}
\begin{array}{|c|c|c|c|c|}\hline
 \MA_{\hat{3}} & \MA_{\hat{6}} & \MA_{{\hat{8}},{\hat{1}}} & \MA_{{\hat{9}},{\hat{3}}} & \MA_{{\hat{10}},{\hat{1}},{\hat{1}}} \\\hline
 \text{M2} & \text{M5} & \underset{\text{\tiny(MKK)}\vphantom{|}}{6^1} & 5^3 & \underset{\text{\tiny(M9)}\vphantom{|}}{8^{(1,0)}\vphantom{{}^{\big|}}} \\\hline
\end{array} \ .
\end{align}

\subsection{Type IIA supergravity}

In order to obtain type IIA supergravity, we consider the standard 11D--10D map,
\begin{align}
\begin{split}
 \Mg_{ij}\,\rmd x^i\,\rmd x^j &= \Exp{-\frac{2}{3}\,\APhi} \Ag_{mn}\,\rmd x^m\,\rmd x^n +\Exp{\frac{4}{3}\,\APhi}\,(\rmd x^z + \AC_1)^2 \,,
\\
 \MA_{\hat{3}} &= \AC_3 + \AB_2\wedge \rmd x^{\Az} \qquad (m,n=0,\dotsc,9)\,, 
\end{split}
\label{eq:11D-10D}
\end{align}
where $x^{\Az}$ is the coordinate along the M-theory circle. 
Then, we obtain the type IIA Lagrangian
\begin{align}
\begin{split}
 \cL_{\text{IIA}} &= \Exp{-2\APhi}\bigl(* R + 4\,\rmd\APhi\wedge*\rmd\APhi-\tfrac{1}{2}\,\AH_3\wedge *\AH_3\bigr) 
\\
 &\quad - \tfrac{1}{2}\,\bigl(\cG_2\wedge *\cG_2 + \cG_4\wedge *\cG_4 + \AB_2\wedge \rmd \AC_3 \wedge \rmd \AC_3\bigr) \,,
\end{split}
\end{align}
where we have defined
\begin{align}
 \AH_3 \equiv \rmd \AB_2 \,, \qquad 
 \cG_2 \equiv \rmd \AC_1 \,, \qquad 
 \cG_4 \equiv \rmd \AC_3 -\AH_3\wedge \AC_1\,,
\end{align}
and used the identity
\begin{align}
 \MF_{\hat{4}} = \cG_4 + \AH_3\wedge (\rmd x^{\Az}+\AC_1)\,.
\end{align}
Again, the equations of motion for $\AB_2$ and $\AC_3$ are expressed as Bianchi identities
\begin{align}
 \rmd \AH_7 - \cG_2\wedge * \cG_4 - \tfrac{1}{2}\,\cG_4\wedge \cG_4 = 0 \,,\qquad 
 \rmd \cG_6 - \AH_3\wedge \cG_4 =0 \,,
\end{align}
where the dual field strengths are defined by
\begin{align}
 \AH_7\equiv \Exp{-2\APhi}* \AH_3\,,\qquad \cG_6 \equiv - * \cG_4 \,.
\end{align}
Then, we can introduce the dual potentials $\AC_5$ and $\AB_6$ as follows:
\begin{align}
 \cG_6 \equiv \rmd \AC_5 -\AH_3\wedge \AC_3\,, \qquad
 \AH_7 \equiv \rmd \AB_6 - \cG_6 \wedge \AC_1 + \tfrac{1}{2}\,\cG_4\wedge \AC_3 - \tfrac{1}{2}\,\AH_3 \wedge \AC_3 \wedge \AC_1 \,.
\end{align}
Through the electric-magnetic duality, we obtain the following 11D--10D map:
\begin{align}
 \MA_{\hat{6}} = \AB_6 + \bigl(\AC_5 - \tfrac{1}{2!}\, \AC_3\wedge \AB_2\bigr)\wedge \rmd x^{\Az} \,,\qquad 
 \MF_{\hat{7}} = \AH_7 + \cG_6 \wedge (\rmd x^{\Az} + \AC_1) \,.
\end{align}

On the other hand, the equation of motion for $\AC_1$ is expressed as
\begin{align}
 \rmd \cG_8 - \AH_3\wedge \cG_6 =0 \,,\qquad \cG_8 \equiv * \cG_2 \,,
\end{align}
and we can introduce the 7-form potential $\AC_7$ as
\begin{align}
 \cG_8 \equiv \rmd \AC_7 -\AH_3\wedge \AC_3\,. 
\label{eq:G-8-def}
\end{align}
The 11D uplift of this 8-form field strength is discussed in section \ref{sec:gauge} [see Eq.~\eqref{eq:cG9-cG11}]. 

In 11D supergravity, we have introduced non-standard potentials $\MA_{{\hat{8}},{\hat{1}}}$, $\MA_{{\hat{10}},{\hat{1}},{\hat{1}}}$, and $\MA_{{\hat{9}},{\hat{3}}}$. 
Here, we consider the following simple 11D--10D map for these potentials:\footnote{We do not consider overlined potentials, such as $\overline{\AA}_{8}$, which do not couple to supersymmetric branes.}
\begin{align}
\begin{split}
\begin{alignedat}{2}
 \MA_{\hat{8}, 1} &= \AA_{8, 1} + \AA_{7, 1}\wedge \rmd x^{\Az} \,, \qquad&
 \MA_{\hat{8}, \Az} &= \overline{\AA}_{8} + \AA_7 \wedge\rmd x^{\Az} \,,
\\
 \MA_{\hat{9}, 3} &= \AA_{9, 3} + \AA_{8, 3} \wedge \rmd x^{\Az} \,,\qquad&
 \MA_{\hat{9}, 2\Az} &= \overline{\AA}_{9, 2} + \AA_{8, 2} \wedge \rmd x^{\Az} \,,
\\
 \MA_{\hat{10},1,1} &= \AA_{10,1,1} + \AA_{9,1,1}\wedge \rmd x^{\Az} \,, \qquad&
 \MA_{\hat{10},\Az,\Az} &= \overline{\AA}_{10} + \AA_9\wedge \rmd x^{\Az} \,.
\end{alignedat}
\end{split}
\end{align}
The potential $\AA_7$ is related to the R--R 7-form $\AC_7$ introduced in \eqref{eq:G-8-def} and the 9-form $\AA_9$ is related to the standard R--R 9-form $\AC_9$ (see section \ref{sec:standard-T}). 

In the type IIA case, the correspondence between the potentials and the supersymmetric branes are summarized as follows:\footnote{We here ignore the component $\AA_{9, 3}$, which couples to the $5^3_4$-brane.}
\begin{align}
\begin{array}{|c|c|c|c|c|c|c|c|c|c|c|c|c|}\hline
 \AB_2 & \AC_1 & \AC_3 & \AC_5 & \AB_6 & \AC_7 & \AA_{7, 1} & \AA_{8, 1} & \AC_9 & \AA_{8, 2} & \AA_{8, 3} & \AA_{9,1,1} & \AA_{10,1,1} \\\hline
 \text{F1} & \text{D0} & \text{D2} & \text{D4} & \underset{\text{\tiny(NS5)}\vphantom{|}}{5_2} & \text{D6} & \underset{\text{\tiny(KK5A)}\vphantom{|}}{5_2^1} & \underset{\text{\tiny(KK6A)}\vphantom{|}}{6_3^{1}\vphantom{{}^{\big|}}} & \text{D8} & 5^2_2 & 4^3_3 & \underset{\text{\tiny(KK8A)}\vphantom{|}}{7^{(1,0)}_3} & \underset{\text{\tiny(NS9A)}\vphantom{|}}{8^{(1,0)}_4} \\\hline
\end{array}\,. 
\end{align}

\subsection{Type IIB supergravity}

The standard $\SL(2)$ $S$-duality-invariant (pseudo) Lagrangian for type IIB supergravity is
\begin{align}
 \cL_{\text{IIB}} &= *_{\rmE} \sfR + \tfrac{1}{4}\,\BF_{1 \SLa\SLb} \wedge *_{\rmE}\BF^{\SLa\SLb}_1 - \tfrac{1}{2}\,\Bm_{\SLa\SLb}\,\BF^{\SLa}_3\wedge *_{\rmE}\BF^{\SLb}_3
 - \tfrac{1}{4}\, \BF_5\wedge *_{\rmE}\BF_5 + \tfrac{1}{4}\,\epsilon_{\SLa\SLb}\,\BA_4\wedge \BF^{\SLa}_3\wedge \BF^{\SLb}_3\,,
\end{align}
where $\SLa=\SLE{1},\SLE{2}$ are indices of $\SL(2)$ doublets and $(\epsilon_{\SLa\SLb})=(\epsilon^{\SLa\SLb})=\bigl(\begin{smallmatrix} 0 & 1 \\ -1 & 0 \end{smallmatrix}\bigr)$. 
The fundamental fields are $\{\Bg_{mn},\,\Bm_{\SLa\SLb},\,\BA^{\SLa}_2,\,\BA_4\}$, and $\Bg_{mn}$ is the Einstein-frame metric, for which the Hodge star operator is denoted by $*_{\rmE}$\,. 
The scalar field $\Bm_{\SLa\SLb}$ is symmetric $\Bm_{\SLa\SLb}=\Bm_{(\SLa\SLb)}$ and satisfies
\begin{align}
 \Bm^{\SLa}{}_{\SLc}\,\Bm^{\SLc}{}_{\SLb} = - \Bm^{\SLa\SLc}\,\Bm_{\SLc\SLb} = - \delta^{\SLa}_{\SLb} \,,
\end{align}
where we have raised or lowered the $\SL(2)$ indices as $v^{\SLa}=\epsilon^{\SLa\SLb}\,v_{\SLb}$ and $v_{\SLa}= v^{\SLb}\,\epsilon_{\SLb\SLa}$\,. 
The field strengths are defined by
\begin{align}
 \BF^{\SLa\SLb}_1 \equiv \Bm^{\SLa\SLc}\, \rmd \Bm_{\SLc}{}^{\SLb} = \BF^{(\SLa\SLb)}_1 \,,\qquad
 \BF^{\SLa}_3 \equiv \rmd \BA^{\SLa}_2 \,, \qquad
 \BF_5 \equiv \rmd \BA_4 + \tfrac{1}{2}\, \epsilon_{\SLa\SLb}\, \BF^{\SLa}_3\wedge \BA^{\SLb}_2\,,
\end{align}
which satisfy the Bianchi identities
\begin{align}
 \rmd \BF^{\SLa\SLb}_1 + \epsilon_{\SLc\SLd}\,\BF^{\SLa\SLc}_1 \wedge \BF^{\SLd\SLb}_1 = 0 \,,\qquad 
 \rmd \BF^{\SLa}_3 = 0\,,\qquad 
 \rmd \BF_5+ \tfrac{1}{2}\,\epsilon_{\SLa\SLb}\, \BF^{\SLa}_3\wedge \BF^{\SLb}_3 = 0 \,.
\end{align}
The self-duality relation for the 5-form field strength,
\begin{align}
 \BF_5 = *_{\rmE}\BF_5\,,
\label{eq:F5-self}
\end{align}
should be imposed at the level of equations of motion. 

As is well-known, under the self-duality relation \eqref{eq:F5-self} the equation of motion for $\BA_4$ is equivalent to the last Bianchi identity. 
If we additionally define the dual field strengths as\footnote{As noted in \cite{hep-th/9908094,hep-th/0506013}, the triplet $\BF^{\SLa\SLb}_1$ has only two independent components because it satisfies $\Bm_{\SLa\SLb}\,\BF^{\SLa\SLb}_1 = 0$. Then, the duality \eqref{eq:hodge-7-3-9-1} shows that the triplet $\BF^{\SLa\SLb}_9$ also has only two independent components.}
\begin{align}
 \BF^{\SLa}_7 \equiv \Bm^{\SLa}{}_{\SLb}*_{\rmE} \BF^{\SLb}_3 \,, \qquad 
 \BF^{\SLa\SLb}_9 \equiv *_{\rmE} \BF^{\SLa\SLb}_1 \,,
\label{eq:hodge-7-3-9-1}
\end{align}
the equations of motion for $\Bm_{\SLa\SLb}$ and $\BA^{\SLa}_2$ also can be expressed as the Bianchi identities
\begin{align}
 \rmd\BF^{\SLa\SLb}_9 - \BF^{(\SLa}_3\wedge \BF^{\SLb)}_7 = 0\,,\qquad
 \rmd\BF^{\SLa}_7 - \BF^{\SLa}_3\wedge \BF_5 = 0\,.
\end{align}
They suggest us to introduce the higher potentials, $\BA^{\SLa}_6$ and $\BA^{\SLa\SLb}_8$ as
\begin{align}
 \BF^{\SLa}_7 &\equiv \rmd\BA^{\SLa}_6 - \BF^{\SLa}_3\wedge \BA_4 + \tfrac{1}{3!}\,\epsilon_{\SLc\SLd}\,\BF^{\SLc}_3 \wedge\BA^{\SLd}_2\wedge \BA^{\SLa}_2\,,
\\
 \BF^{\SLa\SLb}_9 &\equiv \rmd\BA^{\SLa\SLb}_8 - \BF_3^{(\SLa} \wedge \BA_6^{\SLb)} + \tfrac{1}{4!}\,\epsilon_{\SLc\SLd}\,\BF^{\SLc}_3\wedge \BA^{\SLd}_2\wedge \BA^{\SLa}_2\wedge \BA^{\SLb}_2\,.
\end{align}
In \cite{hep-th/0506013,hep-th/0602280,hep-th/0611036,1004.1348}, a 10-form potential was also introduced by considering the supersymmetry algebra (which is also predicted by $E_{11}$ \cite{hep-th/0511153}),\footnote{Additional 10-form potential $\overline{\BA}^{\SLa}_{10}$ was also introduced there, but here we do not consider this potential because this does not couple to supersymmetric branes.} and the field strength, in our convention, is defined as
\begin{align}
 &\BF^{\SLa\SLb\SLc}_{11} \equiv \rmd \BA^{\SLa\SLb\SLc}_{10} - \BF^{(\SLa}_3\wedge \BA^{\SLb\SLc)}_8 + \tfrac{1}{5!}\,\epsilon_{\SLf\SLg}\,\BF^{\SLf}_3\wedge \BA^{\SLg}_2\wedge \BA^{\SLa}_2\wedge \BA^{\SLb}_2 \wedge \BA^{\SLc}_2 \ (= 0)\,.
\end{align}
This satisfies the Bianchi identity (without considering the dimensionality)
\begin{align}
 \rmd \BF^{\SLa\SLb\SLc}_{11} - \BF^{(\SLa}_3\wedge \BF^{\SLb\SLc)}_9 =0\,.
\end{align}

In this paper, we consider the following set of type IIB fields:
\begin{align}
 \{\Bg_{mn},\,\Bm_{\SLa\SLb},\,\BA^{\SLa}_2,\,\BA_4,\,\BA^{\SLa}_6,\,\BA_{7,1},\,\BA^{\SLa\SLb}_8,\,\BA^{\SLa}_{8,2},\,\BA^{\SLa\SLb\SLc}_{10},\,\BA_{9,2,1}\}\,,
\end{align}
which transform covariantly under $\SL(2)$ $S$-duality transformations. 
At the present stage, definitions of $\BA_{7,1}$, $\BA^{\SLa}_{8,2}$, $\BA_{9,2,1}$, and $\BA^{\SLa\SLb\SLc}_{10}$ are not specified. 
They are defined in section \ref{sec:1-form}. 

In the following discussion, a redefinition of the 6-form potential
\begin{align}
 \bBA^{\SLa}_6 \equiv \BA^{\SLa}_6 - \BA_4\wedge \BA^{\SLa}_2\,, 
\end{align}
makes the $T$-duality rules slightly shorter. 
Thus, the 6-form $\bBA^{\SLa}_6$ rather than $\BA^{\SLa}_6$ is mainly used in this paper. 
For notational consistency, other fields also may be denoted by bold typeface,
\begin{align*}
 \bBA^{\SLa}_2\equiv \BA^{\SLa}_2,\ 
 \bBA_4\equiv \BA_4,\ 
 \bBA_{7,1}\equiv \BA_{7,1},\ 
 \bBA^{\SLa\SLb}_8\equiv \BA^{\SLa\SLb}_8,\ 
 \bBA^{\SLa}_{8,2}\equiv \BA^{\SLa}_{8,2},\ 
 \bBA^{\SLa\SLb\SLc}_{10}\equiv \BA^{\SLa\SLb\SLc}_{10},\ 
 \bBA_{9,2,1}\equiv \BA_{9,2,1}\,.
\end{align*}
The relation between the potentials and the supersymmetric branes are as follows:
\begin{align}
\begin{array}{|c|c|c|c|c|c|c|c|}\hline
 \bBA^{\SLa}_2 & \bBA_4 & \bBA^{\SLa}_6 & \bBA_{7,1} & \bBA^{\SLa\SLb}_{8} & \bBA^{\SLa}_{8,2} & \bBA^{\SLa\SLb\SLc}_{10}& \bBA_{9,2,1} \\\hline
 \text{F1/D1} & \text{D3} & \text{D5/}\!\!\underset{\text{\tiny(NS5)}\vphantom{|}}{5_2\vphantom{{}^{\big|}}} & \underset{\text{\tiny(KK5B)}\vphantom{|}}{5_2^1} & \text{D7}/\!\!\underset{\text{\tiny(NS7B)}\vphantom{|}}{7_3} & 5^2_2/5^2_3 & \text{D9}/\!\!\!\underset{\text{\tiny(NS9B)}\vphantom{|}}{9_4} & \underset{\text{\tiny(KK7B)}\vphantom{|}}{6_3^{(1,1)}} \\\hline
\end{array}\,.
\end{align}
It is noted that an $\SL(2)$ $n$-plet $\bBA_{\cdots}^{\SLa_1\cdots \SLa_{n-1}}$ $(n\geq 2)$ always couples to only two supersymmetric branes. 
The components which couple to supersymmetric branes are $\bBA_{\cdots}^{\SLE{1}\cdots \SLE{1}}$ and $\bBA_{\cdots}^{\SLE{2}\cdots \SLE{2}}$ as discussed in \cite{hep-th/0611036,1009.4657} (see also \cite{1907.07177}).

\subsection{Parameterization of the $U$-duality-covariant 1-form}
\label{sec:1-form}

When 11D supergravity or type II supergravity is compactified to $d$-dimensions, the bosonic fields with one external index $\mu\ (=0,\dotsc,d-1)$ are packaged into the 1-form field $\cA_\mu^I$, where $I$ is the index for the so-called the vector representation of the $U$-duality group $E_{n(n)}$ ($n=11-d$)\,. 
Under the compactification, we decompose the indices in M-theory/type IIB theory as
\begin{align}
 \text{M-theory:}\quad &\{i\}=\{\mu,\underline{i}\} \qquad (\underline{i}=d,\dotsc,8,\Ay,\Az)\,,\
\\
 \text{Type IIB theory:}\quad &\{m\}=\{\mu,\underline{m}\} \qquad (\underline{m}=d,\dotsc,8,\By)\,.
\end{align}
Then the vector index $I$ in M-theory is decomposed into indices of $\SL(n)$ tensors as \cite{hep-th/0104081}
\begin{align}
 (\cA_\mu^I) = \bigl(\cA_\mu^{\underline{i}},\ \tfrac{\cA_{\mu;\underline{i}_1\underline{i}_2}}{\sqrt{2!}},\ \tfrac{\cA_{\mu;\underline{i}_1\cdots \underline{i}_5}}{\sqrt{5!}},\ \tfrac{\cA_{\mu;\underline{i}_1\cdots \underline{i}_7, \underline{k}}}{\sqrt{7!}},\ \tfrac{\cA_{\mu;\underline{i}_1\cdots \underline{i}_8,\underline{k}_1\underline{k}_2\underline{k}_3}}{\sqrt{8!\,3!}},\ \tfrac{\cA_{\mu;\underline{i}_1\cdots \underline{i}_9, \underline{k}, \underline{l}}}{\sqrt{9!}},\cdots \bigr) \,,
\end{align}
where only the relevant components are shown. 
In type IIB theory, the vector index is denoted by $\sfI$ and it is decomposed into indices of $\SL(n-1)\times \SL(2)$ tensors as \cite{hep-th/0107181}
\begin{align}
\begin{split}
 (\bm{\cA}_\mu^\sfI) = \bigl(&\bm{\cA}_\mu^{\underline{m}},\ \bm{\cA}^{\SLa}_{\mu;\underline{m}},\ \tfrac{\bm{\cA}_{\mu;\underline{m}_1\underline{m}_2\underline{m}_3}}{\sqrt{3!}},\ \tfrac{\bm{\cA}^{\SLa}_{\mu;\underline{m}_1\cdots \underline{m}_5}}{\sqrt{5!}},\ \tfrac{\bm{\cA}_{\mu;\underline{m}_1\cdots \underline{m}_6,\underline{p}}}{\sqrt{6!}},
\\
 &\tfrac{\bm{\cA}^{\SLa\SLb}_{\mu;\underline{m}_1\cdots \underline{m}_7}}{\sqrt{7!}},\ \tfrac{\bm{\cA}^{\SLa}_{\mu; \underline{m}_1\cdots \underline{m}_7, \underline{p}_1\underline{p}_2}}{\sqrt{7!\,2!}},\ \tfrac{\bm{\cA}^{\SLa\SLb\SLc}_{\mu; \underline{m}_1\cdots \underline{m}_9}}{\sqrt{9!}},\ \tfrac{\bm{\cA}_{\mu; \underline{m}_1\cdots \underline{m}_8,\underline{p}_1\underline{p}_2,\underline{q}}}{\sqrt{8!\,2!}} ,\cdots \bigr)\,.
\end{split}
\end{align}

Now, we parameterize each component of the 1-form field in terms of the bosonic fields introduced in the last subsections. 
In fact, the 1-form has the universal form \cite{1909.01335}
\begin{align}
 \cA_\mu^I = \MN_{\mu}{}^I + \MA_\mu^{\underline{j}}\, \MN_{\underline{j}}{}^I\quad (\text{M-theory})\,,\qquad 
 \bm{\cA}_\mu^{\sfI} = \BN_{\mu}{}^{\sfI} + \BA_\mu^{\underline{n}}\, \BN_{\underline{n}}{}^{\sfI}\quad (\text{type IIB theory})\,,
\end{align}
where $\MN$/$\BN$ are the 11D/10D fields and $\MA_\mu^{\underline{i}}$/$\BA_\mu^{\underline{m}}$ are the graviphoton, which are defined by
\begin{align}
 \MA_\mu^{\underline{i}}\equiv \bm{g}_{\mu\nu}\,\Mg^{\nu\underline{i}} \qquad \bigl[ (\bm{g}_{\mu\nu}) \equiv (\Mg^{\mu\nu})^{-1} \bigr]\,,
\end{align}
and similar for $\BA_\mu^{\underline{m}}$. 
Therefore, the parameterization of the 1-form field is equivalent to the parameterization of the 11D- or 10D-covariant field $\MN$ or $\BN$\,. 

In this paper, we parameterize the 11D tensors $\{\MN\}=\{\MN_{j}{}^i,\ \tfrac{\MN_{j;i_1i_2}}{\sqrt{2!}},\cdots \}$ as follows:
\begin{align}
 \MN_{j}{}^i &=\delta_j^i \,,
\\
 \MN_{j;i_1i_2} &=\MA_{j i_1i_2} \,,
\\
 \MN_{j;i_1\cdots i_5} &= \MA_{j i_1\cdots i_5} - 5\,\MA_{j[i_1i_2}\, \MA_{i_3i_4i_5]}\,,
\\
 \MN_{j;i_1\cdots i_7, k}
 &\simeq \MA_{j i_1\cdots i_7, k} - 21\, \MA_{j [i_1\cdots i_5}\,\MA_{i_6i_7] k} 
 +35\,\MA_{j [i_1i_2}\,\MA_{i_3i_4i_5}\,\MA_{i_6i_7]k} \,,
\\
\begin{split}
 \MN_{j;i_1\cdots i_8, k_1k_2k_3} &\simeq 
 \MA_{j i_1\cdots i_8, k_1k_2k_3}
 + 3\,\MA_{j \bar{k}_1\bar{k}_2}\, \MA_{i_1\cdots i_8, \bar{k}_3} 
 + 14\,\MA_{j [i_1i_2}\,\MA_{i_3\cdots i_8]} \,\MA_{k_1k_2k_3}
\\
 &\quad - 84\,\MA_{j [i_1|\bar{k}_1|}\,\MA_{i_2 |\bar{k}_2\bar{k}_3|}\,\MA_{i_3\cdots i_8]} 
 - 42\,\MA_{j \bar{k}_1\bar{k}_2}\,\MA_{[i_1i_2|\bar{k}_3|}\,\MA_{i_3\cdots i_8]} 
\\
 &\quad
 + 28\,\MA_{j [i_1\cdots i_5}\,\MA_{i_6i_7i_8] k_1k_2k_3}
 -210\,\MA_{j [i_1i_2}\,\MA_{i_3i_4i_5}\,\MA_{i_6i_7|\bar{k}_1|}\,\MA_{i_8] \bar{k}_2\bar{k}_3} \,,
\end{split}
\label{eq:N-M-9-3}
\\
\begin{split}
 \MN_{j;i_1\cdots i_9, k, l}
 &\simeq \MA_{j i_1\cdots i_9, k, l}
 - 84\,\MA_{j [i_1\cdots i_6 |k,l|}\, \MA_{i_7i_8i_9]}
 +378\,\MA_{j [i_1\cdots i_5}\,\MA_{i_6i_7 |k|}\, \MA_{i_8i_9] l} 
\\
 &\quad -315\,\MA_{j [i_1i_2}\,\MA_{i_3i_4i_5}\,\MA_{i_6i_7|k|}\,\MA_{i_8i_9] l}\,.
\end{split}
\end{align}
Here, the overlined indices are totally antisymmetrized; e.g.~$\MA_{j \bar{k}_1\bar{k}_2}\, \MA_{i_1\cdots i_8, \bar{k}_3} = \MA_{j [\bar{k}_1\bar{k}_2|}\, \MA_{i_1\cdots i_8, |\bar{k}_3]}$. 
In addition, the equality
\begin{align}
 \MN_{j; i_1\cdots i_p, j_1\cdots j_q, k_1\cdots k_r,\cdots} 
 \simeq (\cdots )_{j i_1\cdots i_p j_1\cdots j_q k_1\cdots k_r \cdots} \,,
\end{align}
denotes that it is valid only for the indices satisfying the restriction rule,
\begin{align}
 \{i_1,\dotsc, i_p\} \supset \{j_1,\dotsc, j_q\} \supset \{k_1,\dotsc, k_r\} \supset \cdots\,.
\label{eq:M-restriction}
\end{align}
Since the 1-form $\cA_\mu^I$ is uniquely defined, the above parameterizations uniquely define our bosonic fields, in particular, the mixed-symmetry potentials, $\MA_{\hat{8},\hat{1}}$, $\MA_{\hat{9},\hat{3}}$, and $\MA_{\hat{10},\hat{1},\hat{1}}$\,. 

The detailed procedure, how to determine the above parameterization of $\MN$ is explained in section 2 of \cite{1909.01335}. 
By considering the consistency between the M-theory and the type IIB parameterizations, the above parameterizations are uniquely determined (up to redefinitions of mixed-symmetry potentials). 
The same parameterization can be obtained also by constructing a matrix representation of the $E_{11}$ generators as discussed in section 3 of \cite{1909.01335}. 
In the second approach based on $E_{11}$, the parameterizations will be completely determined without requiring the restriction rule \eqref{eq:M-restriction} (see \cite{1909.01335} for more details). 

Now, let us turn to the type IIB parameterization. 
In type IIB theory, we parameterize the $\SL(2)$-covariant 10D tensors $\{\BN\}$ as follows:
\begin{align}
 \BN_{n}{}^m &= \delta_n^m \,,
\\
 \BN_{n;m}^{\SLa} &= \bBA_{nm}^{\SLa} \,,
\\
 \BN_{n;m_1m_2m_3} &= \bBA_{nm_1m_2m_3} - \tfrac{3}{2}\,\epsilon_{\SLc\SLd}\, \bBA^{\SLc}_{n[m_1}\,\bBA^{\SLd}_{m_2m_3]} \,,
\\
 \BN^{\SLa}_{n;m_1 \cdots m_5} &= \bBA^{\SLa}_{nm_1 \cdots m_5} +5\, \bBA^{\SLa}_{n[m_1}\,\bBA_{m_2\cdots m_6]}
 + 5\,\epsilon_{\SLc\SLd}\, \bBA^{\SLc}_{n[m_1}\, \bBA^{\SLd}_{m_2m_3}\, \bBA^{\SLa}_{m_4m_5]} \,,
\\
\begin{split}
 \BN_{n;m_1\cdots m_6, p} &\simeq \bBA_{n m_1\cdots m_6, p} + \epsilon_{\SLc\SLd}\, \bBA^{\SLc}_{np}\,\bBA^{\SLd}_{m_1\cdots m_6}+10\,\bBA_{n[m_1m_2m_3}\,\bBA_{m_4m_5m_6]p}
\\
 &\quad -30\,\epsilon_{\SLa\SLb}\,\bBA^{\SLa}_{n[m_1}\,\bBA^{\SLb}_{m_2m_3}\,\bBA_{m_4m_5m_6]p} 
\\
 &\quad +\tfrac{15}{2}\,\epsilon_{\SLa\SLb}\,\epsilon_{\SLc\SLd}\,\bBA^{\SLa}_{n[m_1}\,\bBA^{\SLb}_{m_2m_3}\,\bBA^{\SLc}_{m_4m_5}\,\bBA^{\SLd}_{m_6]p} \,,
\end{split}
\\
\begin{split}
 \BN^{\SLa\SLb}_{n;m_1\cdots m_7} &\simeq \bBA^{\SLa\SLb}_{n m_1\cdots m_7}
 - 21\,\bBA^{(\SLa}_{n[m_2\cdots m_5}\, \bBA^{\SLb)}_{m_6m_7]}
 -105\, \bBA_{n[m_2m_2m_3}\, \bBA^{\SLa}_{m_4m_5}\,\bBA^{\SLb}_{m_6m_7]}
\\
 &\quad 
 -105\, \bBA^{(\SLa}_{n[m_1}\,\bBA^{\SLb)}_{m_2m_3}\,\bBA_{m_4\cdots m_7]}
 -\tfrac{105}{4}\, \epsilon_{\SLc\SLd}\,\bBA^{\SLc}_{n[m_1}\,\bBA^{\SLd}_{m_2m_3}\,\bBA^{\SLa}_{m_4m_5}\,\bBA^{\SLb}_{m_6m_7]} \,,
\end{split}
\\
\begin{split}
 \BN^{\SLa}_{n; m_1\cdots m_7, p_1p_2}
 &\simeq \bBA^{\SLa}_{n m_1\cdots m_7, p_1p_2}
 -2\,\bBA^{\SLa}_{n \bar{p}_1}\, \bBA_{m_1\cdots m_7 , \bar{p}_2} 
 \rlap{$\displaystyle{} -35\,\bBA^{\SLa}_{n[m_1m_2m_3| p_1p_2|}\,\bBA_{m_4\cdots m_7]}$}
\\
 &\quad +\tfrac{21}{2}\, \epsilon_{\SLc\SLd}\,\bBA^{\SLa}_{n[m_1\cdots m_5}\,\bBA^{\SLc}_{m_6m_7]}\,\bBA^{\SLd}_{p_1p_2} 
 + 7\, \epsilon_{\SLc\SLd}\, \bBA^{\SLc}_{n\bar{p}_1}\,\bBA^{\SLd}_{[m_1|\bar{p}_2|} \,\bBA^{\SLa}_{m_2\cdots m_7]}
\\
 &\quad +210\,\bBA_{n[m_1m_2|\bar{p}_1|}\,\bBA_{m_3m_4m_5|\bar{p}_2|}\, \bBA^{\SLa}_{m_6m_7]}
\\
 &\quad 
 -105\,\epsilon_{\SLc\SLd}\, \bBA^{\SLc}_{n[m_1}\,\bBA^{\SLd}_{m_2m_3}\,\bBA_{m_4m_5|p_1p_2|}\,\bBA^{\SLa}_{m_6m_7]}
\\
 &\quad 
 +\tfrac{1575}{8}\,\epsilon_{\SLc\SLd}\,\epsilon_{\SLf\SLg}\, \bBA^{\SLc}_{n[m_1}\,\bBA^{\SLd}_{m_2m_3}\,\bBA^{\SLf}_{m_4m_5}\,\bBA^{\SLg}_{|p_1p_2|}\,\bBA^{\SLa}_{m_6m_7]} 
\\
 &\quad +\tfrac{1365}{8} \, \epsilon_{\SLc\SLd}\,\epsilon_{\SLf\SLg}\,\bBA^{\SLc}_{n\bar{p}_1}\,\bBA^{\SLd}_{[m_1m_2}\,\bBA^{\SLf}_{m_3m_4}\,\bBA^{\SLg}_{m_5|\bar{p}_2|}\,\bBA^{\SLa}_{m_6m_7]} 
\\
 &\quad -1470\, \epsilon_{\SLc\SLd}\,\epsilon_{\SLf\SLg}\,\bBA^{\SLc}_{n[m_1}\,\bBA^{\SLd}_{m_2|\bar{p}_1|}\, \bBA^{\SLf}_{m_3m_4}\,\bBA^{\SLg}_{m_5|\bar{p}_2|}\,\bBA^{\SLa}_{m_6m_7]}\,,
\end{split}
\\
\begin{split}
 \BN^{\SLa\SLb\SLc}_{n;m_1\cdots m_9}
 &\simeq \bBA^{\SLa\SLb\SLc}_{n m_1\cdots m_9} 
 -36\, \bBA^{(\SLa\SLb}_{n[m_1\cdots m_7}\, \bBA^{\SLc)}_{m_8m_9]} 
 +378\,\bBA^{(\SLa}_{n[m_1\cdots m_5}\, \bBA^{\SLb}_{m_6m_7}\,\bBA^{\SLc)}_{m_8m_9]}
\\
 &\quad
 +2520\, \bBA_{n[m_1m_2m_3}\, \bBA^{(\SLa}_{m_4m_5}\,\bBA^{\SLb}_{m_6m_7}\,\bBA^{\SLc)}_{m_8m_9]} 
\\
 &\quad
 +1890\, \bBA^{(\SLa}_{n[m_1}\,\bBA^{\SLb}_{m_2m_3}\,\bBA^{\SLc)}_{m_4m_5}\, \bBA_{m_6\cdots m_9]} 
\\
 &\quad
 +189\, \epsilon_{\SLc\SLd}\,\bBA^{\SLc}_{n[m_1}\,\bBA^{\SLd}_{m_2m_3}\,\bBA^{(\SLa}_{m_4m_5}\,\bBA^{\SLb}_{m_6m_7}\,\bBA^{\SLc)}_{m_8m_9]}\,,
\end{split}
\\
\begin{split}
 \BN_{n;m_1\cdots m_8,p_1p_2,q}
 &\simeq
 \bBA_{nm_1\cdots m_8 ,p_1p_2 ,q}
 +\epsilon_{\SLc\SLd}\,\bBA^{\SLc}_{nq}\,\bBA^{\SLd}_{m_1\cdots m_8 , p_1p_2} 
 +56\,\bBA_{n[m_1\cdots m_6| , \bar{p}_1|}\, \bBA_{m_7m_8] \bar{p}_2 q} 
\\
 &\quad
 +168\,\epsilon_{\SLc\SLd}\,\bBA_{n[m_1\cdots m_5| \bar{p}_1 , \bar{p}_2|}\,\bBA^{\SLc}_{m_6m_7}\,\bBA^{\SLd}_{m_8]q} 
 \rlap{$\displaystyle{} 
 -70\,\epsilon_{\SLc\SLd}\,\bBA^{\SLc}_{nq}\,\bBA^{\SLd}_{[m_1\cdots m_4|p_1p_2|}\, \bBA_{m_5\cdots m_8]}$}
\\
 &\quad
 +84\,\epsilon_{\SLa\SLb}\,\epsilon_{\SLc\SLd}\,\bBA^{\SLa}_{nq}\,\bBA^{\SLb}_{[m_1\cdots m_5|\bar{p}_1|}\,\bBA^{\SLc}_{m_6m_7}\,\bBA^{\SLd}_{m_8]\bar{p}_2} 
\\
 &\quad
 -\tfrac{560}{3} \,\bBA_{n[m_1m_2|\bar{p}_1|}\, \bBA_{m_3m_4m_5|\bar{p}_2|}\,\bBA_{m_6m_7m_8]q} 
\\
 &\quad
 +840\, \epsilon_{\SLc\SLd}\,\bBA_{n[m_1m_2|\bar{p}_1|}\,\bBA_{m_3m_4m_5|q|}\, \bBA^{\SLc}_{m_6m_7}\,\bBA^{\SLd}_{m_8]\bar{p}_2} 
\\
 &\quad
 -840\, \epsilon_{\SLc\SLd}\,\bBA_{n[m_1m_2|q|}\,\bBA_{m_3m_4m_5|\bar{p}_1|}\, \bBA^{\SLc}_{m_6m_7}\,\bBA^{\SLd}_{m_8]\bar{p}_2} 
\\
 &\quad
 -210\, \epsilon_{\SLa\SLb}\,\epsilon_{\SLc\SLd}\,\bBA^{\SLa}_{n[m_1}\,\bBA^{\SLb}_{m_2m_3}\, \bBA_{m_4m_5 |p_1 p_2|}\, \bBA^{\SLc}_{m_6m_7}\,\bBA^{\SLd}_{m_8]q} 
\\
 &\quad
 +1470\,\epsilon_{\SLa\SLb}\,\epsilon_{\SLc\SLd}\,\epsilon_{\SLf\SLg}\,\bBA^{\SLa}_{n[m_1}\,\bBA^{\SLb}_{m_2m_3}\,\bBA^{\SLc}_{m_4m_5}\,\bBA^{\SLd}_{|\bar{p}_1\bar{p}_2|}\,\bBA^{\SLf}_{m_6m_7}\,\bBA^{\SLg}_{m_8]q} 
\\
 &\quad
 -23100\,\epsilon_{\SLa\SLb}\,\epsilon_{\SLc\SLd}\,\epsilon_{\SLf\SLg}\,\bBA^{\SLa}_{n[m_1}\,\bBA^{\SLb}_{m_2|\bar{p}_1|}\, \bBA^{\SLc}_{m_3m_4}\,\bBA^{\SLd}_{m_5|\bar{p}_2|}\,\bBA^{\SLf}_{m_6m_7}\,\bBA^{\SLg}_{m_8]q} \,.
\end{split}
\end{align}
Here, the equality
\begin{align}
 \BN^{\SLa_1\cdots\SLa_n}_{n; m_1\cdots m_p, n_1\cdots n_q, p_1\cdots p_r,\cdots} 
 \simeq (\cdots)^{\SLa_1\cdots\SLa_n}_{n m_1\cdots m_p n_1\cdots n_q p_1\cdots p_r \cdots} \,,
\end{align}
denotes that it is valid only for the indices satisfying the restriction rules
\begin{align}
 \{m_1,\dotsc, m_p\} \supset \{n_1,\dotsc, n_q\} \supset \{p_1,\dotsc, p_r\} \supset \cdots\,,\qquad 
 \SLa_1=\cdots=\SLa_n\,.
\label{eq:B-restriction}
\end{align}

Now, let us comment more on the restriction rules given in \eqref{eq:M-restriction} and \eqref{eq:B-restriction}. 
As already mentioned, our parameterizations of $\MN$ and $\BN$ are valid only for the restricted components. 
One of the motivations of this paper is to provide a firm ground to study the worldvolume dynamics of exotic branes. 
For that purpose, it will be enough to consider the restricted components, because the other components, which break the restriction rules, do not couple to any supersymmetric branes \cite{hep-th/0611036,0708.2287,1108.5067,1109.2025,1009.4657}. 
Components satisfying and breaking the rule are contained in different duality orbits, and they can be separated. 
Indeed, in our parameterizations, the restricted components of $\MN$ and $\BN$ are always parameterized by the restricted components of mixed-symmetry potentials. 
For example, in the parameterization of $\MN_{j;i_1\cdots i_8, k_1k_2k_3}$ given in \eqref{eq:N-M-9-3}, as long as the rules $\{i_1,\dotsc, i_8\} \supset\{k_1,k_2,k_3\}$ is satisfied, the dual graviton $\MA_{i_1\cdots i_8, \bar{k}_3}$ appearing on the right-hand side also satisfies the rule $\{i_1,\dotsc, i_8\} \supset\{k_3\}$\,.\footnote{This property can be spoiled by a redefinition; e.g.~$\MA_{i_1\cdots i_9, k_1k_2k_3}\to \MA_{i_1\cdots i_9, k_1k_2k_3}+ \MA_{[i_1i_2i_3}\, \MA_{i_4\cdots i_9]\bar{k}_1\bar{k}_2, \bar{k}_3}$.} 
In this sense, our parameterizations respect the restriction rule, and the $T$-duality rules obtained in the next section connect only the restricted components in type IIA/IIB theories. 
In other words, components breaking the restriction rule do not appear in our $T$-duality rules. 

\section{Duality rules}
\label{sec:duality}

As it has been discussed in \cite{1909.01335}, the two 1-forms $\cA_\mu^I$ and $\bm{\cA}_\mu^\sfI$ are the same object expressed in different bases. 
Indeed, they are related through a constant matrix $S$ as
\begin{align}
 \cA_\mu^I = S^I{}_\sfI\,\bm{\cA}_\mu^\sfI\,,
\end{align}
which is called the linear map \cite{1701.07819} (see also \cite{hep-th/0402140}). 
In order to explain the linear map in more detail, let us further decompose the internal indices in M-theory and type IIB theory as
\begin{align}
 \text{M-theory:}\quad \{\underline{i}\}=\{\sfa,\SLa\}\,,\qquad 
 \text{Type IIB theory:}\quad \{\underline{m}\}=\{\sfa,\By\} \,,
\label{eq:internal-decomposition}
\end{align}
where $\sfa=d,\dotsc,8$ and $\SLa=\Ay,\Az$. 
Under the decomposition, components of the 1-forms $\{\cA_\mu^I\}$ in M-theory and $\{\bm{\cA}_\mu^\sfI\}$ in type IIB theory are decomposed into $\SL(n-2)\times \SL(2)$ tensors. 
Then, the linear maps connect the $\SL(n-2)\times \SL(2)$ tensors in the two theories. 
In this paper, we consider the following linear maps (which are extensions of \cite{1701.07819,1909.01335}):
\begin{align}
\begin{split}
 &\cA_\mu^{\sfa} \overset{\tiny\textcircled{a}}{=} \bm{\cA}_\mu^{\sfa}\,, \quad
 \cA_\mu^{\SLa} \!\overset{\tiny\textcircled{c}\textcircled{e}}{=}\! \bm{\cA}_{\mu;\By}^{\SLa} \,, \quad
 \cA_{\mu;\sfa_1\sfa_2} \overset{\tiny\textcircled{\raisebox{0.3pt}{g}}}{=} \bm{\cA}_{\mu;\sfa_1\sfa_2\By}\,, \quad
 \cA_{\mu;\sfa \SLa} \!\overset{\tiny\textcircled{\raisebox{-0.9pt}{d}}\textcircled{\raisebox{-0.9pt}{f}}}{=}\! \bm{\cA}_{\mu;\sfa}^{\SLb}\,\epsilon_{\SLb\SLa} \,, \quad
 \cA_{\mu;\Ay\Az} \overset{\tiny\textcircled{\raisebox{-0.9pt}{b}}}{=} \bm{\cA}_{\mu}^{\By} \,,
\\
 &\cA_{\mu;\sfa_1\cdots \sfa_5} \overset{\tiny\textcircled{o}}{=} \bm{\cA}_{\mu;\sfa_1\cdots \sfa_5\By,\By} \,, \quad
 \cA_{\mu;\sfa_1\cdots \sfa_4\SLa} \!\overset{\tiny\textcircled{\raisebox{-0.7pt}{i}}\textcircled{n}}{=}\! \bm{\cA}^{\SLb}_{\mu;\sfa_1\cdots \sfa_4\By}\,\epsilon_{\SLb\SLa}\,,\quad
 \cA_{\mu;\sfa_1\sfa_2\sfa_3 \Ay\Az} \overset{\tiny\textcircled{\raisebox{-0.9pt}{h}}}{=} \bm{\cA}_{\mu;\sfa_1\sfa_2\sfa_3} \,,
\\
 &\cA_{\mu;\sfa_1\cdots \sfa_6\SLa,\sfa} \!\overset{\tiny\textcircled{r}\textcircled{t}}{\simeq}\! \bm{\cA}^{\SLb}_{\mu;\sfa_1\cdots \sfa_6\By,\sfa\By}\,\epsilon_{\SLb\SLa} \,,\qquad
 \cA_{\mu;\sfa_1\cdots \sfa_6(\SLa_1,\SLa_2)} \!\overset{\tiny\textcircled{\raisebox{-0.9pt}{k}}\textcircled{x}}{\simeq}\! \bm{\cA}^{\SLb_1\SLb_2}_{\mu;\sfa_1\cdots \sfa_6\By}\,\epsilon_{\SLb_1\SLa_1}\,\epsilon_{\SLb_2\SLa_2} \,,
\\
 &\cA_{\mu;\sfa_1\cdots \sfa_5\Ay\Az,\sfa} \overset{\tiny\textcircled{p}}{\simeq} \bm{\cA}_{\mu;\sfa_1\cdots \sfa_5\By,\sfa}\,,\qquad
 \cA_{\mu;\sfa_1\cdots \sfa_5 \Ay\Az,\SLa} \!\overset{\tiny\textcircled{j}\textcircled{\raisebox{-0.9pt}{\~n}}}{=}\! \bm{\cA}_{\mu;\sfa_1\cdots \sfa_5}^{\SLb}\,\epsilon_{\SLb\SLa} \,,
\\
 &\cA_{\mu;\sfa_1\cdots \sfa_6\Ay\Az,\sfb_1\sfb_2\SLa} \!\overset{\tiny\textcircled{s}\textcircled{u}}{\simeq}\! \bm{\cA}^{\SLb}_{\mu;\sfa_1\cdots \sfa_6\By,\sfb_1\sfb_2}\,\epsilon_{\SLb\SLa}\,,\qquad
 \cA_{\mu;\sfa_1\cdots \sfa_6\Ay\Az,\sfa\Ay\Az} \overset{\tiny\textcircled{q}}{\simeq} \bm{\cA}_{\mu;\sfa_1\cdots \sfa_6,\sfa} \,,
\\
 &\cA_{\mu;\sfa_1\cdots \sfa_7,\sfa} \overset{\tiny\textcircled{v}}{\simeq} \bm{\cA}_{\mu;\sfa_1\cdots \sfa_7\By,\sfa\By,\By} \,,\qquad
 \cA_{\mu;\sfa_1\cdots \sfa_7\Ay\Az,\sfa,\sfb} \overset{\tiny\textcircled{w}}{\simeq} \bm{\cA}_{\mu;\sfa_1\cdots \sfa_7\By,\sfa\By,\sfb} \,,
\\
 &\cA_{\mu;\sfa_1\cdots \sfa_7\Ay\Az, (\SLa_1, \SLa_2)} \!\overset{\tiny\textcircled{l}\textcircled{y}}{\simeq}\! \bm{\cA}^{\SLb_1\SLb_2}_{\mu;\sfa_1\cdots \sfa_7} \,\epsilon_{\SLb_1\SLa_1}\,\epsilon_{\SLb_2\SLa_2} \,,\quad
 \rlap{$\displaystyle \cA_{\mu;\sfa_1\cdots \sfa_8(\SLa_1,\SLa_2,\SLa_3)} \!\overset{\tiny\textcircled{m}\textcircled{z}}{\simeq}\! \bm{\cA}^{\SLb_1\SLb_2\SLb_3}_{\mu;\sfa_1\cdots \sfa_8\By} \,\epsilon_{\SLb_1\SLa_1}\,\epsilon_{\SLb_2\SLa_2}\,\epsilon_{\SLb_3\SLa_3}\,.$}
\end{split}
\label{eq:linear-map}
\end{align}
These relate the M-theory fields (left-hand side) and the type IIB fields (right-hand side), and by rewriting the M-theory fields in terms of type IIA fields, we obtain the $T$-duality rules between type IIA/IIB theories. 
By decomposing the indices $\SLa,\SLb$ into $\Ay\equiv\SLE{1}$ and $\Az\equiv\SLE{2}$, and taking into account of the restriction rule \eqref{eq:M-restriction}, we find that there are 27 linear maps.\footnote{For example, the restriction rule \eqref{eq:M-restriction} for $\cA_{\mu;\sfa_1\cdots \sfa_8(\SLa_1,\SLa_2,\SLa_3)}$ is $\SLa_1=\SLa_2=\SLa_3$ and it gives two linear maps; $\cA_{\mu;\sfa_1\cdots \sfa_8(\Ay,\Ay,\Ay)} \overset{\tiny\textcircled{z}}{=} -\bm{\cA}^{\SLE{222}}_{\mu;\sfa_1\cdots \sfa_8\By}$ and $\cA_{\mu;\sfa_1\cdots \sfa_8(\Az,\Az,\Az)} \overset{\tiny\textcircled{m}}{=} \bm{\cA}^{\SLE{111}}_{\mu;\sfa_1\cdots \sfa_8\By}$.} 
They correspond to the 27 $T$-duality lines \textcircled{a}--\textcircled{z} depicted in Figure \ref{fig:duality-web}. 
In the following subsections, we obtain the 27 $T$-duality rules from the above linear maps. 

Before proceeding, let us comment on the second restriction rule in type IIB theory \eqref{eq:B-restriction}. 
Apparently, it looks different from the first restriction rule, but in fact both of them can be understood as a consequence of the M-theory rule \eqref{eq:M-restriction} \cite{1907.07177}. 
As an example, let us consider the last linear map $\cA_{\mu;\sfa_1\cdots \sfa_8(\SLa_1,\SLa_2,\SLa_3)} \simeq \bm{\cA}^{\SLb_1\SLb_2\SLb_3}_{\mu;\sfa_1\cdots \sfa_8\By} \,\epsilon_{\SLb_1\SLa_1}\,\epsilon_{\SLb_2\SLa_2}\,\epsilon_{\SLb_3\SLa_3}$. 
The M-theory rule requires
\begin{align}
 \{\sfa_1,\cdots, \sfa_8,\SLa_1\}\supset \{\SLa_2\} \supset \{\SLa_3\}\,,
\end{align}
and this leads to $\SLa_1=\SLa_2=\SLa_3$\,. 
Then, the corresponding type IIB field $\bm{\cA}^{\SLb_1\SLb_2\SLb_3}_{\mu;\sfa_1\cdots \sfa_8\By}$ need to satisfy $\SLb_1=\SLb_2=\SLb_3$ and the second restriction rule in type IIB theory \eqref{eq:B-restriction} is derived. 

\subsection{Standard potentials}
\label{sec:standard-T}

Let us begin with a comparison of the linear maps \textcircled{a}--\textcircled{m} with the standard $T$-duality rules. 
For this purpose, we parameterize the type IIA fields and the $S$-duality-covariant type IIB fields by using familiar fields. 
We parameterize the 7-form and 9-form in type IIA theory as
\begin{align}
 \AA_7 = \AC_7 - \tfrac{1}{3!}\,\AC_3\wedge\AB_2\wedge\AB_2\,,\qquad
 \AA_9 = \AC_9 - \tfrac{1}{4!}\,\AC_3\wedge\AB_2\wedge\AB_2\wedge\AB_2 \,,
\label{eq:IIA-7-9}
\end{align}
and parameterize the type IIB fields as follows:
\begin{align}
 (\Bm_{\SLa\SLb}) &\equiv \Exp{\BPhi} \begin{pmatrix} \Exp{-2\,\BPhi} +(\BC_0)^2 & \BC_0 \\ \BC_0 & 1 \end{pmatrix},\quad
 (\BA^{\SLa}_2) \equiv \begin{pmatrix} \BB_2 \\\ -\BC_2\end{pmatrix} ,\quad
 \BA_4 \equiv \BC_4 - \tfrac{1}{2}\,\BC_2\wedge \BB_2\,, 
\label{eq:IIB-param}
\\
 (\BA^{\SLa}_6) &\equiv \begin{pmatrix} \BC_6 - \tfrac{1}{3!}\, \BC_2 \wedge \BB_2\wedge \BB_2 \\ -\bigl(\BB_6 - \tfrac{2}{3!}\,\BB_2\wedge \BC_2 \wedge \BC_2\bigr) \end{pmatrix} ,\quad 
 (\bBA^{\SLa}_6) \equiv \begin{pmatrix} \bBC_6 \\ - \bBD_6 \end{pmatrix},
\\
 (\BA^{\SLa\SLb}_8) &\simeq \begin{pmatrix} \BA^{\SLE{11}}_8 \\ \BA^{\SLE{22}}_8 \end{pmatrix} 
 \equiv \begin{pmatrix}
 \BC_8 - \tfrac{1}{4!}\, \BC_2\wedge \BB_2 \wedge \BB_2 \wedge \BB_2 \\
 \BE_8 - \tfrac{3}{4!}\, \BB_2\wedge \BC_2 \wedge \BC_2 \wedge \BC_2 \end{pmatrix}, \quad
 (\BA^{\SLa}_{8,2}) \equiv \begin{pmatrix} \bBD_{8,2} \\\ -\bBE_{8,2} \end{pmatrix} , 
\\
 (\BA^{\SLa\SLb\SLc}_{10}) &\simeq \begin{pmatrix} \BA^{\SLE{111}}_{10} \\ \BA^{\SLE{222}}_{10} \end{pmatrix} 
 \equiv \begin{pmatrix}
 \BC_{10} - \tfrac{1}{5!}\,\BC_2 \wedge \BB_2 \wedge \BB_2 \wedge \BB_2 \wedge \BB_2 \\
 -\bigl(\BF_{10} - \tfrac{4}{5!}\,\BB_2 \wedge \BC_2 \wedge \BC_2 \wedge \BC_2 \wedge \BC_2 \bigr) \end{pmatrix}.
\label{eq:10-form-param}
\end{align}
These parameterizations are given such that the linear maps \textcircled{a}--\textcircled{m} in \eqref{eq:linear-map} reproduce the standard $T$-duality rules for NS--NS fields,
\begin{align}
\begin{split}
 \AA_\mu^a &\overset{\text{A--B}}{=} \BA_\mu^a\,,\qquad 
 \AA_\mu^{\Ay} \overset{\text{A--B}}{=} \BB_{\mu\By} + \BA_\mu^{p}\,\BB_{p\By} \,, \qquad
 \BA_\mu^{\By} \overset{\text{B--A}}{=} \AB_{\mu \Ay} + \AA_\mu^{p}\,\AB_{p\Ay} \,,
\\
 \AB_{ab} &\overset{\text{A--B}}{=} \BB_{ab} - \tfrac{\BB_{a\By}\,\Bg_{b\By}-\Bg_{a\By}\,\BB_{b\By}}{\Bg_{\By\By}}\,,\qquad 
 \AB_{a\Ay} \overset{\text{A--B}}{=} -\tfrac{\Bg_{a\By}}{\Bg_{\By\By}} \,, 
\\
 \BB_{ab} &\overset{\text{B--A}}{=} \AB_{ab} - \tfrac{\AB_{a\Ay}\,\Ag_{b\Ay}-\Ag_{a\Ay}\,\AB_{b\Ay}}{\Ag_{\Ay\Ay}}\,,\qquad 
 \BB_{ay} \overset{\text{B--A}}{=} -\tfrac{\Ag_{a\Ay}}{\Ag_{\Ay\Ay}} \,, 
\end{split}
\end{align}
and R--R fields,
\begin{align}
\begin{split}
 \AC_{a_1\cdots a_{n-1}\Ay}&\overset{\text{A--B}}{=} \BC_{a_1\cdots a_{n-1}} - \tfrac{(n-1)\,\BC_{[a_1\cdots a_{n-2}|\By|}\,\Bg_{a_{n-1}]\By}}{\Bg_{\By\By}}\,,
\\
 \AC_{a_1\cdots a_n} &\overset{\text{A--B}}{=} \BC_{a_1\cdots a_n\By} - n\, \BC_{[a_1\cdots a_{n-1}}\, \BB_{a_n]\By} - \tfrac{n\,(n-1)\,\BC_{[a_1\cdots a_{n-2}|\By|}\, \BB_{a_{n-1}|\By|}\,\Bg_{a_n]\By}}{\Bg_{\By\By}}\,,
\\
 \BC_{a_1\cdots a_{n-1}\By}&\overset{\text{B--A}}{=} \AC_{a_1\cdots a_{n-1}} - \tfrac{(n-1)\,\AC_{[a_1\cdots a_{n-2}|\Ay|}\,\Ag_{a_{n-1}]\Ay}}{\Ag_{\Ay\Ay}}\,,
\\
 \BC_{a_1\cdots a_n} &\overset{\text{B--A}}{=} \AC_{a_1\cdots a_n\Ay} - n\, \AC_{[a_1\cdots a_{n-1}}\, \AB_{a_n]\Ay} - \tfrac{n\,(n-1)\,\AC_{[a_1\cdots a_{n-2}|\Ay|}\, \AB_{a_{n-1}|\Ay|}\,\Ag_{a_n]\Ay}}{\Ag_{\Ay\Ay}}\,. 
\end{split}\end{align}
Here, the indices $a_1,a_2,\cdots$ are 9D indices, which are orthogonal to the $T$-duality direction $\Ay$ or $\By$\,. 
On the other hand, in the linear map \eqref{eq:linear-map}, the indices are restricted to $\sfa_1,\sfa_2,\cdots$ which run over the internal $(n-2)$ dimensions [recall \eqref{eq:internal-decomposition}]. 
By assuming that the $T$-duality rules have the 9D covariance, we have extended the linear map by replacing $\sfa$ with $a$\,. 
This is always assumed in the following discussion. 

Under the above parameterizations, the field strengths in type IIB supergravity become
\begin{align}
\begin{split}
 &\bigl(\BF^{\SLa\SLb}_1\bigr) = \begin{pmatrix} 1 & 0 \\ -\BC_0 & 1 \end{pmatrix} \begin{pmatrix} \Exp{2\BPhi}\cG_1 & \rmd\BPhi \\ \rmd\BPhi & - \cG_1 \end{pmatrix} \begin{pmatrix} 1 & -\BC_0 \\ 0 & 1 \end{pmatrix} , \quad
 \bigl(\BF^{\SLa}_3\bigr) = \begin{pmatrix} 1 & 0 \\ -\BC_0 & 1 \end{pmatrix} \begin{pmatrix} \BH_3 \\ -\cG_3 \end{pmatrix}, 
\\
 &\BF_5 = \cG_5 \,, \qquad
 \bigl(\BF^{\SLa}_7\bigr) = \begin{pmatrix} 1 & 0 \\ -\BC_0 & 1 \end{pmatrix} \begin{pmatrix} \cG_7 \\ -\BH_7 \end{pmatrix} , \qquad
 \bigl(\BF^{\SLa\SLb}_9\bigr) \simeq \begin{pmatrix} \BF^{\SLE{11}}_9 \\ \BF^{\SLE{22}}_9 \end{pmatrix} = \begin{pmatrix} \cG_9 \\ \BH_9 \end{pmatrix}\,,
\end{split}
\end{align}
where we have defined
\begin{align}
 \BH_3 &\equiv \rmd \BB_2\,,\qquad \cG_{2p+1} \equiv \rmd \BC_{2p} - \BH_3\wedge \BC_{2p-2}\quad (\BC_{-2}\equiv 0)\,,
\\
 \BH_7 &\equiv \rmd \BB_6 - \BC_4\wedge\rmd \BC_2 -\tfrac{1}{2}\,\BC_2\wedge \BC_2\wedge \BH_3 - \BC_0\,\cG_7\,,
\\
 \BH_9 &\equiv \rmd \BE_8 - \BB_6 \wedge \rmd\BC_2 - \tfrac{1}{3!}\,\BH_3\wedge \BC_2\wedge \BC_2\wedge \BC_2\,.
\end{align}
By using the Hodge star operator $*$ in the string frame ($*_\rmE \alpha_p=\Exp{\frac{p-5}{2}\BPhi}*\alpha_p$), we obtain
\begin{align}
 &\cG_p = (-1)^{\frac{p(p-1)}{2}} *\cG_{10-p}\,,\qquad \BH_7 = \Exp{-2\BPhi} * \BH_3 \,,
\\
 &\BH_9 = -2\Exp{-2\BPhi}\BC_0 *\rmd\BPhi + \bigl(\BC_0^2 - \Exp{-2\BPhi}\bigr)\,* \cG_1\,.
\end{align}
The electric-magnetic duality for $\BH_9$ is considered further in section \ref{sec:S-duality}. 

For later convenience, we also define the following 6-form, which has been used in \cite{0712.3235}
\begin{align}
 \cB_6 \equiv \BB_6 - \BC_4\wedge \BC_2 \,.
\end{align}
We also introduce several redefinitions of the dual graviton. 
The type IIB dual graviton $\cN_{7,1}$\footnote{In the literature (e.g.~\cite{hep-th/9806169,0712.3235}), the last index of the dual graviton is supposed to be a particular isometry direction, and it is not written down explicitly. Accordingly, the dual graviton is treated as a 7-form.} introduced in \cite{0712.3235} is related to our $\BA_{7, 1}$ as
\begin{align}
 \cN_{m_1 \cdots m_7, n} &\simeq \BA_{m_1 \cdots m_7, n} -7\,\BB_{[m_1\cdots m_6}\,\BB_{m_7]n} 
 -\tfrac{105}{4}\, \BC_{[m_1\cdots m_4}\,(\BB_{m_5m_6}\,\BC_{m_7]n}-3\,\BC_{m_5m_6}\,\BB_{m_7]n}) 
\nn\\
 &\quad -\tfrac{315}{4}\, \BC_{[m_1m_2}\,\BC_{m_3m_4}\,\BB_{m_5m_6}\,\BB_{m_7]n}\,. 
\end{align}
As we see below, this is useful to simplify the $T$-duality rules (although the $S$-duality covariance is not manifest). 
In M-theory, we introduce
\begin{align}
 \bm{\MA}_{i_1\cdots i_8 , j} \equiv \MA_{i_1\cdots i_8 , j} - \tfrac{56}{3}\, \MA_{[i_1\cdots i_6}\,\MA_{i_7i_8] j}\,,
\end{align}
and define $\pmb{\AA}_{7,1}$ as $\pmb{\AA}_{m_1\cdots m_7,n} \equiv \bm{\MA}_{m_1\cdots m_7\Az,n}$. 
More explicitly, we have
\begin{align}
 \pmb{\AA}_{m_1\cdots m_7,n} &\simeq \AA_{m_1 \cdots m_7, n}
 +\tfrac{14}{3}\, \AB_{[m_1\cdots m_6}\,\AB_{m_7] n} 
\nn\\
 &\quad - 14\, \AC_{[m_1\cdots m_5}\,\AC_{m_6m_7]n}
 + 70\, \AC_{[m_1m_2m_3}\,\AC_{m_4m_5|n|}\,\AB_{m_6m_7]} \,.
\end{align}
The type IIA dual graviton $\pmb{\AA}_{7,n}$ associated with a Killing direction (i.e.~isometry direction) $n$ corresponds to $\cN^{(7)}$ of \cite{0712.3235}, and it is also useful to simplify the $T$-duality rules. 

\subsection{Potentials for solitonic five branes}

In this subsection, we consider the potentials that couple to the solitonic 5-branes (i.e.~5-branes with tensions $T\propto g_s^{-2}$). 
We begin by reproducing the known $T$-duality rules \textcircled{n}--\textcircled{p}. 
There, we demonstrate that redefinitions of potentials can make the $T$-duality rules simpler. 
After reproducing the known results, we obtain new $T$-duality rules \textcircled{q}--\textcircled{s}. 
In section \ref{sec:D-potential}, we find the field redefinitions, which make the $T$-duality rules considerably simple, and discuss the relation to the potentials studied in the context of the double field theory (DFT) \cite{hep-th/9302036,hep-th/9305073,hep-th/9308133,0904.4664,1006.4823}. 

\paragraph{\underline{$T$-duality rule \textcircled{n}: $5_2\ (\text{IIA})\leftrightarrow 5_2\ (\text{IIB})$}\\}
By substituting our parameterizations into the linear map $\cA_{\mu;\sfa_1\cdots \sfa_4\Ay} \!\overset{\tiny\textcircled{n}}{=}\! -\bm{\cA}^{\SLE{2}}_{\mu;\sfa_1\cdots \sfa_4\By}$, we find
\begin{align}
 \AB_{a_1\cdots a_5\Ay}&\overset{\text{A--B}}{=} \BB_{a_1\cdots a_5 \By}
 -5\,\BC_{[a_1a_2a_3|\By|}\,\BC_{a_4a_5]}
 -5\,\bigl(\BC_{[a_1\cdots a_4}-\tfrac{2\,\BC_{[a_1a_2a_3|\By|}\,\Bg_{a_4|\By|}}{\Bg_{\By\By}}\bigr)\,\BC_{a_5]\By} \,,
\\
 \BB_{a_1\cdots a_5\By}
 &\overset{\text{B--A}}{=}\AB_{a_1\cdots a_5\Ay}
 +5\,\AC_{[a_1\cdots a_4|\Ay|}\,\bigl(\AC_{a_5]}-\tfrac{\AC_{|\Ay|}\,\Ag_{a_5]\Ay}}{\Ag_{\Ay\Ay}}\bigr)
\nn\\
 &\quad\ +5\,\bigl(\AC_{[a_1a_2a_3}-\tfrac{3\,\AC_{[a_1a_2|\Ay|}\,\Ag_{a_3|\Ay|}}{\Ag_{\Ay\Ay}}\bigr)\,\AC_{a_4a_5]\Ay} \,.
\end{align}
They have been obtained in \cite{hep-th/9806169} (see Appendix \ref{app:sugra} for their conventions). 

They can be simplified by using the 6-form $\cB_6$ \cite{0712.3235}
\begin{align}
 \AB_{a_1\cdots a_5\Ay}&\overset{\text{A--B}}{=} \cB_{a_1\cdots a_5 \By}
 +5\,\BC_{[a_1a_2a_3|\By|}\, \bigl(\BC_{a_4a_5]}- \tfrac{2\, \BC_{a_4|\By|}\,\Bg_{a_5]\By}}{\Bg_{\By\By}}\bigr) \,,
\\
 \cB_{a_1\cdots a_5\By}
 &\overset{\text{B--A}}{=}\AB_{a_1\cdots a_5\Ay}
 - 5\,\bigl(\AC_{[a_1a_2a_3}-\tfrac{3\,\AC_{[a_1a_2|\Ay|}\,\Ag_{a_3|\Ay|}}{\Ag_{\Ay\Ay}}\bigr)\,\AC_{a_4a_5]\Ay} \,. 
\end{align}

On the other hand, if we employ the $S$-duality-covariant fields, we find
\begin{align}
 \AB_{a_1\cdots a_5\Ay}&\overset{\text{A--B}}{=} \bBD_{a_1\cdots a_5 \By}
 +5\,\bBA_{[a_1a_2a_3|\By|}\,\bigl(\BC_{a_4a_5]} - \tfrac{2\,\BC_{a_4|\By|}\,\Bg_{a_5]\By}}{\Bg_{\By\By}}\bigr)
\nn\\
 &\quad\ +\tfrac{5}{2}\,\epsilon_{\SLc\SLd}\, \bBA^{\SLc}_{[a_1a_2}\,\bBA^{\SLd}_{a_3|\By|}\,\bigl(\BC_{a_4a_5]} - \tfrac{6\,\BC_{a_4|\By|}\,\Bg_{a_5]\By}}{\Bg_{\By\By}}\bigr)\,,
\label{eq:B6-B6-S}
\\
 \bBD_{a_1\cdots a_5 \By}
 &\overset{\text{B--A}}{=} \AB_{a_1\cdots a_5\Ay}
 -5\,\bigl(\AC_{[a_1a_2a_3}-\tfrac{2\,\AC_{[a_1a_2|\Ay|}\,\Ag_{a_3|\Ay|}}{\Ag_{\Ay\Ay}}\bigr)\,\AC_{a_4a_5]\Ay} 
\nn\\
 &\quad\ +10\,\AC_{[a_1a_2|\Ay|}\,\bigl(\AC_{a_3}-\tfrac{\AC_{|\Ay|}\,\Ag_{a_3|\Ay|}}{\Ag_{\Ay\Ay}}\bigr)\,\AB_{a_4a_5]}\,.
\end{align}
Since our linear maps \eqref{eq:linear-map} have the $S$-duality covariance, the $T$-duality rules are covariant under $S$-duality. 
In the present example, we can uplift the $T$-duality rule \eqref{eq:B6-B6-S} into
\begin{align}
 \MA_{a_1\cdots a_5\SLa}&\overset{\text{M--B}}{=} \Bigl[\bBA^{\SLb}_{a_1\cdots a_5 \By}
 +5\,\bBA_{[a_1a_2a_3|\By|}\,\bigl(\BA^{\SLb}_{a_4a_5]} - \tfrac{2\,\BA^{\SLb}_{a_4|\By|}\,\Bg_{a_5]\By}}{\Bg_{\By\By}}\bigr)
\nn\\
 &\qquad +\tfrac{5}{2}\,\epsilon_{\SLc\SLd}\, \bBA^{\SLc}_{[a_1a_2}\,\bBA^{\SLd}_{a_3|\By|}\,\bigl(\BA^{\SLb}_{a_4a_5]} - \tfrac{6\,\BA^{\SLb}_{a_4|\By|}\,\Bg_{a_5]\By}}{\Bg_{\By\By}}\bigr)\Bigr]\,\epsilon_{\SLb\SLa}\,,
\end{align}
which indeed reproduce \eqref{eq:B6-B6-S} by choosing $\SLa=\Ay$. 
On the other hand, $\SLa=\Az$ gives
\begin{align}
 \AC_{a_1\cdots a_5} - 5\,\AC_{[a_1a_2a_3}\,\AB_{a_4a_5]} &\overset{\text{A--B}}{=} \bBC_{a_1\cdots a_5 \By}
 +5\,\bBA_{[a_1a_2a_3|\By|}\,\bigl(\BB_{a_4a_5]} - \tfrac{2\,\BB_{a_4|\By|}\,\Bg_{a_5]\By}}{\Bg_{\By\By}}\bigr)
\nn\\
 &\qquad +\tfrac{5}{2}\,\epsilon_{\SLc\SLd}\, \bBA^{\SLc}_{[a_1a_2}\,\bBA^{\SLd}_{a_3|\By|}\,\bigl(\BB_{a_4a_5]} - \tfrac{6\,\BB_{a_4|\By|}\,\Bg_{a_5]\By}}{\Bg_{\By\By}}\bigr) \,,
\end{align}
which corresponds to the $T$-duality rule for R--R potentials obtained before. 
Each of the $T$-duality rules obtained in this paper has this kind of $S$-dual partner. 

\paragraph{\underline{$T$-duality rule \textcircled{\raisebox{-0.9pt}{\~n}}: $5^1_2\ (\text{IIA})\leftrightarrow 5_2\ (\text{IIB})$}\\}
From the linear map $\cA_{\mu;\sfa_1\cdots \sfa_5 \Ay\Az,\Ay} \!\overset{\tiny\textcircled{\raisebox{-0.9pt}{\~n}}}{=}\! -\bm{\cA}_{\mu;\sfa_1\cdots \sfa_5}^{\SLE{2}}$, we obtain \cite{hep-th/9806169}
\begin{align}
 \AA_{a_1\cdots a_6\Ay, \Ay}&\overset{\text{A--B}}{=} \BB_{a_1\cdots a_6}
 - \tfrac{6\,\BB_{[a_1\cdots a_5|\By|}\,\Bg_{a_6]\By}}{\Bg_{\By\By}}
 - 30\, \BB_{[a_1a_2}\,\BC_{a_3a_4}\, \bigl(\BC_{a_5a_6]} -\tfrac{4\,\BC_{a_5|\By|}\,\Bg_{a_6]\By}}{\Bg_{\By\By}} \bigr)
\nn\\
 &\quad\ + \tfrac{20\, \BC_{[a_1a_2a_3|\By|}\, \BC_{a_4a_5}\,\Bg_{a_6]\By}}{\Bg_{\By\By}} \,,
\\
 \BB_{a_1\cdots a_6}
 &\overset{\text{B--A}}{=}
 \AA_{a_1\cdots a_6\Ay,\Ay}
 -6\,\AB_{[a_1\cdots a_5|\Ay|}\,\AB_{a_6]\Ay}
 -30\,\AC_{[a_1\cdots a_4|\Ay|}\,\bigl(\AC_{a_5}-\tfrac{\AC_{|\Ay|}\,\Ag_{a_5]\Ay}}{\Ag_{\Ay\Ay}} \bigr)\,\AB_{a_6]\Ay}
\nn\\
 &\quad -10\,\AC_{[a_1a_2a_3}\,\AC_{a_4a_5|\Ay|}\,\AB_{a_6]\Ay}
 +30\,\AC_{[a_1a_2|\Ay|}\,\AC_{a_3a_4|\Ay|}\,\bigl(\AB_{a_5a_6]} -\tfrac{3\,\AB_{a_5|\Ay|}\,\Ag_{a_6]\Ay}}{\Ag_{\Ay\Ay}}\bigr) \,.
\end{align}
By using the potentials $\pmb{\AA}_{7,1}$ and $\cB_6$, we can simplify the duality rules as \cite{0712.3235}
\begin{align}
 \pmb{\AA}_{a_1\cdots a_6\Ay, \Ay}&\overset{\text{A--B}}{=} \cB_{a_1\cdots a_6}
 - \tfrac{2\,\cB_{[a_1\cdots a_5|\By|}\,\Bg_{a_6]\By}}{\Bg_{\By\By}}
 + 5\,\BC_{[a_1\cdots a_4}\, \bigl(\BC_{a_5a_6} - \tfrac{2\,\BC_{a_5|\By|}\,\Bg_{a_6]\By}}{\Bg_{\By\By}}\bigr) \,,
\\
 \cB_{a_1\cdots a_6}
 &\overset{\text{B--A}}{=}
 \pmb{\AA}_{a_1\cdots a_6\Ay,\Ay}
 -2\,\AB_{[a_1\cdots a_5|\Ay|}\,\AB_{a_6]\Ay}
 -5\,\AC_{[a_1\cdots a_4|\Ay|}\, \AC_{a_5a_6]\Ay}
\nn\\
 &\quad\ 
 +30\,\bigl(\AC_{[a_1a_2a_3}-\tfrac{3\,\AC_{[a_1a_2|\Ay|}\,\Ag_{a_3|\Ay|}}{\Ag_{\Ay\Ay}}\bigr)\,\AC_{a_4a_5|\Ay|}\,\AB_{a_6]\Ay} \,.
\end{align}
Instead, if we use the $S$-duality covariant fields, we find
\begin{align}
 \AA_{a_1\cdots a_6\Ay, \Ay}&\overset{\text{A--B}}{=} \bBD_{a_1\cdots a_6}
 - \tfrac{6\,\bBD_{[a_1\cdots a_5|\By|}\,\Bg_{a_6]\By}}{\Bg_{\By\By}}
 +15\, \bBA_{[a_1\cdots a_4}\, \bigl(\BC_{a_5a_6]} -\tfrac{2\,\BC_{a_5|\By|}\,\Bg_{a_6]\By}}{\Bg_{\By\By}} \bigr)
\nn\\
 &\quad\ 
 - \tfrac{40\, \bBA_{[a_1a_2a_3|\By|}\,\BC_{a_4a_5}\,\Bg_{a_6]\By}}{\Bg_{\By\By}}
 - \tfrac{30\, \epsilon_{\SLc\SLd}\,\BC_{[a_1a_2} \, \bBA^{\SLc}_{a_3a_4}\,\bBA^{\SLd}_{a_5|\By|}\,\Bg_{a_6]\By}}{\Bg_{\By\By}}\,,
\\
 \bBD_{a_1\cdots a_6}
 &\overset{\text{B--A}}{=}
 \AA_{a_1\cdots a_6\Ay,\Ay}
 -6 \,\AB_{[a_1\cdots a_5|\Ay|}\,\AB_{a_6]\Ay} 
 -15\,\AC_{[a_1\cdots a_4|\Ay|}\, \AC_{a_5a_6]\Ay} 
\nn\\
 &\quad\ 
 +45\,\AC_{[a_1a_2|\Ay|}\,\AC_{a_3a_4|\Ay|}\,\bigl(\AB_{a_5a_6}-\tfrac{2\,\AB_{a_5|\Ay|}\,\Ag_{a_6]\Ay}}{\Ag_{\Ay\Ay}} \bigr) 
\nn\\
 &\quad\ 
 +50\,\bigl(\AC_{[a_1a_2a_3}-\tfrac{3\,\AC_{[a_1a_2|\Ay|}\,\Ag_{a_3|\Ay|}}{\Ag_{\Ay\Ay}} \bigr)\,\AC_{a_4a_5|\Ay|}\,\AB_{a_6]\Ay} 
\nn\\
 &\quad\ 
 -60\,\AC_{[a_1a_2|\Ay|}\,\bigl(\AC_{a_3}-\tfrac{\AC_{|\Ay|}\,\Ag_{a_3|\Ay|}}{\Ag_{\Ay\Ay}}\bigr)\, \AB_{a_4a_5}\,\AB_{a_6]\Ay}\,.
\end{align}
The $S$-duality counterpart is the $T$-duality rule \textcircled{j}, relating the R--R 7-form and 6-form. 

In the following, we adopt $\{\AB_6,\,\pmb{\AA}_{7,1},\,\cB_6,\,\cN_{7,1}\}$ for the $S$-duality non-covariant expressions while $\{\AB_6,\,\AA_{7,1},\,\bBD_6,\,\bBA_{7,1}\}$ for the $S$-duality covariant expressions. 

\paragraph{\underline{$T$-duality rule \textcircled{o}: $5_2\ (\text{IIA})\leftrightarrow 5^1_2\ (\text{IIB})$}\\}
From $\cA_{\mu;\sfa_1\cdots \sfa_5} \overset{\tiny\textcircled{o}}{=} \bm{\cA}_{\mu;\sfa_1\cdots \sfa_5\By,\By}$, we obtain \cite{0712.3235}
\begin{align}
 \AB_{a_1\cdots a_6}&\overset{\text{A--B}}{=}
 \cN_{a_1 \cdots a_6 \By, \By}
 -6\, \cB_{[a_1\cdots a_5|\By|}\, \BB_{a_6] \By}
 -30\, \BC_{[a_1 a_2 a_3|\By|}\, \bigl(\BC_{a_4a_5}-\tfrac{2\, \BC_{a_4|\By|}\, \Bg_{a_5|\By|}}{\Bg_{\By\By}}\bigr)\, \BB_{a_6] \By} \,,
\\
 \cN_{a_1\cdots a_6\By,\By}
 &\overset{\text{B--A}}{=} \AB_{a_1\cdots a_6}
 -\tfrac{6\,\AB_{[a_1\cdots a_5|\Ay|}\,\Ag_{a_6]\Ay}}{\Ag_{\Ay\Ay}} \,.
\end{align}
For the $S$-duality covariant fields, we find \cite{1909.01335}
\begin{align}
 \AB_{a_1\cdots a_6}&\overset{\text{A--B}}{=}
 \bBA_{a_1 \cdots a_6 \By, \By}
 -15\, \epsilon_{\SLc\SLd}\,\bBA_{[a_1a_2a_3|\By|}\, \bigl(\bBA^{\SLc}_{a_4a_5}\, \bBA^{\SLd}_{a_6] \By}
 +\tfrac{2\,\bBA^{\SLc}_{a_4|\By|}\, \bBA^{\SLd}_{a_5|\By|}\, \Bg_{a_6]\By}}{\Bg_{\By\By}}\bigr)\,,
\\
 \bBA_{a_1\cdots a_6\By,\By}
 &\overset{\text{B--A}}{=} \AB_{a_1\cdots a_6}
 -15\,\AC_{[a_1a_2a_3}\,\AB_{a_4a_5}\,\bigl(\AC_{a_6}-\tfrac{\AC_{|\Ay|}\,\Ag_{a_6]\Ay}}{\Ag_{\Ay\Ay}}\bigr)
 -\tfrac{15\,\AC_{[a_1a_2a_3}\,\AC_{a_4a_5|\Ay|}\,\Ag_{a_6]\Ay}}{\Ag_{\Ay\Ay}} \,,
\end{align}
which are self-dual under $S$-duality. 

\paragraph{\underline{$T$-duality rule \textcircled{p}: $5^1_2\ (\text{IIA})\leftrightarrow 5^1_2\ (\text{IIB})$}\\}
From the linear map $\cA_{\mu;\sfa_1\cdots \sfa_5\Ay\Az,\sfa} \overset{\tiny\textcircled{p}}{\simeq} \bm{\cA}_{\mu;\sfa_1\cdots \sfa_5\By,\sfa}$, we obtain
\begin{align}
 \pmb{\AA}_{a_1\cdots a_6\Ay, b}&\overset{\text{A--B}}{\simeq}
 \cN_{a_1\cdots a_6\By, b} - \tfrac{1}{3}\,\tfrac{\cN_{a_1\cdots a_6\By, \By}\,\Ag_{b\Ay}}{\Ag_{\Ay\Ay}}
 -4\,\cB_{[a_1\cdots a_5|\By|}\,\BB_{a_6] b}
\nn\\
 &\quad\ 
 -\bigl(\cB_{a_1\cdots a_6}-\tfrac{2\,\cB_{[a_1\cdots a_5|\By|}\,\Bg_{a_6]\By}}{\Ag_{\Ay\Ay}}\bigr)\,\BB_{b\By} 
 +5\,\BC_{[a_1\cdots a_4|b\By|}\,\bigl(\BC_{a_5a_6]} -\tfrac{2\,\BC_{a_5|\By|}\,\Bg_{a_6]\By}}{\Ag_{\Ay\Ay}}\bigr) 
\nn\\
 &\quad\ 
 -\tfrac{5}{2}\,\bigl(\BC_{[a_1\cdots a_4} - \tfrac{4\,\BC_{[a_1a_2a_3|\By|}\,\Bg_{a_4|\By|}}{\Ag_{\Ay\Ay}}\bigr) \, \BC_{a_5a_6]b\By} 
\nn\\
 &\quad\ 
 - 5\,\BC_{[a_1\cdots a_4}\,\bigl(\BC_{a_5a_6]} -\tfrac{2\,\BC_{a_5|\By|}\,\Bg_{a_6]\By}}{\Ag_{\Ay\Ay}}\bigr)\,\BB_{b\By} 
\nn\\
 &\quad\ 
 -20\,\BC_{[a_1a_2a_3|\By|}\,\bigl(\BC_{a_4a_5} -\tfrac{2\,\BC_{a_4|\By|}\,\Bg_{a_5|\By|}}{\Ag_{\Ay\Ay}}\bigr)\,\BB_{a_6]b} \,,
\\
 \cN_{a_1\cdots a_6\By, b}
 &\overset{\text{B--A}}{\simeq}
 \pmb{\AA}_{a_1\cdots a_6\Ay, b}
 - \tfrac{\pmb{\AA}_{a_1\cdots a_6\Ay, \Ay}\,\Ag_{b\Ay}}{\Ag_{\Ay\Ay}}
 +4\,\AB_{[a_1\cdots a_5|\Ay|} \, \bigl(\AB_{a_6]b} - \tfrac{\AB_{a_6]\Ay}\,\Ag_{b\Ay}}{\Ag_{\Ay\Ay}}\bigr) 
\nn\\
 &\quad\ 
 -\tfrac{1}{3}\, \AB_{a_1\cdots a_6}\,\AB_{b\Ay}
 +\tfrac{6\,\AB_{[a_1\cdots a_5|\Ay|}\, \Ag_{a_6]\Ay}\,\AB_{b\Ay}}{\Ag_{\Ay\Ay}}
 +2\,\AC_{[a_1\cdots a_5}\,\AC_{a_6]b\Ay} 
\nn\\
 &\quad\ 
 +\tfrac{15\,\AC_{[a_1\cdots a_4|\Ay|}\,\AC_{a_5|b\Ay|}\,\Ag_{a_6]\Ay}}{\Ag_{\Ay\Ay}}
 +\tfrac{5}{2}\, \AC_{[a_1\cdots a_4|\Ay|}\, \bigl(\AC_{a_5a_6]b} - \tfrac{\AC_{a_5a_6]\Ay}\,\Ag_{b\Ay}}{\Ag_{\Ay\Ay}}\bigr) \,.
\end{align}
They are partially obtained in Eq.~(5.13) of \cite{hep-th/9908094} under the truncation $\BB_2 = 0=\BC_2$\,. 
The full result without the truncation is obtained in \cite{1909.01335}. 
The same $T$-duality map seems to be obtained in Eqs.~(3.10) and (3.11) of \cite{0907.3614}, although the relation to our potentials is not clear. 

On the other hand, by using the $S$-duality covariant fields, we obtain 
\begin{align}
 \AA_{a_1\cdots a_6\Ay, b}&\overset{\text{A--B}}{\simeq}
 \bBA_{a_1\cdots a_6\By, b} - \tfrac{\bBA_{a_1 \cdots a_6 \By, \By}\,\Bg_{b\By}}{\Bg_{\By\By}}
 + 6\,\epsilon_{\SLc\SLd}\, \bBA^{\SLc}_{[a_1 \cdots a_5 |\By|}\, \bBA^{\SLd}_{a_6]b} 
 + \tfrac{30\,\epsilon_{\SLc\SLd}\, \bBA^{\SLc}_{[a_1 \cdots a_4 |b\By|}\, \bBA^{\SLd}_{a_5|\By|}\,\Bg_{a_6]\By}}{\Bg_{\By\By}}
\nn\\
 &\quad\ 
 +10\,\bigl(\bBA_{[a_1a_2a_3|b|} + \tfrac{\bBA_{[a_1a_2|b\By|}\,\Bg_{a_3|\By|}}{\Bg_{\By\By}} \bigr)\,\bBA_{a_4a_5a_6]\By} 
 +20\,\epsilon_{\SLc\SLd}\,\bBA_{[a_1a_2a_3|\By|}\,\bBA^{\SLc}_{a_4a_5}\,\bBA^{\SLd}_{a_6]b} 
\nn\\
 &\quad\ 
 -30\,\epsilon_{\SLc\SLd}\,\bBA_{[a_1a_2a_3|b|}\,\bBA^{\SLc}_{a_4a_5}\,\bBA^{\SLd}_{a_6]\By} 
 -\tfrac{10\,\epsilon_{\SLc\SLd}\,\bBA_{[a_1a_2a_3|\By|}\,\bBA^{\SLc}_{a_4a_5}\,\bBA^{\SLd}_{a_6]\By}\,\Bg_{b\By}}{\Bg_{\By\By}}
\nn\\
 &\quad\ 
 +\tfrac{30\,\epsilon_{\SLc\SLd}\,\bBA_{[a_1a_2a_3|\By|}\,\bBA^{\SLc}_{a_4a_5}\,\bBA^{\SLd}_{|b\By|}\,\Bg_{a_6]\By}}{\Bg_{\By\By}}
 +\tfrac{20\,\epsilon_{\SLc\SLd}\,\bBA_{[a_1a_2a_3|\By|}\,\bBA^{\SLc}_{a_4|b|}\,\bBA^{\SLd}_{a_5|\By|}\,\Bg_{a_6]\By}}{\Bg_{\By\By}}
\nn\\
 &\quad\ 
 -\tfrac{60\,\epsilon_{\SLc\SLd}\,\bBA_{[a_1a_2a_3|b|}\,\bBA^{\SLc}_{a_4|\By|}\,\bBA^{\SLd}_{a_5|\By|}\,\Bg_{a_6]\By}}{\Bg_{\By\By}}
 -\tfrac{15}{2}\,\epsilon_{\SLc\SLd}\,\epsilon_{\SLf\SLg}\,\bBA^{\SLc}_{[a_1a_2}\,\bBA^{\SLd}_{a_3|b|}\,\bBA^{\SLf}_{a_4a_5}\,\bBA^{\SLg}_{a_6]\By} \nn\\
 &\quad\ 
 +\tfrac{45}{2}\,\tfrac{\epsilon_{\SLc\SLd}\,\epsilon_{\SLf\SLg}\,\bBA^{\SLc}_{[a_1a_2}\,\bBA^{\SLd}_{a_3|\By|}\,\bBA^{\SLf}_{a_4a_5}\,\bBA^{\SLg}_{|b\By|}\,\Bg_{a_6]\By}}{\Bg_{\By\By}}
 -\tfrac{15}{2}\,\tfrac{\epsilon_{\SLc\SLd}\,\epsilon_{\SLf\SLg}\,\bBA^{\SLc}_{[a_1a_2}\,\bBA^{\SLd}_{a_3|b|}\,\bBA^{\SLf}_{a_4|\By|}\,\bBA^{\SLg}_{a_5|\By|}\,\Bg_{a_6]\By}}{\Bg_{\By\By}}\,,
\\
 \bBA_{a_1\cdots a_6\By, b}
 &\overset{\text{B--A}}{\simeq}
 \AA_{a_1\cdots a_6\Ay, b}
 -\AB_{a_1\cdots a_6}\,\AB_{b\Ay}
 -6\,\AB_{[a_1\cdots a_5|\Ay|}\,\AB_{a_6]b}
 +6\,\AC_{[a_1\cdots a_5}\,\AC_{a_6]b\Ay}
\nn\\
 &\quad\ 
 -10\,\AC_{[a_1a_2a_3|b\Ay|}\,\bigl(\AC_{a_4a_5a_6]}+\tfrac{3}{2}\,\tfrac{\AC_{a_4a_5|\Ay|}\,\Ag_{a_6]\Ay}}{\Ag_{\Ay\Ay}}\bigr)
\nn\\
 &\quad\ 
 -15\,\AC_{[a_1a_2a_3|b\Ay|}\,\bigl(\AC_{a_4}- \tfrac{\AC_{|\Ay|}\,\Ag_{a_4|\Ay|}}{\Ag_{\Ay\Ay}}\bigr)\,\AB_{a_5a_6]}
\nn\\
 &\quad\ 
 +20\,\AC_{[a_1a_2a_3}\,\bigl(\AC_{a_4a_5|b}- \tfrac{\AC_{a_4a_5|\Ay}\,\Ag_{b\Ay}}{\Ag_{\Ay\Ay}}\bigr)\,\AB_{|a_6]\Ay}
 -50\,\AC_{[a_1a_2a_3}\,\AC_{a_4|b\Ay|}\,\AB_{a_5a_6]}
\nn\\
 &\quad\ 
 +\tfrac{15}{2}\,\AC_{[a_1a_2|b|}\,\AC_{a_3a_4|\Ay|}\,\AB_{a_5a_6]}
 +\tfrac{20\,\AC_{[a_1a_2a_3}\,\AC_{a_4a_5|\Ay|}\,\AB_{a_6]\Ay}\,\Ag_{b\Ay}}{\Ag_{\Ay\Ay}} 
\nn\\
 &\quad\ 
 +\tfrac{15\,\AC_{[a_1a_2a_3}\,\AC_{a_4a_5|\Ay|}\,\Ag_{a_6]\Ay}\,\AB_{b\Ay}}{\Ag_{\Ay\Ay}} 
 +\tfrac{45\,\AC_{[a_1a_2|\Ay|}\,\AC_{a_3a_4|\Ay|}\,\AC_{a_5|b|}\,\Ag_{a_6]\Ay}}{\Ag_{\Ay\Ay}} 
\nn\\
 &\quad\ 
 +15\,\AC_{[a_1a_2a_3}\,\bigl(\AC_{a_4}-\tfrac{\AC_{|\Ay|}\,\Ag_{a_4|\Ay|}}{\Ag_{\Ay\Ay}}\bigr)\,\AB_{a_5a_6]}\,\AB_{b\Ay}
\nn\\
 &\quad\ 
 -45\,\AC_{[a_1a_2|\Ay|}\,\bigl(\AC_{a_3}-\tfrac{\AC_{|\Ay|}\,\Ag_{a_3|\Ay|}}{\Ag_{\Ay\Ay}}\bigr)\,\AB_{a_4a_5}\,\AB_{a_6]b} \,.
\end{align}
They are again self-dual under $S$-duality.

\paragraph{A short comment\\}
In the following, we present new results. 
The $T$-duality rules obtained below are rather lengthy, and we determine the maps only from type IIA fields to type IIB fields. 
However, in section \ref{sec:D-potential}, we find a redefinition of mixed-symmetry potentials, which transforms our potentials into the potentials $\TmD_{6}$, $\TmD_{7,1}$, and $\TmD_{8,2}$\,. 
The $T$-duality rules for the new fields are very simple, and one can easily find the inverse map, if necessary.

\paragraph{\underline{$T$-duality rule \textcircled{q}: $5^2_2\ (\text{IIA})\leftrightarrow 5^1_2\ (\text{IIB})$}\\}
The linear map $\cA_{\mu;\sfa_1\cdots \sfa_6\Ay\Az,\sfa\Ay\Az} \overset{\tiny\textcircled{q}}{\simeq} \bm{\cA}_{\mu;\sfa_1\cdots \sfa_6,\sfa}$ gives
\begin{align}
 \AA_{a_1\cdots a_7 \By, b\By}
 &\overset{\text{A--B}}{\simeq} \bBA_{a_1\cdots a_7, b}
 -\tfrac{35 \,\bBA_{[a_1a_2a_3|b|}\,\bBA_{a_4a_5a_6|\By|}\,\Bg_{a_7]\By}}{\Bg_{\By\By}}
\nn\\
 &\quad\ -\tfrac{105}{2}\,\tfrac{\epsilon_{\SLc\SLd}\,\bBA_{[a_1a_2a_3|b|}\,\bBA^{\SLc}_{a_4a_5}\,\bBA^{\SLd}_{a_6|\By|}\,\Bg_{a_7]\By}}{\Bg_{\By\By}}
 -\tfrac{105}{2}\,\tfrac{\epsilon_{\SLc\SLd}\,\bBA_{[a_1a_2a_3|\By|}\,\bBA^{\SLc}_{a_4a_5}\,\bBA^{\SLd}_{a_6|b|}\,\Bg_{a_7]\By}}{\Bg_{\By\By}}
\nn\\
 &\quad\ 
 +\tfrac{315}{4}\,\tfrac{\epsilon_{\SLc\SLd}\,\epsilon_{\SLf\SLg}\,\bBA^{\SLc}_{[a_1a_2}\,\bBA^{\SLd}_{a_3|b|}\,\bBA^{\SLf}_{a_4a_5}\,\bBA^{\SLg}_{a_6|\By|}\,\Bg_{a_7]\By}}{\Bg_{\By\By}}\,,
\end{align}
which is self-dual under $S$-duality. 

\paragraph{\underline{$T$-duality rule \textcircled{r}: $5^1_2\ (\text{IIA})\leftrightarrow 5^2_2\ (\text{IIB})$}\\}
From the linear map $\cA_{\mu;\sfa_1\cdots \sfa_6\Az,\sfa} \!\overset{\tiny\textcircled{r}}{\simeq}\! \bm{\cA}^{\SLE{1}}_{\mu;\sfa_1\cdots \sfa_6\By,\sfa\By}$, we obtain
\begin{align}
 \AA_{a_1\cdots a_7, b} &\overset{\text{A--B}}{\simeq} \bBD_{a_1\cdots a_7\By, b\By}
 -7\,\bBA_{[a_1\cdots a_6 |\By ,\By|}\,\BB_{a_7]b}
 -\tfrac{42\,\bBA_{[a_1\cdots a_5 |b\By, \By|}\,\BB_{a_6|\By|} \,\Bg_{a_7]\By}}{\Bg_{\By\By}}\nn\\
 &\quad\ 
 -\tfrac{21}{2}\,\epsilon_{\SLc\SLd}\,\bBC_{[a_1\cdots a_5|b|}\,\bBA^{\SLc}_{a_6|\By|}\,\bBA^{\SLd}_{a_7]\By} 
 + 21\,\epsilon_{\SLc\SLd}\,\bBC_{[a_1\cdots a_5|\By|}\,\bBA^{\SLc}_{a_6|b|}\,\bBA^{\SLd}_{a_7]\By} 
\nn\\
 &\quad\ 
 -\tfrac{105\,\epsilon_{\SLc\SLd}\,\bBC_{[a_1\cdots a_4 |b\By|}\,\bBA^{\SLc}_{a_5|\By|}\,\bBA^{\SLd}_{a_6|\By|}\,\Bg_{a_7]\By}}{\Bg_{\By\By}} 
 +140\,\bBA_{[a_1a_2a_3 |\By|}\,\bBA_{a_4a_5 |b\By|}\,\BB_{a_6a_7]} 
\nn\\
 &\quad\ 
 - 70\,\bBA_{[a_1a_2a_3 |\By|}\,\bBA_{a_4a_5a_6 |b|}\,\BB_{a_7]\By}
 - \tfrac{70\,\bBA_{[a_1a_2a_3 |\By|}\,\bBA_{a_4a_5 |b\By|}\,\BB_{a_6|\By|} \,\Bg_{a_7]\By}}{\Bg_{\By\By}}
\nn\\
 &\quad\ 
 - 70\,\epsilon_{\SLc\SLd}\,\bBA_{[a_1a_2a_3 |\By|}\,\BB_{a_4a_5}\,\bBA^{\SLc}_{a_6a_7]}\,\bBA^{\SLd}_{b\By} 
 +140\,\epsilon_{\SLc\SLd}\,\bBA_{[a_1a_2a_3|\By|}\,\BB_{a_4a_5}\,\bBA^{\SLc}_{a_6|b|}\,\bBA^{\SLd}_{a_7]\By} 
\nn\\
 &\quad\ 
 +105\,\epsilon_{\SLc\SLd}\,\bBA_{[a_1a_2 |b\By|}\,\BB_{a_3a_4}\,\bBA^{\SLc}_{a_5a_6}\,\bBA^{\SLd}_{a_7]\By} 
 +\tfrac{140\,\epsilon_{\SLc\SLd}\,\bBA_{[a_1a_2a_3 |\By|}\,\BB_{a_4|\By|}\bBA^{\SLc}_{a_5a_6}\,\bBA^{\SLd}_{|b\By|}\,\Bg_{a_7]\By}}{\Bg_{\By\By}} 
\nn\\
 &\quad\ 
 -\tfrac{140\,\epsilon_{\SLc\SLd}\,\bBA_{[a_1a_2a_3 |\By|}\,\BB_{a_4|b|}\,\bBA^{\SLc}_{a_5|\By|}\,\bBA^{\SLd}_{a_6|\By|}\,\Bg_{a_7]\By}}{\Bg_{\By\By}} 
 -\tfrac{105\,\epsilon_{\SLc\SLd}\,\bBA_{[a_1a_2 |b\By|}\,\BB_{a_3a_4}\,\bBA^{\SLc}_{a_5|\By|}\,\bBA^{\SLd}_{a_6|\By|}\,\Bg_{a_7]\By}}{\Bg_{\By\By}} 
\nn\\
 &\quad\ 
 -\tfrac{1575}{8}\,\epsilon_{\SLc\SLd}\,\epsilon_{\SLf\SLg}\,\BB_{[a_1a_2}\,\bBA^{\SLc}_{a_3a_4}\,\bBA^{\SLd}_{a_5|\By|}\,\bBA^{\SLf}_{a_6a_7]}\,\bBA^{\SLg}_{b\By} 
\nn\\
 &\quad\ 
 +\tfrac{2835}{4}\,\epsilon_{\SLc\SLd}\,\epsilon_{\SLf\SLg}\,\BB_{[a_1a_2}\,\bBA^{\SLc}_{a_3a_4}\,\bBA^{\SLd}_{a_5|b|}\,\bBA^{\SLf}_{a_6|\By|}\,\bBA^{\SLg}_{a_7]\By} 
\nn\\
 &\quad\ 
 + \tfrac{315}{2}\,\tfrac{\epsilon_{\SLc\SLd}\,\epsilon_{\SLf\SLg}\,\BB_{[a_1a_2}\,\bBA^{\SLc}_{a_3a_4}\,\bBA^{\SLd}_{a_5|\By|}\,\bBA^{\SLf}_{a_6|\By}\,\bBA^{\SLg}_{b\By}\,\Bg_{|a_7]\By}}{\Bg_{\By\By}} \,.
\label{eq:A71-D82}
\end{align}
Under the simplifying assumption $\BB_2=\BC_2=0$\,, this map has been obtained in \cite{hep-th/9908094} [the last line of Eq.~(5.12)], where $N^{(8)}$ corresponds to our $\bBD_{8, b\By}$ (up to $\BB_2=\BC_2=0$). 
The $S$-dual counterpart of this $T$-duality rule is obtained later in \eqref{eq:A81-E82}. 

\paragraph{\underline{$T$-duality rule \textcircled{s}: $5^2_2\ (\text{IIA})\leftrightarrow 5^2_2\ (\text{IIB})$}\\}
From the linear map $\cA_{\mu;\sfa_1\cdots \sfa_6\Ay\Az,\sfb_1\sfb_2\Az} \!\overset{\tiny\textcircled{s}}{\simeq}\! \bm{\cA}^{\SLE{1}}_{\mu;\sfa_1\cdots \sfa_6\By,\sfb_1\sfb_2}$, we obtain
\begin{align}
 \AA_{a_1\cdots a_7\Ay, b_1b_2}
 &\overset{\text{A--B}}{=} \bBD_{a_1\cdots a_7\By, b_1b_2}
 -7\,\bBC_{[a_1\cdots a_6}\,\bBA_{a_7] b_1b_2\By} 
 -\tfrac{7}{2} \,\epsilon_{\SLc\SLd}\,\bBC_{[a_1\cdots a_6}\,\bBA^{\SLc}_{|b_1b_2|}\,\bBA^{\SLd}_{a_7]\By} 
\nn\\
 &\quad\ 
 -\tfrac{35}{2} \,\bBC_{[a_1a_2a_3 |b_1b_2\By|}\, \bigl(\bBA_{a_4\cdots a_7]} + \tfrac{2\,\bBA_{a_4a_5a_6|\By|} \,\Bg_{a_7]\By}}{\Bg_{\By\By}}\bigr)
\nn\\
 &\quad\ 
 +\tfrac{105}{2}\,\epsilon_{\SLc\SLd}\,\bBC_{[a_1\cdots a_4 |\overline{b}_1\By|}\,\bBA^{\SLc}_{a_5a_6}\,\bBA^{\SLd}_{a_7]\overline{b}_2} 
 -\tfrac{105}{2} \,\tfrac{\epsilon_{\SLc\SLd}\,\bBC_{[a_1a_2a_3 |b_1b_2\By|}\,\bBA^{\SLc}_{a_4a_5}\,\bBA^{\SLd}_{a_6|\By|} \,\Bg_{a_7]\By}}{\Bg_{\By\By}}
\nn\\
 &\quad\ 
 +35\,\bBA_{[a_1a_2a_3|\overline{b}_1|}\,\bBA_{a_4a_5a_6|\By|}\,\BB_{a_7]\overline{b}_2} 
 -\tfrac{315}{2} \,\bBA_{[a_1a_2a_3|\overline{b}_1|}\,\bBA_{a_4a_5|\overline{b}_2\By|}\,\BB_{a_6a_7]} 
\nn\\
 &\quad\ 
 + 70\,\bBA_{[a_1a_2a_3|\overline{b}_1|}\,\bBA_{a_4a_5a_6|\overline{b}_2|}\,\BB_{a_7]\By} 
 + \tfrac{105\,\bBA_{[a_1a_2a_3|\By|}\,\bBA_{a_4 |b_1b_2\By|}\,\BB_{a_5a_6} \,\Bg_{a_7]\By}}{\Bg_{\By\By}}
\nn\\
 &\quad\ 
 + \tfrac{105\,\bBA_{[a_1a_2|b_1b_2|}\,\bBA_{a_3a_4a_5|\By|}\,\BB_{a_6|\By|} \,\Bg_{a_7]\By}}{\Bg_{\By\By}}
 -\tfrac{105}{4} \,\epsilon_{\SLc\SLd}\,\bBA_{[a_1a_2|b_1b_2|}\,\bBA^{\SLc}_{a_3a_4}\,\bBA^{\SLd}_{a_5|\By|}\,\BB_{a_6a_7]} 
\nn\\
 &\quad\ 
 +35\,\epsilon_{\SLc\SLd}\,\bBA_{[a_1a_2a_3|\By|}\,\bBA^{\SLc}_{a_4a_5}\,\bBA^{\SLd}_{a_6|\overline{b}_1|}\,\BB_{a_7]\overline{b}_2} 
 +\tfrac{35}{4} \,\epsilon_{\SLc\SLd}\,\bBA_{[a_1a_2a_3|\By|}\,\bBA^{\SLc}_{a_4a_5}\,\bBA^{\SLd}_{|b_1b_2|}\,\BB_{a_6a_7]} 
\nn\\
 &\quad\ 
 -\tfrac{315}{2} \,\tfrac{\epsilon_{\SLc\SLd}\,\bBA_{[a_1a_2 |\overline{b}_1\By|}\,\bBA^{\SLc}_{a_3a_4}\,\bBA^{\SLd}_{a_5|\By|}\,\BB_{a_6a_7]}\,\Bg_{\overline{b}_2\By}}{\Bg_{\By\By}}
 -\tfrac{70\,\epsilon_{\SLc\SLd}\,\bBA_{[a_1a_2a_3|\By|}\,\bBA^{\SLc}_{a_4a_5}\,\bBA^{\SLd}_{a_6|\By}\,\BB_{b_1b_2} \,\Bg_{|a_7]\By}}{\Bg_{\By\By}}
\nn\\
 &\quad\ 
 +\tfrac{35\,\epsilon_{\SLc\SLd}\,\bBA_{[a_1a_2a_3|\By|}\,\bBA^{\SLc}_{a_4a_5}\,\bBA^{\SLd}_{|\overline{b}_1\By|}\,\BB_{a_6|\overline{b}_2|} \,\Bg_{a_7]\By}}{\Bg_{\By\By}}
 -\tfrac{280\,\epsilon_{\SLc\SLd}\,\bBA_{[a_1a_2a_3|\By|}\,\bBA^{\SLc}_{a_4|\overline{b}_1|}\,\bBA^{\SLd}_{a_5|\By|}\, \BB_{a_6|\overline{b}_2|} \,\Bg_{a_7]\By}}{\Bg_{\By\By}}
\nn\\
 &\quad\ 
 -\tfrac{140\,\epsilon_{\SLc\SLd}\,\bBA_{[a_1a_2a_3|\By|}\,\bBA^{\SLc}_{a_4|\overline{b}_1}\,\bBA^{\SLd}_{\overline{b}_2\By|}\,\BB_{a_5a_6} \,\Bg_{a_7]\By}}{\Bg_{\By\By}}
 -\tfrac{35}{2} \,\tfrac{\epsilon_{\SLc\SLd}\,\bBA_{[a_1a_2a_3|\By}\,\bBA^{\SLc}_{b_1b_2|}\,\bBA^{\SLd}_{a_4|\By|}\,\BB_{a_5a_6} \,\Bg_{a_7]\By}}{\Bg_{\By\By}}
\nn\\
 &\quad\ 
 -\tfrac{315}{4} \,\tfrac{\epsilon_{\SLc\SLd}\,\bBA_{[a_1a_2|b_1b_2|}\,\bBA^{\SLc}_{a_3|\By|}\,\bBA^{\SLd}_{a_4|\By|}\,\BB_{a_5a_6} \,\Bg_{a_7]\By}}{\Bg_{\By\By}}
\nn\\
 &\quad\ 
 -315\,\epsilon_{\SLc\SLd}\,\epsilon_{\SLf\SLg}\,\bBA^{\SLc}_{[a_1a_2}\,\bBA^{\SLd}_{a_3|\overline{b}_1|}\,\bBA^{\SLf}_{a_4a_5}\,\bBA^{\SLg}_{a_6|\overline{b}_2|}\,\BB_{a_7]\By} 
\nn\\
 &\quad\ 
 +\tfrac{2835}{8}\,\epsilon_{\SLc\SLd}\,\epsilon_{\SLf\SLg}\,\bBA^{\SLc}_{[a_1a_2}\,\bBA^{\SLd}_{a_3|\overline{b}_1|}\,\bBA^{\SLf}_{a_4a_5}\,\bBA^{\SLg}_{|\overline{b}_2\By|}\,\BB_{a_6a_7]} 
\nn\\
 &\quad\ 
 +\tfrac{1575}{2}\,\epsilon_{\SLc\SLd}\,\epsilon_{\SLf\SLg}\,\bBA^{\SLc}_{[a_1a_2}\,\bBA^{\SLd}_{a_3|\overline{b}_1|}\,\bBA^{\SLf}_{a_4|\overline{b}_2|}\,\bBA^{\SLg}_{a_5|\By|}\,\BB_{a_6a_7]} 
\nn\\
 &\quad\ 
 +\tfrac{315}{4}\,\tfrac{\epsilon_{\SLc\SLd}\,\epsilon_{\SLf\SLg}\,\bBA^{\SLc}_{[a_1a_2}\,\bBA^{\SLd}_{a_3|\overline{b}_1|}\,\bBA^{\SLf}_{a_4|\By}\,\bBA^{\SLg}_{\overline{b}_2\By}\,\BB_{|a_5a_6} \,\Bg_{a_7]\By}}{\Bg_{\By\By}}
\nn\\
 &\quad\ 
 +\tfrac{105}{8}\,\tfrac{\epsilon_{\SLc\SLd}\,\epsilon_{\SLf\SLg}\,\bBA^{\SLc}_{[a_1a_2}\,\bBA^{\SLd}_{a_3|\By|}\,\bBA^{\SLf}_{a_4a_5}\,\bBA^{\SLg}_{|\overline{b}_1\By|}\,\BB_{a_6a_7]} \,\Bg_{\overline{b}_2\By}}{\Bg_{\By\By}}
\nn\\
 &\quad\ 
 +\tfrac{105}{2}\,\tfrac{\epsilon_{\SLc\SLd}\,\epsilon_{\SLf\SLg}\,\bBA^{\SLc}_{[a_1a_2}\,\bBA^{\SLd}_{a_3|\By|}\,\bBA^{\SLf}_{a_4|\overline{b}_1|}\,\bBA^{\SLg}_{a_5|\By|}\,\BB_{a_6a_7]} \,\Bg_{\overline{b}_2\By}}{\Bg_{\By\By}}
\nn\\
 &\quad\ 
 -\tfrac{525}{4}\,\tfrac{\epsilon_{\SLc\SLd}\,\epsilon_{\SLf\SLg}\,\bBA^{\SLc}_{[a_1a_2}\,\bBA^{\SLd}_{a_3|\By|}\,\bBA^{\SLf}_{a_4a_5}\,\bBA^{\SLg}_{|\overline{b}_1\By|}\,\BB_{a_6|\overline{b}_2|}\,\Bg_{a_7]\By}}{\Bg_{\By\By}}
\nn\\
 &\quad\ 
 -\tfrac{210\,\epsilon_{\SLc\SLd}\,\epsilon_{\SLf\SLg}\,\bBA^{\SLc}_{[a_1a_2}\,\bBA^{\SLd}_{a_3|\By|}\,\bBA^{\SLf}_{a_4|\overline{b}_1|}\,\bBA^{\SLg}_{a_5|\By|}\,\BB_{a_6|\overline{b}_2|}\,\Bg_{a_7]\By}}{\Bg_{\By\By}}\,.
\label{eq:D82A-D82B}
\end{align}
The $S$-dual counterpart of this $T$-duality rule is found later in \eqref{eq:D83A-E82B}. 
It may be possible to make the expression simpler by finding some $S$-duality-covariant field redefinitions, but here do not attempt to find such redefinitions. 

\subsubsection{$T$-dual-manifest redefinitions}
\label{sec:D-potential}

Now, let us consider the field redefinition that makes the $T$-duality rules very simple. 
In the case of the R--R fields in type IIA/IIB theory, the R--R polyform in the $C$-basis is defined as
\begin{align}
 \AC \equiv \AC_1 + \AC_3 + \AC_5 + \AC_7 + \AC_9 \,, \qquad
 \BC \equiv \BC_0 + \BC_2 + \BC_4 + \BC_6 + \BC_8 + \BC_{10} \,.
\end{align}
By considering a redefinition into the $A$-basis \cite{hep-th/0103233},
\begin{align}
 \TmC \equiv \Exp{-\AB_2\wedge}\AC\,,\qquad
 \TmC \equiv \Exp{-\BB_2\wedge}\BC\,,
\label{eq:A-def}
\end{align}
we find that the $T$-duality rules for the new fields are simple \cite{hep-th/0103149}
\begin{align}
 \TmC_{a_1\cdots a_p} \overset{\text{A--B}}{=} \TmC_{a_1\cdots a_p\By}\,,\qquad 
 \TmC_{a_1\cdots a_p\Ay} \overset{\text{A--B}}{=} \TmC_{a_1\cdots a_p}\,.
\label{eq:A-T-dual}
\end{align}
This is according to the fact that the $A$-basis transforms as an $\OO(10,10)$ spinor. 
As studied in \cite{hep-th/9907132,1106.5452,1107.0008}, if we define the (real) gamma matrices $\{\Gamma^M\} = \{\Gamma^m,\,\Gamma_m\}$ that satisfy
\begin{align}
 \bigl\{\Gamma^M,\, \Gamma^N \bigr\} = \eta^{MN}\,,\qquad 
 (\eta^{MN})\equiv \begin{pmatrix} 0 & \SLd^m_n \\ \SLd_m^n & 0 \end{pmatrix},\qquad
 (\Gamma_m)^\rmT = \Gamma^m \,,\qquad
 (\Gamma^m)^\rmT = \Gamma_m \,,
\end{align}
and also define the Clifford vacuum $\ket{0}$ as
\begin{align}
\begin{split}
 \Gamma_m\ket{0} =0\,,\qquad \langle 0\vert 0\rangle = 1 \,, \qquad \Gamma^{11}\ket{0} = \ket{0}\,, 
\end{split}\end{align}
where $\bra{0}\equiv \ket{0}^\rmT$ and $\Gamma^{11}\equiv (-1)^{N_F}$ ($N_F\equiv \Gamma^m\,\Gamma_m$), we find that
\begin{align}
 \ket{\TmC}\equiv \sum_p \tfrac{1}{p!}\,\TmC_{m_1\dots m_p}\,\Gamma^{m_1\cdots m_p}\ket{0}\,,\qquad 
 \Gamma^{M_1\cdots M_p}\equiv \Gamma^{[M_1}\cdots \Gamma^{M_p]}\,,
\end{align}
transforms as an $\OO(10,10)$ spinor. 
Here, the R--R field $\ket{\TmC}$ is defined to have a definite chirality
\begin{align}
 \Gamma^{11}\ket{\TmC} = \mp \ket{\TmC} \quad (\text{IIA/IIB}) \,. 
\end{align}
Under the factorized $T$-duality along the $x^y$-direction, it transforms as
\begin{align}
 \ket{\TmC} \to \ket{\TmC'} = \bigl(\Gamma^y -\Gamma_y\bigr)\,\Gamma^{11}\,\ket{\TmC} \,,
\end{align}
and in terms of the components, this transformation rule gives the rules \eqref{eq:A-T-dual}. 

Similarly, the potentials which couple to the solitonic 5-branes also constitute an $\OO(10,10)$-covariant potential denoted by $\TmD_{M_1\cdots M_4}$ \cite{1102.0934}, where the 20D indices are totally antisymmetric. 
This tensor can be generally decomposed into $\SL(10)$ tensors:
\begin{align}
\begin{split}
 \TmD^{m_1m_2m_3m_4} &= \tfrac{1}{6!}\,\epsilon^{m_1\cdots m_4n_1\cdots n_6}\,\TmD_{n_1\cdots n_6} \,,
\\
 \TmD^{m_1m_2m_3}{}_{m_4} &= \tfrac{1}{7!}\,\epsilon^{m_1m_2m_3n_1\cdots n_7}\, \TmD_{n_1\cdots n_7,m_4}+\cdots \,,
\\
 \TmD^{m_1m_2}{}_{m_3m_4} &= \tfrac{1}{8!}\,\epsilon^{m_1m_2n_1\cdots n_8}\, \TmD_{n_1\cdots n_8,m_3m_4}+\cdots \,,
\\
 \TmD^{m_1}{}_{m_2m_3m_4} &= \tfrac{1}{9!}\,\epsilon^{m_1n_1\cdots n_9}\, \TmD_{n_1\cdots n_9, m_2m_3m_4}+\cdots \,,
\\
 \TmD_{m_1m_2m_3m_4} &= \tfrac{1}{10!}\,\epsilon^{n_1\cdots n_{10}}\,\TmD_{n_1\cdots n_{10},m_1\cdots m_4}\,,
\end{split}
\end{align}
where the ellipsis denote the irrelevant contribution from the potentials that do not couple to supersymmetric branes. 
Under the $T$-duality along the $x^y$-direction, this transforms as
\begin{align}
 \TmD'_{M_1\cdots M_4} = \Lambda_{M_1}{}^{N_1}\cdots \Lambda_{M_4}{}^{N_4}\,\TmD_{N_1\cdots N_4} \,, \qquad
 (\Lambda_M{}^N) \equiv 
 \begin{pmatrix}
 \bm{1}- e_y & e_y \\
 e_y & \bm{1}- e_y
\end{pmatrix} ,
\end{align}
where $e_y$ is a $10\times 10$ matrix, $e_y=\diag(0,\dotsc,0,\underbrace{1}_y,0,\dotsc,0)$. 
By rewriting the transformation rule in terms of the component fields $\TmD_6$, $\TmD_{7,1}$, $\TmD_{8,2}$, $\TmD_{9,3}$, and $\TmD_{10,4}$\,, we obtain
\begin{align}
\begin{split}
 \TmD_{a_1\cdots a_6b_1\cdots b_n,b_1\cdots b_n} &\overset{\text{A--B}}{=} \TmD_{a_1\cdots a_6b_1\cdots b_n\By,b_1\cdots b_n\By}\quad (n=0,\dotsc,3)\,,
\\
 \TmD_{a_1\cdots a_5b_1\cdots b_n\Ay ,b_1\cdots b_n} &\overset{\text{A--B}}{=} \TmD_{a_1\cdots a_5b_1\cdots b_n\By ,b_1\cdots b_n}\quad (n=0,\dotsc ,4)\,,
\\
 \TmD_{a_1\cdots a_6b_1\cdots b_n\Ay ,b_1\cdots b_n\Ay} &\overset{\text{A--B}}{=} \TmD_{a_1\cdots a_6b_1\cdots b_n,b_1\cdots b_n}\quad (n=0,\dotsc ,3)\,.
\end{split}
\label{eq:D-T-duality}
\end{align}

Similar to the case of the R--R potential $\TmC_{p}$, which is related to our potentials as \eqref{eq:A-def}, it is natural to expect that $\TmD_{6+n,n}$ are also obtained by considering a redefinition of our mixed-symmetry potentials. 
Indeed, if we redefine the type IIA fields as
\begin{align}
 \TmD_{m_1\cdots m_6} &\equiv \AB_{m_1\cdots m_6} -3\,\AC_{[m_1\cdots m_5}\,\AC_{m_6]}\,,
\\
 \TmD_{m_1\cdots m_7, n} 
 &\simeq \AA_{m_1\cdots m_7, n}
 + 7\, \AB_{[m_1\cdots m_6} \,\AB_{m_7]n}
 - \tfrac{1}{2}\, \AC_{m_1\cdots m_7} \,\AC_{n} 
 - \tfrac{21}{2}\, \AC_{[m_1\cdots m_5}\,\AC_{m_6m_7]n}
\nn\\
 &\quad + 70\, \AC_{[m_1m_2m_3}\, \AC_{m_4m_5 |n|}\, \AB_{m_6m_7]} 
 + 21\, \AC_{[m_1\cdots m_5}\, \AB_{m_6|n|}\, \AC_{m_7]} \,,
\label{eq:D71A}
\\
 \TmD_{m_1\cdots m_8,\,n_1n_2} 
 &\simeq \AA_{m_1\cdots m_8,\,n_1n_2} 
 - 4\,\AC_{[m_1\cdots m_7}\,\AC_{m_8]\bar{n}_1\bar{n}_2} 
 - 4\,\AC_{[m_1\cdots m_7}\,\AC_{m_8]}\,\AB_{n_1n_2} 
\nn\\
 &\quad
 - 56\,\AC_{[m_1\cdots m_6|\bar{n}_1|}\,\AC_{m_7}\,\AB_{m_8]\bar{n}_2} 
 -168\,\AC_{[m_1\cdots m_5}\,\AC_{m_6}\,\AB_{m_7|\bar{n}_1|}\,\AB_{m_8]\bar{n}_2} 
\nn\\
 &\quad
 + 84\,\AC_{[m_1\cdots m_5}\,\AB_{m_6m_7}\,\AC_{m_8]\bar{n}_1\bar{n}_2} 
 +140\,\AC_{[m_1m_2m_3}\,\AB_{m_4m_5}\,\AC_{m_6m_7m_8]\bar{n}_1\bar{n}_2} 
\nn\\
 &\quad
 -210\,\AC_{[m_1m_2m_3}\,\AB_{m_4m_5}\,\AB_{m_6m_7}\,\AC_{m_8]n_1n_2}\,,
\end{align}
and type IIB fields as
\begin{align}
 \TmD_{m_1\cdots m_6} &\equiv \BB_{m_1\cdots m_6}-\tfrac{15}{2}\,\BC_{[m_1\cdots m_4}\,\BC_{m_5m_6]}-\tfrac{1}{2}\,\BC_0\, \BC_{m_1\cdots m_6}\,,
\label{eq:D-B-6}
\\
 \TmD_{m_1\cdots m_7, n} 
 &\simeq \BA_{m_1\cdots m_7, n}
 + \tfrac{7}{2}\, \BC_{[m_1\cdots m_6}\, \BC_{m_7] n} 
 - \tfrac{7}{2}\, \BC_0\, \BC_{[m_1\cdots m_6}\, \BB_{m_7] n} 
 - \tfrac{105}{4}\, \BC_{[m_1\cdots m_4}\,\BB_{m_5m_6}\,\BC_{m_7]n}
\nn\\
 &\quad
 + \tfrac{105}{4}\, \BC_{[m_1\cdots m_4}\,\BC_{m_5m_6}\,\BB_{m_7]n}
 + \tfrac{315}{4}\, \BB_{[m_1m_2}\,\BB_{m_3m_4}\,\BC_{m_5m_6}\,\BC_{m_7]n}\,,
\label{eq:D-B-71}
\\
 \TmD_{m_1\cdots m_8, n_1n_2} 
 &\simeq \bBD_{m_1\cdots m_8, n_1n_2} 
 - 4\,\BC_{[m_1\cdots m_7, |\bar{n}_1|}\,\BC_{m_8]\bar{n}_2}
 + 4\,\BC_0\,\BC_{[m_1\cdots m_7, |\bar{n}_1|}\,\BB_{m_8]\bar{n}_2}
\nn\\
 &\quad
 + 28\,\BC_{[m_1\cdots m_6}\,\BB_{m_7|\bar{n}_1|}\,\BC_{m_8]\bar{n}_2}
 - 28\,\BC_0\,\BC_{[m_1\cdots m_6}\,\BB_{m_7|\bar{n}_1|}\,\BB_{m_8]\bar{n}_2}
\nn\\
 &\quad
 - 28\,\BC_{[m_1\cdots m_5|\bar{n}_1|}\,\BC_{m_6m_7m_8]\bar{n}_2}
 + 84\,\BC_{[m_1\cdots m_5 |\bar{n}_1|}\,\BC_{m_6m_7}\,\BB_{m_8]\bar{n}_2}
\nn\\
 &\quad
 + 420\,\BC_{[m_1\cdots m_4}\,\BC_{m_5m_6 |n_1n_2|}\,\BB_{m_7m_8]}
 - \tfrac{35}{2}\,\BC_{[m_1\cdots m_4}\,\BC_{m_5\cdots m_8]}\,\BB_{n_1n_2}
\nn\\
 &\quad
 -2520\,\BC_{[m_1m_2m_3|\bar{n}_1|}\,\BB_{m_4m_5}\,\BB_{m_6|\bar{n}_2|} \,\BC_{m_7m_8]}
\nn\\
 &\quad
 - 840\,\BC_{[m_1m_2m_3|\bar{n}_1|}\,\BB_{m_4m_5}\,\BB_{m_6m_7} \,\BC_{m_8]\bar{n}_2}
\nn\\
 &\quad
 - 420\,\BC_{[m_1\cdots m_4}\,\BB_{m_5|\bar{n}_1|}\,\BB_{m_6|\bar{n}_2|}\,\BC_{m_7m_8]}
\nn\\
 &\quad
 +1365\,\BB_{[m_1m_2}\,\BB_{m_3|\bar{n}_1|}\,\BB_{m_4|\bar{n}_2|}\,\BC_{m_5m_6}\,\BC_{m_7m_8]}
\nn\\
 &\quad
 -2765\,\BB_{[m_1m_2}\,\BB_{m_3m_4}\,\BB_{m_5m_6}\,\BC_{m_7|\bar{n}_1|}\,\BC_{m_8]\bar{n}_2}
\nn\\
 &\quad
 + 805\,\BB_{[m_1m_2}\,\BB_{m_3m_4}\,\BB_{m_5m_6}\,\BC_{m_7m_8]}\,\BC_{n_1n_2} \,,
\label{eq:D-B-82}
\end{align}
the complicated $T$-duality rules obtained in this subsection are surprisingly simplified as \eqref{eq:D-T-duality}. 

As a consistency check, let us express the 7-form field strength in type IIA/IIB theory by using the new 6-form $\TmD_6$\,.
Then, we obtain
\begin{align}
\begin{split}
 \AH_7 &= \rmd \TmD_6 - \tfrac{1}{2}\,\bigl(\cG_6 \wedge \AC_1 - \cG_4\wedge \AC_3 + \cG_2 \wedge \AC_5\bigr) \,, 
\\
 \BH_7 &= \rmd \TmD_6 - \tfrac{1}{2}\,\bigl(\cG_7\,\BC_0 - \cG_5\wedge \BC_2 + \cG_3\wedge \BC_4 - \cG_1\wedge \BC_6\bigr) \,,
\end{split}
\end{align}
and they precisely coincide with the expression given in \cite{1102.0934} up to conventions. 
This shows that our $\TmD_{6+n,n}$ $(n=0,1,2)$ are precisely the same as $\TmD_{6+n,n}$ studied there, and they can be straightforwardly extended also to $n=3,4$. 
As shown in \cite{1903.05601}, the field strength, $\AH_7$ or $\BH_7$, can be regarded as a particular component of\footnote{In terms of DFT, by supposing that $D_{M_1\cdots M_4}$ is a generalized tensor with weight 1, we can show $\delta_V H_{MNP}= \hat{\Lie}_V H_{MNP} + 2\,\partial^R\partial_{[M}V^S\,D_{NP]RS}$ under generalized diffeomorphisms. The anomalous term vanishes under the assumptions that lower indices of $H^{mn}{}_p$, $H^m{}_{np}$, and $H_{mnp}$ are associated with Killing directions.}
\begin{align}
 H_{MNP} \equiv \partial^Q \TmD_{MNP Q} - \tfrac{1}{2}\,\overline{\bra{\TmC}} \Gamma_{MNP} \ket{F}\,, \qquad 
 \ket{F} \equiv \Gamma^M\,\partial_M \ket{\TmC} \,,
\label{eq:HMNP}
\end{align}
where $\overline{\bra{A}}\equiv \bra{A}\,C^\rmT\equiv (\ket{A})^\rmT\,C^\rmT$ with $C \equiv (\Gamma^0+ \Gamma_0) \cdots (\Gamma^9 + \Gamma_9)$. 
The indices $M,N,\cdots$ are raised/lowered by using $\eta_{MN}$ and the derivative $\partial_M$ can be understood as $(\partial_M)=(\partial_m,0)$. 
Then, we can show that $\AH_7$ or $\BH_7$ in type IIA or IIB theory is reproduced from
\begin{align}
 H_7 \equiv \tfrac{1}{7!\,3!}\,\epsilon_{m_1\cdots m_7n_1n_2n_3}\, H^{n_1n_2n_3} \,\rmd x^{m_1}\wedge\cdots\wedge \rmd x^{m_7}\,.
\end{align}
Other components are also easily computed. 
For example, the component $H^{a_1a_2}{}_n$ associated with a Killing direction $n$ satisfying $n\not\in \{a_1,\,a_2\}$ gives the field strength of the dual graviton,
\begin{align}
 \iota_n H_{8, n} \equiv \tfrac{1}{7!\,2!}\,\epsilon_{m_1\cdots m_7 a_1a_2 n}\, H^{a_1a_2}{}_n \,\rmd x^{m_1}\wedge\cdots\wedge\rmd x^{m_7}\,.
\end{align}
In type IIA/IIB theory, this reproduces
\begin{align}
\begin{split}
 \iota_n \AH_{8, n} &= \rmd \iota_n \TmD_{7, n} + \tfrac{1}{2}\,\bigl(\iota_n F_8 \, \iota_n\TmC_1 - \iota_n F_6\wedge \iota_n\TmC_3 + \iota_n F_4 \wedge \iota_n \TmC_5 - \iota_n F_2 \wedge \iota_n\TmC_7\bigr) \,,
\label{eq:def-AH7}
\\
 \iota_n \BH_{8, n} &= \rmd \iota_n \TmD_{7, n} + \tfrac{1}{2}\,\bigl(\iota_n F_7\wedge \iota_n\TmC_2 - \iota_n F_5\wedge \iota_n \TmC_4 + \iota_n F_3\wedge \iota_n\TmC_6 - \iota_n F_1\,\iota_n \TmC_8\bigr)\,,
\end{split}
\end{align}
where $F_{p+1}\equiv \rmd\TmC_p$ (note that $\iota_n F_1=\Lie_n \TmC_0=0$)\,. 
The 11D uplift or the $S$-duality-invariant expression is given respectively in section \ref{sec:gauge-IIA} or \ref{sec:gauge-IIB}. 
We can compute the other components as well, yielding the field strengths for mixed-symmetry potentials $\TmD_{8,2}$ and $\TmD_{9,3}$\,. 

Here, it will be useful to comment on the notion of the level $n$. 
If we look at, for example, the right-hand side of \eqref{eq:D-B-82}, terms like $\BC_{....}\,\BC_{..}\,\BB_{..}\,\BB_{..}$ appear, but $\BC_{....}\,\BB_{..}\,\BB_{..}\,\BB_{..}$ never appears. 
This can be understood by considering the level, which has been introduced in the study of the $E_{11}$ conjecture \cite{hep-th/0207267,hep-th/0212291}. 
In type II theories, a potential which couples to a brane with the tension $T\propto g_s^{-n}$ has the level $n$ \cite{0805.4451}. 
For example, $\BB_2$ has level $0$ while the R--R potentials have level $1$. 
Since the potentials $\TmD_{6+n,n}$ have level $2$, the level on the right-hand side of \eqref{eq:D-B-82} must be summed up to $2$, and it is the reason why $\BC_{....}\,\BB_{..}\,\BB_{..}\,\BB_{..}$ does not appear. 
The level is always respected in various equations, such as the parameterization of $\BN$ given in section \ref{sec:1-form}, the $T$-duality rules, the field strengths such as \eqref{eq:HMNP}, and the field redefinitions, and it helps when we find the complicated redefinitions such as \eqref{eq:D-B-82}. 

\subsection{Potentials for exotic branes}

Here, we find the $T$-duality rules for mixed-symmetry potentials that couple to exotic branes with tensions $T\propto g_s^{-3}$ and $g_s^{-4}$. 
Since the potentials have many indices, the $T$-duality rules are generally more complicated than before. 
Thus, we again find only the $T$-duality rules, each of which maps a type IIA potential to type IIB potentials, and by using those, we identify the relation between our potentials and the manifestly $T$-duality-covariant potentials. 

\paragraph{\underline{$T$-duality rule \textcircled{t}: $6^1_3\ (\text{IIA})\leftrightarrow 5^2_3\ (\text{IIB})$:}\\}
From the linear map $\cA_{\mu;\sfa_1\cdots \sfa_6\Ay,\sfa} \!\overset{\tiny\textcircled{t}}{\simeq}\! -\bm{\cA}^{\SLE{2}}_{\mu;\sfa_1\cdots \sfa_6\By,\sfa\By}$, we obtain
\begin{align}
 \AA_{a_1\cdots a_7\Ay, b} &\overset{\text{A--B}}{\simeq} \bBE_{a_1\cdots a_7\By, b\By}
 -7\,\bBA_{[a_1\cdots a_6 |\By ,\By|}\,\BC_{a_7]b}
 -\tfrac{42\,\bBA_{[a_1\cdots a_5 |b\By, \By|}\,\BC_{a_6|\By|} \,\Bg_{a_7]\By}}{\Bg_{\By\By}}\nn\\
 &\quad\ 
 -\tfrac{21}{2}\,\epsilon_{\SLc\SLd}\,\bBD_{[a_1\cdots a_5|b|}\,\bBA^{\SLc}_{a_6|\By|}\,\bBA^{\SLd}_{a_7]\By} 
 + 21\,\epsilon_{\SLc\SLd}\,\bBD_{[a_1\cdots a_5|\By|}\,\bBA^{\SLc}_{a_6|b|}\,\bBA^{\SLd}_{a_7]\By} 
\nn\\
 &\quad\ 
 -\tfrac{105\,\epsilon_{\SLc\SLd}\,\bBD_{[a_1\cdots a_4 |b\By|}\,\bBA^{\SLc}_{a_5|\By|}\,\bBA^{\SLd}_{a_6|\By|}\,\Bg_{a_7]\By}}{\Bg_{\By\By}} 
 +140\,\bBA_{[a_1a_2a_3 |\By|}\,\bBA_{a_4a_5 |b\By|}\,\BC_{a_6a_7]} 
\nn\\
 &\quad\ 
 - 70\,\bBA_{[a_1a_2a_3 |\By|}\,\bBA_{a_4a_5a_6 |b|}\,\BC_{a_7]\By}
 - \tfrac{70\,\bBA_{[a_1a_2a_3 |\By|}\,\bBA_{a_4a_5 |b\By|}\,\BC_{a_6|\By|} \,\Bg_{a_7]\By}}{\Bg_{\By\By}}
\nn\\
 &\quad\ 
 - 70\,\epsilon_{\SLc\SLd}\,\bBA_{[a_1a_2a_3 |\By|}\,\BC_{a_4a_5}\,\bBA^{\SLc}_{a_6a_7]}\,\bBA^{\SLd}_{b\By} 
 +140\,\epsilon_{\SLc\SLd}\,\bBA_{[a_1a_2a_3|\By|}\,\BC_{a_4a_5}\,\bBA^{\SLc}_{a_6|b|}\,\bBA^{\SLd}_{a_7]\By} 
\nn\\
 &\quad\ 
 +105\,\epsilon_{\SLc\SLd}\,\bBA_{[a_1a_2 |b\By|}\,\BC_{a_3a_4}\,\bBA^{\SLc}_{a_5a_6}\,\bBA^{\SLd}_{a_7]\By} 
 +\tfrac{140\,\epsilon_{\SLc\SLd}\,\bBA_{[a_1a_2a_3 |\By|}\,\BC_{a_4|\By|}\bBA^{\SLc}_{a_5a_6}\,\bBA^{\SLd}_{|b\By|}\,\Bg_{a_7]\By}}{\Bg_{\By\By}} 
\nn\\
 &\quad\ 
 -\tfrac{140\,\epsilon_{\SLc\SLd}\,\bBA_{[a_1a_2a_3 |\By|}\,\BC_{a_4|b|}\,\bBA^{\SLc}_{a_5|\By|}\,\bBA^{\SLd}_{a_6|\By|}\,\Bg_{a_7]\By}}{\Bg_{\By\By}} 
 -\tfrac{105\,\epsilon_{\SLc\SLd}\,\bBA_{[a_1a_2 |b\By|}\,\BC_{a_3a_4}\,\bBA^{\SLc}_{a_5|\By|}\,\bBA^{\SLd}_{a_6|\By|}\,\Bg_{a_7]\By}}{\Bg_{\By\By}} 
\nn\\
 &\quad\ 
 -\tfrac{1575}{8}\,\epsilon_{\SLc\SLd}\,\epsilon_{\SLf\SLg}\,\BC_{[a_1a_2}\,\bBA^{\SLc}_{a_3a_4}\,\bBA^{\SLd}_{a_5|\By|}\,\bBA^{\SLf}_{a_6a_7]}\,\bBA^{\SLg}_{b\By} 
\nn\\
 &\quad\ 
 +\tfrac{2835}{4}\,\epsilon_{\SLc\SLd}\,\epsilon_{\SLf\SLg}\,\BC_{[a_1a_2}\,\bBA^{\SLc}_{a_3a_4}\,\bBA^{\SLd}_{a_5|b|}\,\bBA^{\SLf}_{a_6|\By|}\,\bBA^{\SLg}_{a_7]\By} 
\nn\\
 &\quad\ 
 + \tfrac{315}{2}\,\tfrac{\epsilon_{\SLc\SLd}\,\epsilon_{\SLf\SLg}\,\BC_{[a_1a_2}\,\bBA^{\SLc}_{a_3a_4}\,\bBA^{\SLd}_{a_5|\By|}\,\bBA^{\SLf}_{a_6|\By}\,\bBA^{\SLg}_{b\By}\,\Bg_{|a_7]\By}}{\Bg_{\By\By}} \,.
\label{eq:A81-E82}
\end{align}
This is $S$-dual to the $T$-dual rule \eqref{eq:A71-D82}. 
Under $\BB_2=\BC_2=0$\,, this map has been obtained in \cite{hep-th/9908094} [the middle line of Eq.~(5.12)], where $\cN^{(8)}$ corresponds to $\bBE_{8, b\By}$ (under $\BB_2=\BC_2=0$). 

\paragraph{\underline{$T$-duality rule \textcircled{u}: $4^3_3\ (\text{IIA})\leftrightarrow 5^2_3\ (\text{IIB})$}\\}
From the linear map $\cA_{\mu;\sfa_1\cdots \sfa_6\Ay\Az,\sfb_1\sfb_2 \Ay} \!\overset{\tiny\textcircled{u}}{\simeq}\! -\bm{\cA}^{\SLE{2}}_{\mu;\sfa_1\cdots \sfa_6\By,\sfb_1\sfb_2}$, we obtain
\begin{align}
 \AA_{a_1\cdots a_7\Ay, b_1b_2\Ay}
 &\overset{\text{A--B}}{\simeq} \bBE_{a_1\cdots a_7\By, b_1b_2}
 -7\,\bBD_{[a_1\cdots a_6}\,\bBA_{a_7] b_1b_2\By} 
 -\tfrac{7}{2} \,\epsilon_{\SLc\SLd}\,\bBD_{[a_1\cdots a_6}\,\bBA^{\SLc}_{|b_1b_2|}\,\bBA^{\SLd}_{a_7]\By} 
\nn\\
 &\quad\ 
 -\tfrac{35}{2} \,\bBD_{[a_1a_2a_3 |b_1b_2\By|}\, \bigl(\bBA_{a_4\cdots a_7]} + \tfrac{2\,\bBA_{a_4a_5a_6|\By|} \,\Bg_{a_7]\By}}{\Bg_{\By\By}}\bigr)
\nn\\
 &\quad\ 
 +\tfrac{105}{2}\,\epsilon_{\SLc\SLd}\,\bBD_{[a_1\cdots a_4 |\overline{b}_1\By|}\,\bBA^{\SLc}_{a_5a_6}\,\bBA^{\SLd}_{a_7]\overline{b}_2} 
 -\tfrac{105}{2} \,\tfrac{\epsilon_{\SLc\SLd}\,\bBD_{[a_1a_2a_3 |b_1b_2\By|}\,\bBA^{\SLc}_{a_4a_5}\,\bBA^{\SLd}_{a_6|\By|} \,\Bg_{a_7]\By}}{\Bg_{\By\By}}
\nn\\
 &\quad\ 
 +35\,\bBA_{[a_1a_2a_3|\overline{b}_1|}\,\bBA_{a_4a_5a_6|\By|}\,\BC_{a_7]\overline{b}_2} 
 -\tfrac{315}{2} \,\bBA_{[a_1a_2a_3|\overline{b}_1|}\,\bBA_{a_4a_5|\overline{b}_2\By|}\,\BC_{a_6a_7]} 
\nn\\
 &\quad\ 
 + 70\,\bBA_{[a_1a_2a_3|\overline{b}_1|}\,\bBA_{a_4a_5a_6|\overline{b}_2|}\,\BC_{a_7]\By} 
 + \tfrac{105\,\bBA_{[a_1a_2a_3|\By|}\,\bBA_{a_4 |b_1b_2\By|}\,\BC_{a_5a_6} \,\Bg_{a_7]\By}}{\Bg_{\By\By}}
\nn\\
 &\quad\ 
 + \tfrac{105\,\bBA_{[a_1a_2|b_1b_2|}\,\bBA_{a_3a_4a_5|\By|}\,\BC_{a_6|\By|} \,\Bg_{a_7]\By}}{\Bg_{\By\By}}
 -\tfrac{105}{4} \,\epsilon_{\SLc\SLd}\,\bBA_{[a_1a_2|b_1b_2|}\,\bBA^{\SLc}_{a_3a_4}\,\bBA^{\SLd}_{a_5|\By|}\,\BC_{a_6a_7]} 
\nn\\
 &\quad\ 
 +35\,\epsilon_{\SLc\SLd}\,\bBA_{[a_1a_2a_3|\By|}\,\bBA^{\SLc}_{a_4a_5}\,\bBA^{\SLd}_{a_6|\overline{b}_1|}\,\BC_{a_7]\overline{b}_2} 
 +\tfrac{35}{4} \,\epsilon_{\SLc\SLd}\,\bBA_{[a_1a_2a_3|\By|}\,\bBA^{\SLc}_{a_4a_5}\,\bBA^{\SLd}_{|b_1b_2|}\,\BC_{a_6a_7]} 
\nn\\
 &\quad\ 
 -\tfrac{315}{2} \,\tfrac{\epsilon_{\SLc\SLd}\,\bBA_{[a_1a_2 |\overline{b}_1\By|}\,\bBA^{\SLc}_{a_3a_4}\,\bBA^{\SLd}_{a_5|\By|}\,\BC_{a_6a_7]}\,\Bg_{\overline{b}_2\By}}{\Bg_{\By\By}}
 -\tfrac{70\,\epsilon_{\SLc\SLd}\,\bBA_{[a_1a_2a_3|\By|}\,\bBA^{\SLc}_{a_4a_5}\,\bBA^{\SLd}_{a_6|\By}\,\BC_{b_1b_2} \,\Bg_{|a_7]\By}}{\Bg_{\By\By}}
\nn\\
 &\quad\ 
 +\tfrac{35\,\epsilon_{\SLc\SLd}\,\bBA_{[a_1a_2a_3|\By|}\,\bBA^{\SLc}_{a_4a_5}\,\bBA^{\SLd}_{|\overline{b}_1\By|}\,\BC_{a_6|\overline{b}_2|} \,\Bg_{a_7]\By}}{\Bg_{\By\By}}
 -\tfrac{280\,\epsilon_{\SLc\SLd}\,\bBA_{[a_1a_2a_3|\By|}\,\bBA^{\SLc}_{a_4|\overline{b}_1|}\,\bBA^{\SLd}_{a_5|\By|}\, \BC_{a_6|\overline{b}_2|} \,\Bg_{a_7]\By}}{\Bg_{\By\By}}
\nn\\
 &\quad\ 
 -\tfrac{140\,\epsilon_{\SLc\SLd}\,\bBA_{[a_1a_2a_3|\By|}\,\bBA^{\SLc}_{a_4|\overline{b}_1|}\,\bBA^{\SLd}_{|\overline{b}_2\By|}\,\BC_{a_5a_6} \,\Bg_{a_7]\By}}{\Bg_{\By\By}}
 -\tfrac{35}{2} \,\tfrac{\epsilon_{\SLc\SLd}\,\bBA_{[a_1a_2a_3|\By}\,\bBA^{\SLc}_{b_1b_2|}\,\bBA^{\SLd}_{a_4|\By|}\,\BC_{a_5a_6} \,\Bg_{a_7]\By}}{\Bg_{\By\By}}
\nn\\
 &\quad\ 
 -\tfrac{315}{4} \,\tfrac{\epsilon_{\SLc\SLd}\,\bBA_{[a_1a_2|b_1b_2|}\,\bBA^{\SLc}_{a_3|\By|}\,\bBA^{\SLd}_{a_4|\By|}\,\BC_{a_5a_6} \,\Bg_{a_7]\By}}{\Bg_{\By\By}}
\nn\\
 &\quad\ 
 -315\,\epsilon_{\SLc\SLd}\,\epsilon_{\SLf\SLg}\,\bBA^{\SLc}_{[a_1a_2}\,\bBA^{\SLd}_{a_3|\overline{b}_1|}\,\bBA^{\SLf}_{a_4a_5}\,\bBA^{\SLg}_{a_6|\overline{b}_2|}\,\BC_{a_7]\By} 
\nn\\
 &\quad\ 
 +\tfrac{2835}{8}\,\epsilon_{\SLc\SLd}\,\epsilon_{\SLf\SLg}\,\bBA^{\SLc}_{[a_1a_2}\,\bBA^{\SLd}_{a_3|\overline{b}_1|}\,\bBA^{\SLf}_{a_4a_5}\,\bBA^{\SLg}_{|\overline{b}_2\By|}\,\BC_{a_6a_7]} 
\nn\\
 &\quad\ 
 +\tfrac{1575}{2}\,\epsilon_{\SLc\SLd}\,\epsilon_{\SLf\SLg}\,\bBA^{\SLc}_{[a_1a_2}\,\bBA^{\SLd}_{a_3|\overline{b}_1|}\,\bBA^{\SLf}_{a_4|\overline{b}_2|}\,\bBA^{\SLg}_{a_5|\By|}\,\BC_{a_6a_7]} 
\nn\\
 &\quad\ 
 +\tfrac{315}{4}\,\tfrac{\epsilon_{\SLc\SLd}\,\epsilon_{\SLf\SLg}\,\bBA^{\SLc}_{[a_1a_2}\,\bBA^{\SLd}_{a_3|\overline{b}_1|}\,\bBA^{\SLf}_{a_4|\By}\,\bBA^{\SLg}_{\overline{b}_2\By}\,\BC_{|a_5a_6} \,\Bg_{a_7]\By}}{\Bg_{\By\By}}
\nn\\
 &\quad\ 
 +\tfrac{105}{8}\,\tfrac{\epsilon_{\SLc\SLd}\,\epsilon_{\SLf\SLg}\,\bBA^{\SLc}_{[a_1a_2}\,\bBA^{\SLd}_{a_3|\By|}\,\bBA^{\SLf}_{a_4a_5}\,\bBA^{\SLg}_{|\overline{b}_1\By|}\,\BC_{a_6a_7]} \,\Bg_{\overline{b}_2\By}}{\Bg_{\By\By}}
\nn\\
 &\quad\ 
 +\tfrac{105}{2}\,\tfrac{\epsilon_{\SLc\SLd}\,\epsilon_{\SLf\SLg}\,\bBA^{\SLc}_{[a_1a_2}\,\bBA^{\SLd}_{a_3|\By|}\,\bBA^{\SLf}_{a_4|\overline{b}_1|}\,\bBA^{\SLg}_{a_5|\By|}\,\BC_{a_6a_7]} \,\Bg_{\overline{b}_2\By}}{\Bg_{\By\By}}
\nn\\
 &\quad\ 
 -\tfrac{525}{4}\,\tfrac{\epsilon_{\SLc\SLd}\,\epsilon_{\SLf\SLg}\,\bBA^{\SLc}_{[a_1a_2}\,\bBA^{\SLd}_{a_3|\By|}\,\bBA^{\SLf}_{a_4a_5}\,\bBA^{\SLg}_{|\overline{b}_1\By|}\,\BC_{a_6|\overline{b}_2|}\,\Bg_{a_7]\By}}{\Bg_{\By\By}}
\nn\\
 &\quad\ 
 -\tfrac{210\,\epsilon_{\SLc\SLd}\,\epsilon_{\SLf\SLg}\,\bBA^{\SLc}_{[a_1a_2}\,\bBA^{\SLd}_{a_3|\By|}\,\bBA^{\SLf}_{a_4|\overline{b}_1|}\,\bBA^{\SLg}_{a_5|\By|}\,\BC_{a_6|\overline{b}_2|}\,\Bg_{a_7]\By}}{\Bg_{\By\By}}\,.
\label{eq:D83A-E82B}
\end{align}
This is $S$-dual to the $T$-dual rule \eqref{eq:D82A-D82B}. 

\paragraph{\underline{$T$-duality rule \textcircled{v}: $6^1_3\ (\text{IIA})\leftrightarrow 6^{(1,1)}_3\ (\text{IIB})$}\\}
From the linear map $\cA_{\mu;\sfa_1\cdots \sfa_7,\sfa} \overset{\tiny\textcircled{v}}{\simeq} \bm{\cA}_{\mu;\sfa_1\cdots \sfa_7\By,\sfa\By,\By}$, we obtain
\begin{align}
 \AA_{a_1\cdots a_8, b} &\overset{\text{A--B}}{\simeq} \bBA_{a_1\cdots a_8\By, b\By, \By}
 -84\,\epsilon_{\SLc\SLd}\,\bBA_{[a_1\cdots a_5| b\By ,\By|}\,\bigl(\bBA^{\SLc}_{a_6a_7} -\tfrac{2\,\bBA^{\SLc}_{a_6|\By|}\,\Bg_{a_7|\By|}}{\Bg_{\By\By}}\bigr)\,\bBA^{\SLd}_{a_8]\By} 
\nn\\
 &\quad\ +280\,\epsilon_{\SLc\SLd}\,\bBA_{[a_1a_2a_3|\By|}\,\bigl(\bBA_{a_4a_5a_6|b|} - \tfrac{\bBA_{a_4a_5|b\By|}\,\Bg_{a_6|\By|}}{\Bg_{\By\By}}\bigr)\,\bBA^{\SLc}_{a_7|\By|}\,\bBA^{\SLd}_{a_8]\By} 
\nn\\
 &\quad\ -560\,\epsilon_{\SLc\SLd}\,\bBA_{[a_1a_2a_3|\By|}\,\bBA_{a_4a_5|b\By|}\,\bBA^{\SLc}_{a_6a_7}\,\bBA^{\SLd}_{a_8]\By}
\nn\\
 &\quad\ +140\,\epsilon_{\SLa\SLb}\,\epsilon_{\SLc\SLd}\,\bBA_{[a_1a_2a_3|\By|}\,\bBA^{\SLa}_{a_4a_5}\,\bBA^{\SLb}_{a_6|\By|}\,\bigl(\bBA^{\SLc}_{a_7a_8]} - \tfrac{8\,\bBA^{\SLc}_{a_7|\By}\,\Bg_{|a_8]\By}}{\Bg_{\By\By}} \bigr)\,\bBA^{\SLd}_{b\By}
\nn\\
 &\quad\ +140\,\epsilon_{\SLa\SLb}\,\epsilon_{\SLc\SLd}\,\bBA_{[a_1a_2a_3|\By|}\,\bBA^{\SLa}_{a_4a_5}\, \bBA^{\SLb}_{a_6|b|}\,\bBA^{\SLc}_{a_7|\By|}\,\bBA^{\SLd}_{a_8]\By} \,,
\end{align}
which is self-dual under $S$-duality. 
Under $\BB_2=\BC_2=0$\,, this map has been obtained in \cite{hep-th/9908094} [the first line of Eq.~(5.12)], where $N^{(9)}$ may be related to $\bBA_{9,n\By,\By}$. 

\paragraph{\underline{$T$-duality rule \textcircled{w}: $7^{(1,0)}_3\ (\text{IIA})\leftrightarrow 6^{(1,1)}_3\ (\text{IIB})$}\\}
From the linear map $\cA_{\mu;\sfa_1\cdots \sfa_7\Ay\Az,\sfa,\sfb} \overset{\tiny\textcircled{w}}{\simeq} \bm{\cA}_{\mu;\sfa_1\cdots \sfa_7\By,\sfa\By,\sfb}$, we obtain
\begin{align}
 \AA_{a_1\cdots a_8\Ay, b, c} &\overset{\text{A--B}}{\simeq} 
 \bBA_{a_1\cdots a_8\By, b\By, c}
 -\tfrac{8\,\bBA_{[a_1\cdots a_7|b\By , c\By , \By} \,\Bg_{|a_8]\By}}{\Bg_{\By\By}} 
\nn\\
 &\quad\ 
 +8\,\epsilon_{\SLc\SLd}\, \bBA^{\SLc}_{[a_1\cdots a_7|\By , b\By|} \,\bBA^{\SLd}_{a_8]c} 
 +\tfrac{56\,\epsilon_{\SLc\SLd}\, \bBA^{\SLc}_{[a_1\cdots a_6|b\By , c\By|} \,\bBA^{\SLd}_{a_7|\By|}\, \Bg_{a_8]\By}}{\Bg_{\By\By}} 
\nn\\
 &\quad\ 
 -28\,\epsilon_{\SLc\SLd}\, \bBA_{[a_1\cdots a_6|\By , \By|}\, \bBA^{\SLc}_{a_7|b} \,\bigl(\bBA^{\SLd}_{|a_8]c}-\tfrac{2\,\bBA^{\SLd}_{\By c}\, \Bg_{|a_8]\By}}{\Bg_{\By\By}}\bigr)
\nn\\
 &\quad\ 
 -84\,\epsilon_{\SLc\SLd}\, \bBA_{[a_1\cdots a_5|b\By , c|}\, \bigl(\bBA^{\SLc}_{a_6a_7} - \tfrac{2\,\bBA^{\SLc}_{a_6|\By|} \, \Bg_{a_7|\By|}}{\Bg_{\By\By}} \bigr) \,\bBA^{\SLd}_{a_8]\By} 
\nn\\
 &\quad\ 
 +\tfrac{28\,\epsilon_{\SLc\SLd}\,\bBA_{[a_1\cdots a_6|\By,\By|}\, \bBA^{\SLc}_{a_7a_8]} \,\bBA^{\SLd}_{b\By} \,\Bg_{c\By}}{\Bg_{\By\By}} 
 -\tfrac{84\,\epsilon_{\SLc\SLd}\,\bBA_{[a_1\cdots a_5|b,\By,\By|}\, \bBA^{\SLc}_{a_6a_7} \,\bBA^{\SLd}_{a_8]\By} \,\Bg_{c\By}}{\Bg_{\By\By}} 
\nn\\
 &\quad\ 
 - 28 \,\epsilon_{\SLa\SLb}\,\epsilon_{\SLc\SLd}\, \bigl(\bBA^{\SLa}_{[a_1\cdots a_6} - \tfrac{12\,\bBA^{\SLa}_{[a_1\cdots a_5|\By|}\,\Bg_{a_6|\By|}}{\Bg_{\By\By}} \bigr) \,\bBA^{\SLb}_{a_7|b|}\, \bBA^{\SLc}_{a_8]\By} \,\bBA^{\SLd}_{c\By} 
\nn\\
 &\quad\ 
 +168 \,\epsilon_{\SLa\SLb}\,\epsilon_{\SLc\SLd}\, \bBA^{\SLa}_{[a_1\cdots a_5|\By|} \,\bigl(\bBA^{\SLb}_{a_6|b|} - \tfrac{\bBA^{\SLb}_{|\By b|}\,\Bg_{a_6|\By|}}{\Bg_{\By\By}} \bigr)\, \bBA^{\SLc}_{a_7|c|} \,\bBA^{\SLd}_{a_8]\By} 
\nn\\
 &\quad\ 
 +\tfrac{168\,\epsilon_{\SLa\SLb}\,\epsilon_{\SLc\SLd}\,\bBA^{\SLa}_{[a_1\cdots a_5|b|}\,\bBA^{\SLb}_{a_6|\By|}\,\bBA^{\SLc}_{a_7|\By}\,\bBA^{\SLd}_{c\By} \,\Bg_{|a_8]\By}}{\Bg_{\By\By}} 
\nn\\
 &\quad\ 
 -\tfrac{ 84\,\epsilon_{\SLa\SLb}\,\epsilon_{\SLc\SLd}\,\bBA^{\SLa}_{[a_1\cdots a_5|\By|}\,\bBA^{\SLb}_{a_6|b|} \,\bBA^{\SLc}_{a_7|\By|}\,\bBA^{\SLd}_{a_8]\By} \,\Bg_{c\By}}{\Bg_{\By\By}} 
\nn\\
 &\quad\ 
 +70\,\bigl(\bBA_{[a_1\cdots a_4} -\tfrac{\bBA_{[a_1a_2a_3|\By|}\, \Bg_{a_4|\By|}}{\Bg_{\By\By}} \bigr)\,\bBA_{a_5a_6|b\By|}\,\bBA_{a_7a_8]c\By} 
\nn\\
 &\quad\ 
 +630\,\epsilon_{\SLc\SLd}\, \bBA_{[a_1a_2a_3|\By|}\, \bBA_{a_4a_5|b\By|}\, \bBA^{\SLc}_{a_6a_7} \,\bBA^{\SLd}_{a_8]c} 
\nn\\
 &\quad\ 
 +560\,\epsilon_{\SLc\SLd}\, \bBA_{[a_1a_2a_3|b|}\, \bBA_{a_4a_5a_6|\By|}\, \bBA^{\SLc}_{a_7|c|} \,\bBA^{\SLd}_{a_8]\By} 
\nn\\
 &\quad\ 
 -280\,\epsilon_{\SLc\SLd}\, \bBA_{[a_1a_2a_3|b|} \, \bBA_{a_4a_5a_6|\By|}\, \bBA^{\SLc}_{a_7a_8]} \,\bBA^{\SLd}_{c\By} 
\nn\\
 &\quad\ 
 +\tfrac{1155\,\epsilon_{\SLc\SLd}\,\bBA_{[a_1a_2a_3|\By|}\, \bBA_{a_4a_5|b\By|}\, \bBA^{\SLc}_{a_6a_7}\,\bBA^{\SLd}_{|c\By|} \,\Bg_{a_8]\By}}{\Bg_{\By\By}} 
\nn\\
 &\quad\ 
 -\tfrac{ 525\,\epsilon_{\SLc\SLd}\,\bBA_{[a_1a_2a_3|\By|}\, \bBA_{a_4a_5|b\By|}\, \bBA^{\SLc}_{a_6a_7}\,\bBA^{\SLd}_{a_8]\By} \,\Bg_{c\By}}{\Bg_{\By\By}} 
\nn\\
 &\quad\ 
 +\tfrac{ 210\,\epsilon_{\SLc\SLd}\,\bBA_{[a_1a_2a_3|\By|}\, \bBA_{a_4a_5|b\By|}\, \bBA^{\SLc}_{a_6|c|}\,\bBA^{\SLd}_{a_7|\By|} \,\Bg_{a_8]\By}}{\Bg_{\By\By}} 
\nn\\
 &\quad\ 
 -\tfrac{ 560\,\epsilon_{\SLc\SLd}\,\bBA_{[a_1a_2a_3|b|}\, \bBA_{a_4a_5a_6|\By|}\, \bBA^{\SLc}_{a_7|\By|}\,\bBA^{\SLd}_{a_8]\By} \,\Bg_{c\By}}{\Bg_{\By\By}} 
\nn\\
 &\quad\ 
 +\tfrac{1120\,\epsilon_{\SLc\SLd}\,\bBA_{[a_1a_2a_3|b|}\, \bBA_{a_4a_5a_6|\By|}\, \bBA^{\SLc}_{a_7|\By}\,\bBA^{\SLd}_{c\By} \,\Bg_{|a_8]\By}}{\Bg_{\By\By}} 
\nn\\
 &\quad\ 
 +105\,\epsilon_{\SLa\SLb}\,\epsilon_{\SLc\SLd}\, \bBA_{[a_1\cdots a_4}\, \bBA^{\SLa}_{a_5a_6} \,\bBA^{\SLb}_{a_7|b|}\, \bBA^{\SLc}_{a_8]\By} \,\bBA^{\SLd}_{c\By} 
\nn\\
 &\quad\ 
 -105\,\epsilon_{\SLa\SLb}\,\epsilon_{\SLc\SLd}\, \bBA_{[a_1\cdots a_4}\, \bBA^{\SLa}_{a_5|b|} \,\bBA^{\SLb}_{a_6|c|}\, \bBA^{\SLc}_{a_7|\By|} \,\bBA^{\SLd}_{a_8]\By} 
\nn\\
 &\quad\ 
 -210\,\epsilon_{\SLa\SLb}\,\epsilon_{\SLc\SLd}\, \bBA_{[a_1a_2a_3|\By|}\, \bBA^{\SLa}_{a_4a_5} \,\bBA^{\SLb}_{a_6|b|}\, \bigl(\bBA^{\SLc}_{a_7a_8]}-\tfrac{3\,\bBA^{\SLc}_{a_7|\By|}\,\Bg_{|a_8]\By}}{\Bg_{\By\By}} \bigr) \,\bBA^{\SLd}_{c\By} 
\nn\\
 &\quad\ 
 +210\,\epsilon_{\SLa\SLb}\,\epsilon_{\SLc\SLd}\, \bBA_{[a_1a_2a_3|b|}\, \bBA^{\SLa}_{a_4a_5} \,\bBA^{\SLb}_{a_6|\By|}\, \bigl(\bBA^{\SLc}_{a_7a_8]}-\tfrac{8\,\bBA^{\SLc}_{a_7|\By|}\,\Bg_{|a_8]\By}}{\Bg_{\By\By}} \bigr) \,\bBA^{\SLd}_{c\By} 
\nn\\
 &\quad\ 
 -\tfrac{ 315\,\epsilon_{\SLa\SLb}\,\epsilon_{\SLc\SLd}\,\bBA_{[a_1a_2a_3|\By|}\, \bBA^{\SLa}_{a_4a_5}\,\bBA^{\SLb}_{|b\By|}\,\bBA^{\SLc}_{a_6a_7}\,\bBA^{\SLd}_{|c\By|} \,\Bg_{a_8]\By}}{\Bg_{\By\By}} 
\nn\\
 &\quad\ 
 +\tfrac{315}{2} \, \tfrac{\epsilon_{\SLa\SLb}\,\epsilon_{\SLc\SLd}\,\bBA_{[a_1a_2a_3|\By|}\,\bBA^{\SLa}_{a_4a_5}\,\bBA^{\SLb}_{a_6|\By|}\,\bBA^{\SLc}_{a_7a_8]}\,\bBA^{\SLd}_{b\By}\,\Bg_{c\By}}{\Bg_{\By\By}} 
\nn\\
 &\quad\ 
 -\tfrac{315}{2} \, \tfrac{\epsilon_{\SLa\SLb}\,\epsilon_{\SLc\SLd}\,\bBA_{[a_1a_2a_3|\By|}\,\bBA^{\SLa}_{a_4a_5}\,\bBA^{\SLb}_{a_6|b|}\,\bBA^{\SLc}_{a_7|\By|}\,\bBA^{\SLd}_{a_8]\By}\,\Bg_{c\By}}{\Bg_{\By\By}} 
\nn\\
 &\quad\ 
 +\tfrac{315}{2} \, \tfrac{\epsilon_{\SLa\SLb}\,\epsilon_{\SLc\SLd}\,\bBA_{[a_1a_2|b\By|}\,\bBA^{\SLa}_{a_3a_4}\,\bBA^{\SLb}_{a_5|\By|}\,\bBA^{\SLc}_{a_6a_7}\,\bBA^{\SLd}_{|c\By|}\,\Bg_{a_8]\By}}{\Bg_{\By\By}} 
\nn\\
 &\quad\ 
 -\tfrac{315}{2} \, \tfrac{\epsilon_{\SLa\SLb}\,\epsilon_{\SLc\SLd}\,\bBA_{[a_1a_2|b\By|}\,\bBA^{\SLa}_{a_3a_4}\,\bBA^{\SLb}_{a_5|c|}\,\bBA^{\SLc}_{a_6|\By|}\,\bBA^{\SLd}_{a_7|\By|}\,\Bg_{a_8]\By}}{\Bg_{\By\By}} 
\nn\\
 &\quad\ 
 +\tfrac{3045}{2}\,\epsilon_{\SLa\SLb}\,\epsilon_{\SLc\SLd}\,\epsilon_{\SLf\SLg}\, \bBA^{\SLa}_{[a_1a_2} \,\bBA^{\SLb}_{a_3|b|}\,\bBA^{\SLc}_{a_4a_5} \,\bBA^{\SLd}_{a_6|\By|}\, \,\bBA^{\SLf}_{a_7a_8]} \,\bBA^{\SLg}_{c\By} 
\nn\\
 &\quad\ 
 -\tfrac{11865\,\epsilon_{\SLa\SLb}\,\epsilon_{\SLc\SLd}\,\epsilon_{\SLf\SLg}\, \bBA^{\SLa}_{[a_1a_2}\,\bBA^{\SLb}_{a_3|b|}\,\bBA^{\SLc}_{a_4a_5}\,\bBA^{\SLd}_{a_6|\By|}\, \bBA^{\SLf}_{a_7|\By}\,\bBA^{\SLg}_{c\By}\, \Bg_{|a_8]\By}}{\Bg_{\By\By}} 
\nn\\
 &\quad\ 
 -\tfrac{6405}{4} \,\tfrac{\epsilon_{\SLa\SLb}\,\epsilon_{\SLc\SLd}\,\epsilon_{\SLf\SLg}\, \bBA^{\SLa}_{[a_1a_2}\,\bBA^{\SLb}_{a_3|\By|}\,\bBA^{\SLc}_{a_4a_5}\,\bBA^{\SLd}_{|b\By|}\,\bBA^{\SLf}_{a_6a_7}\,\bBA^{\SLg}_{|c\By|} \,\Bg_{a_8]\By}}{\Bg_{\By\By}} \,.
\end{align}

\paragraph{\underline{$T$-duality rule \textcircled{x}: $6^1_3\ (\text{IIA})\leftrightarrow 7_3\ (\text{IIB})$}\\}
From the linear map $\cA_{\mu;\sfa_1\cdots \sfa_6\Ay,\Ay} \!\overset{\tiny\textcircled{x}}{=}\! \bm{\cA}^{\SLE{22}}_{\mu;\sfa_1\cdots \sfa_6\By}$, we obtain
\begin{align}
 \AA_{a_1\cdots a_7\Ay, \Ay}
 &\overset{\text{A--B}}{=} \BE_{a_1\cdots a_7\By}
 -7\,\bigl(\BB_{[a_1\cdots a_6}-\tfrac{6\,\BB_{[a_1\cdots a_5|\By|}\,\Bg_{a_6|\By|}}{\Bg_{\By\By}}\bigr)\,\BC_{a_7]\By} 
\nn\\
 &\quad\ 
 -35 \,\BC_{[a_1a_2a_3|\By|}\,\bigl(\BC_{a_4a_5}\,\BC_{a_6a_7]} - \tfrac{4\,\BC_{a_4a_5}\,\BC_{a_6|\By|}\,\Bg_{a_7]\By}}{\Bg_{\By\By}}\bigr) \,.
\label{eq:A81-BNS7}
\end{align}
The inverse map has been obtained in Eq.~(4.7) of \cite{hep-th/9908094},\footnote{The correspondent of \eqref{eq:A81-BNS7} has been given in Eq.~(3.3) of \cite{hep-th/9908094}, but there seems to be a small discrepancy regarding the terms including $\BB_2\,(\BC_2)^2$.} and in our convention, we have
\begin{align}
 \BE_{a_1\cdots a_7\By}
 &\overset{\text{B--A}}{=} 
 \AA_{a_1\cdots a_7\Ay, \Ay}
 +7\, \AA_{[a_1\cdots a_6|\Ay , \Ay}\, \bigl(\AC_{|a_7]} - \tfrac{\AC_{|\Ay|}\,\Ag_{|a_7]\Ay}}{\Ag_{\Ay\Ay}}\bigr)
\nn\\
 &\quad\ 
 +35\, \bigl(\AC_{[a_1a_2a_3}-\tfrac{3\,\AC_{[a_1a_2|\Ay|}\, \Ag_{a_3|\Ay|}}{\Ag_{\Ay\Ay}} \bigr)\,\AC_{a_4a_5|\Ay|}\,\AC_{a_6a_7]\Ay}
\nn\\
 &\quad\ 
 +70\,\bigl(2\,\AC_{[a_1a_2a_3}\,\AB_{a_4|\Ay|} + 3\,\AC_{[a_1a_2|\Ay|}\,\AB_{a_3a_4}\bigr)\,\AC_{a_5a_6|\Ay|}\, \bigl(\AC_{a_7]} - \tfrac{\AC_{|\Ay|}\,\Ag_{a_7]\Ay}}{\Ag_{\Ay\Ay}}\bigr)\,. 
\end{align}
The $S$-dual counterpart of this $T$-duality is the map \textcircled{l} connecting $\AC_9$ and $\BC_8$\,. 

\paragraph{\underline{$T$-duality rule \textcircled{y}: $7^{(1,0)}_3\ (\text{IIA})\leftrightarrow 7_3\ (\text{IIB})$}\\}
From the linear map $\cA_{\mu;\sfa_1\cdots \sfa_7\Ay\Az,\Ay,\Ay} \!\overset{\tiny\textcircled{y}}{=}\! \bm{\cA}^{\SLE{22}}_{\mu;\sfa_1\cdots \sfa_7}$, we obtain
\begin{align}
 \AA_{a_1\cdots a_8\Ay, \Ay, \Ay}
 &\overset{\text{A--B}}{=} \BE_{a_1\cdots a_8} -\tfrac{8\,\BE_{[a_1\cdots a_7|\By|}\,\Bg_{a_8]\By}}{\Bg_{\By\By}}
 +\tfrac{210\,\BC_{[a_1a_2a_3|\By|}\,\BC_{a_4a_5}\,\BC_{a_6a_7}\,\Bg_{a_8]\By}}{\Bg_{\By\By}}
\nn\\
 &\quad\ - 315 \,\BC_{[a_1a_2}\,\BC_{a_3a_4}\,\bigl(\BB_{a_5a_6}\,\BC_{a_7a_8]} - \tfrac{6\,\BB_{a_5a_6}\,\BC_{a_7|\By|}\,\Bg_{a_8]\By}}{\Bg_{\By\By}}\bigr) \,.
\end{align}
This is $S$-dual to the $T$-duality is the map \textcircled{m} connecting $\AC_9$ and $\BC_{10}$\,. 
We can also find the inverse map as
\begin{align}
 \BE_{a_1\cdots a_8} 
 &\overset{\text{B--A}}{=} \AA_{a_1\cdots a_8\Ay, \Ay, \Ay} 
 -8\, \AA_{[a_1\cdots a_7|\Ay , \Ay|}\, \AB_{a_8]\Ay}
 + 56\, \AA_{[a_1\cdots a_6|\Ay , \Ay|}\,\AB_{a_7|\Ay|} \, \bigl(\AC_{a_8]}- \tfrac{\AC_{|\Ay|}\,\Ag_{a_8]\Ay}}{\Ag_{\Ay\Ay}} \bigr)
\nn\\
 &\quad\ -70\, \bigl(\AC_{[a_1a_2a_3}-\tfrac{3\, \AC_{[a_1a_2|\Ay|}\,\Ag_{a_3|\Ay|}}{\Ag_{\Ay\Ay}}\bigr)\, \AC_{a_4a_5|\Ay|}\, \AC_{a_6a_7|\Ay|}\,\AB_{a_8]\Ay} 
\nn\\
 &\quad\ +315\,\AC_{[a_1a_2|\Ay|}\,\AC_{a_3a_4|\Ay|}\,\AC_{a_5a_6|\Ay|}\,\bigl(\AB_{a_7a_8}-\tfrac{2\,\AB_{a_7|\Ay|}\,\Ag_{a_8]\Ay}}{\Ag_{\Ay\Ay}}\bigr) 
\nn\\
 &\quad\ +1680\, \AB_{[a_1a_2}\,\AB_{a_3|\Ay|}\,\AC_{a_4a_5|\Ay|}\,\AC_{a_6a_7|\Ay|}\,\bigl(\AC_{a_8]}- \tfrac{\AC_{|\Ay|}\,\Ag_{a_8]\Ay}}{\Ag_{\Ay\Ay}}\bigr) \,.
\end{align}

So far, we have considered the potentials which couple to exotic $(7-p+n)^{(n,p-n)}_3$-branes. 
Finally, let us consider the only map, which is associated with branes with tension $T\propto g_s^{-4}$\,. 

\paragraph{\underline{$T$-duality rule \textcircled{z}: $8^{(1,0)}_4\ (\text{IIA})\leftrightarrow 9_4\ (\text{IIB})$}\\}
From the linear map $\cA_{\mu;\sfa_1\cdots \sfa_8\Ay,\Ay,\Ay} \!\overset{\tiny\textcircled{z}}{=}\! -\bm{\cA}^{\SLE{222}}_{\mu;\sfa_1\cdots \sfa_8\By}$, we find
\begin{align}
 \AA_{a_1\cdots a_9\Ay, \Ay, \Ay}
 &\overset{\text{A--B}}{=} \BF_{a_1\cdots a_9\By}
 -9\,\bigl(\BE_{[a_1\cdots a_8} - \tfrac{8\,\BE_{[a_1\cdots a_7|\By|}\,\Bg_{a_8|\By|}}{\Bg_{\By\By}} \bigr)\,\BC_{a_9]\By} 
\nn\\
 &\quad\ -315\,\BC_{[a_1a_2a_3|\By|}\,\BC_{a_4a_5}\,\BC_{a_6a_7}\,\bigl(\BC_{a_8a_9]} - \tfrac{6\,\BC_{a_8|\By|}\,\Bg_{a_9]\By}}{\Bg_{\By\By}}\bigr)\,.
\label{eq:F1011A-F10B}
\end{align}
The potentials $\AA_{10,1,1}$ and $\BF_{10}$ have level 4 while $\BE_8$ has level 3, and again we can see that the levels indeed match on both sides.

\subsubsection{$T$-dual-manifest redefinitions}
\label{sec:EF-potential}

Having obtained the $T$-duality rules, let us find the field redefinitions that map our mixed-symmetry potentials to the $T$-duality-covariant potentials. 
As studied in \cite{1108.5067}, potentials $\TmE_{8+n,p,n}$ ($n=0,1,2$ and $p=\text{odd/even}$ in type IIA/IIB theory) that couple to the exotic $(7-p+n)^{(n,p-n)}_3$-branes constitute the $T$-duality-covariant potential $\TmE_{MN\dot{a}}$ $(\dot{a}=1,\dotsc,512)$. 
This transforms in the 87040-dimensional tensor-spinor representation of the $\OO(10,10)$ group. 
By using the notation of the $\OO(10,10)$ spinor given in section \ref{sec:D-potential}, we denote it as $\ket{\TmE_{MN}}$ that satisfies the following conditions:
\begin{align}
 \ket{\TmE_{MN}} = - \ket{\TmE_{NM}}\,,\qquad \Gamma^N\,\ket{\TmE_{NM}} = 0\,,\qquad \Gamma^{11}\ket{\TmE_{MN}} = \mp \ket{\TmE_{MN}} \quad (\text{IIA/IIB})\,.
\end{align}
As discussed in \cite{1903.05601}, if we truncate the components which do not couple to supersymmetric branes, $\ket{E_{MN}}$ can be parameterized as
\begin{align}
 \ket{\TmE^{mn}} &= \sum_p \tfrac{1}{8!\,p!}\,\epsilon^{mn q_1\cdots q_8}\,\TmE_{q_1\cdots q_8,r_1\cdots r_p}\,\Gamma^{r_1\cdots r_p}\,\ket{0}\,,
\\
 \ket{\TmE^m{}_n} &= \sum_p \tfrac{1}{9!\,p!}\,\epsilon^{m q_1\cdots q_9}\,\TmE_{q_1\cdots q_9,r_1\cdots r_p,n}\,\Gamma^{r_1\cdots r_p}\,\ket{0}\,,
\\
 \ket{\TmE_{mn}} &= \sum_p \tfrac{1}{10!\,p!}\,\epsilon^{q_1\cdots q_{10}}\,\TmE_{q_1\cdots q_{10},r_1\cdots r_p,mn}\,\Gamma^{r_1\cdots r_p}\,\ket{0}\,.
\end{align}
The constraint $\Gamma^N\,\ket{\TmE_{NM}} = 0$ is automatically satisfied under the restriction rule for the indices. 
Under the factorized $T$-duality along the $x^y$-direction, it transforms as
\begin{align}
 \ket{\TmE'_{M_1M_2}} = \Lambda_{M_1}{}^{N_1}\,\Lambda_{M_2}{}^{N_2}\, \bigl(\Gamma^y -\Gamma_y\bigr)\,\Gamma^{11}\,\ket{\TmE_{N_1N_2}} \,,
\end{align}
and in terms of the components, we have
\begin{align}
\begin{split}
 \TmE_{a_1\cdots a_{8+n}, b_1\cdots b_p, c_1\cdots c_n} &\overset{\text{A--B}}{\simeq} \TmE_{a_1\cdots a_{8+n} \By , b_1\cdots b_p \By , c_1\cdots c_n \By} \,,
\\
 \TmE_{a_1\cdots a_{7+n}\Ay, b_1\cdots b_p, c_1\cdots c_n} &\overset{\text{A--B}}{\simeq} \TmE_{a_1\cdots a_{7+n} \By , b_1\cdots b_p \By , c_1\cdots c_n} \,,
\\
 \TmE_{a_1\cdots a_{7+n}\Ay, b_1\cdots b_{p-1} \Ay, c_1\cdots c_n} &\overset{\text{A--B}}{\simeq} \TmE_{a_1\cdots a_{7+n} \By , b_1\cdots b_{p-1} , c_1\cdots c_n} \,,
\\
 \TmE_{a_1\cdots a_{8+n}\Ay, b_1\cdots b_{p-1} \Ay, c_1\cdots c_n\Ay} &\overset{\text{A--B}}{\simeq} \TmE_{a_1\cdots a_{8+n} , b_1\cdots b_{p-1} , c_1\cdots c_n} \,,
\end{split}
\label{eq:E-T-duality}
\end{align}
where $n=0,1,2$ and $p=1,3,5,7$. 
The $T$-duality web for the family of potentials $\TmE_{8+n,p,n}$, which contains our $T$-dualities \textcircled{t}--\textcircled{y}, can be summarized as follows:
\begin{align}
\vcenter{\xymatrix@C=15pt@R=18pt{
 \TmE_{8,7} \ar@{<.>}[d]\ar@{<.>}[r] & \TmE_{8,6} \ar@{<.>}[d]\ar@{<.>}[r] & \TmE_{8,5} \ar@{<.>}[d]\ar@{<.>}[r] & \TmE_{8,4} \ar@{<.>}[d]\ar@{<.>}[r] & \TmE_{8,3} \ar@{<.>}[d]\ar@{<->}[r]^{\tiny\textcircled{u}} & \TmE_{8,2} \ar@{<.>}[d]\ar@{<->}[r]^{\tiny\textcircled{t}} & \TmE_{8,1} \ar@{<->}[d]^{\tiny\textcircled{v}}\ar@{<->}[r]^{\tiny\textcircled{x}} & \TmE_{8} \ar@{<->}[d]^{\tiny\textcircled{y}} \\
 \TmE_{9,8,1} \ar@{<.>}[d]\ar@{<.>}[r] & \TmE_{9,7,1} \ar@{<.>}[d]\ar@{<.>}[r] & \TmE_{9,6,1} \ar@{<.>}[d]\ar@{<.>}[r] & \TmE_{9,5,1} \ar@{<.>}[d]\ar@{<.>}[r] & \TmE_{9,4,1} \ar@{<.>}[d]\ar@{<.>}[r] & \TmE_{9,3,1} \ar@{<.>}[d]\ar@{<.>}[r] & \TmE_{9,2,1} \ar@{<.>}[d]\ar@{<->}[r]^{\tiny\textcircled{w}} & \TmE_{9,1,1} \ar@{<.>}[d] \\
 \TmE_{10,9,2} \ar@{<.>}[r] & \TmE_{10,8,2} \ar@{<.>}[r] & \TmE_{10,7,2} \ar@{<.>}[r] & \TmE_{10,6,2} \ar@{<.>}[r] & \TmE_{10,5,2} \ar@{<.>}[r] & \TmE_{10,4,2} \ar@{<.>}[r] & \TmE_{10,3,2} \ar@{<.>}[r] & \TmE_{10,2,2} 
}}.
\end{align}

Through trial and error, we have found that the following redefinitions indeed map our mixed-symmetry potentials to the $T$-duality-covariant potentials $\TmE_{8+n,p,n}$:
\begin{align}
\text{\underline{Type IIA:}}
\nn\\
 \TmE_{m_1\cdots m_8 , n}
 &\simeq \AA_{m_1\cdots m_8, n}
 - 56 \,\bigl(\AB_{[m_1\cdots m_5|n|} + \tfrac{5}{3} \,\AC_{[m_1\cdots m_4|n|}\,\AC_{m_5}\bigr)\, \AC_{m_6m_7m_8]} \,,
\\
 \TmE_{m_1\cdots m_8 , n_1n_2n_3}
 &\simeq \AA_{m_1\cdots m_8 , n_1n_2n_3}
 +\tfrac{280}{3} \,\AC_{[m_1\cdots m_4|n_1n_2n_3|}\,\AC_{m_5m_6m_7}\,\AC_{m_8]} 
\nn\\
 &\quad -28\,\AB_{[m_1m_2m_3 |n_1n_2n_3|}\,\bigl(\AC_{m_4\cdots m_8]} -5\, \AC_{m_4m_5m_6}\,\AB_{m_7m_8]}\bigr) 
\nn\\
 &\quad -280\,\AC_{[m_1m_2m_3|\overline{n}_1\overline{n}_2|}\,\AC_{m_4m_5m_6}\,\AC_{m_7m_8]\overline{n}_3}
 +56\,\AC_{[m_1\cdots m_5}\,\AC_{m_6m_7|n_1n_2n_3|}\,\AC_{m_8]} 
\nn\\
 &\quad +56\,\AC_{[m_1\cdots m_5}\,\bigl(\AC_{m_6m_7m_8]}\, \AC_{\overline{n}_1} -3\, \AC_{m_6m_7|\overline{n}_1|}\,\AC_{m_8]} \bigr)\, \AB_{\overline{n}_2\overline{n}_3}
\nn\\
 &\quad -210\,\AC_{[m_1m_2m_3}\,\AC_{m_4m_5|\overline{n}_1|}\, \AC_{m_6|\overline{n}_2\overline{n}_3|}\,\AB_{m_7m_8]} \,,
\\
 \TmE_{m_1\cdots m_9 , n , p}
 &\simeq \AA_{m_1\cdots m_9 , n , p} 
 -36\,\AA_{[m_1\cdots m_7| n, p|} \,\AB_{m_8m_9]} 
 -84\,\AA_{[m_1\cdots m_6| n , p|}\, \AC_{m_7m_8m_9]}
\nn\\
 &\quad -252\,\AB_{[m_1\cdots m_6}\, \AC_{m_7m_8|n|}\, \AB_{m_9]p} 
 +21\, \AC_{[m_1\cdots m_5}\, \AC_{m_6\cdots m_9] n}\, \AC_{p} 
\nn\\
 &\quad 
 +420\,\AC_{[m_1\cdots m_4|n|}\,\AC_{m_5m_6m_7}\, \bigl(\AC_{m_8m_9]p} -2\,\AB_{m_8|p|}\,\AC_{m_9]} \bigr)
\nn\\
 &\quad -1575\,\AC_{[m_1m_2m_3}\, \AC_{m_4m_5|n|}\,\AC_{m_6m_7|p|}\,\AB_{m_8m_9]} \,,
\\
\text{\underline{Type IIB:}}
\nn\\
 \TmE_{m_1\cdots m_8}
 &= \BE_{m_1 \cdots m_8}
 -28\,\BB_{[m_1\cdots m_6}\,\BC_{m_7m_8]} 
 +140\,\BC_{[m_1\cdots m_4}\,\BC_{m_5m_6}\,\BC_{m_7m_8]} 
\nn\\
 &\quad +\tfrac{35}{3}\,\BC_{[m_1\cdots m_4}\,\BC_{m_5\cdots m_8]}\,\BC_0 \,,
\\
 \TmE_{m_1\cdots m_8 , n_1n_2}
 &\simeq
 \bBE_{m_1\cdots m_8, n_1n_2} 
 +420\,\BB_{[m_1\cdots m_4 |n_1n_2|}\,\BB_{m_5m_6}\,\BC_{m_7m_8]} 
 + 56\,\BB_{[m_1\cdots m_6}\,\BB_{m_7|\overline{n}_1|}\,\BC_{m_8]\overline{n}_2} 
\nn\\
 &\quad 
 -28\,\BB_{[m_1\cdots m_6}\,\bigl(\BC_{m_7m_8] n_1n_2} + \BB_{m_7m_8]}\,\BC_{n_1n_2}\bigr)
\nn\\
 &\quad
 +\tfrac{70}{3} \,\BC_{[m_1\cdots m_4|n_1n_2|}\,\BC_{m_5\cdots m_8]}\,\BC_0 
 -70\,\BC_{[m_1\cdots m_4|n_1n_2|}\,\BC_{m_5m_6}\,\BC_{m_7m_8]} 
\nn\\
 &\quad 
 -\tfrac{35}{6} \,\BC_{[m_1\cdots m_4}\,\BC_{m_5\cdots m_8]}\,\bigl(\BC_{n_1n_2}+2\,\BB_{n_1n_2}\,\BC_0\bigr)
\nn\\
 &\quad 
 +490\,\BC_{[m_1\cdots m_4}\,\BC_{m_5m_6 |n_1n_2|}\,\BC_{m_7m_8]} 
 +1260\, \BC_{[m_1\cdots m_4}\,\BC_{m_5m_6}\,\BC_{m_7 |\overline{n}_1|}\,\BB_{m_8]\overline{n}_2} 
\nn\\
 &\quad
 - 630\,\BC_{[m_1m_2|n_1n_2|}\,\BC_{m_3m_4}\,\BC_{m_5m_6}\,\BB_{m_7m_8]} 
 - 455\,\BC_{[m_1\cdots m_4}\,\BC_{m_5m_6}\,\BC_{m_7m_8]}\,\BB_{n_1n_2} 
\nn\\
 &\quad 
 +\tfrac{1995}{2}\,\BC_{[m_1m_2}\,\BC_{m_3m_4}\,\BC_{m_5m_6}\,\BB_{m_7m_8]}\,\BB_{n_1n_2} 
\nn\\
 &\quad
 -\tfrac{1155}{4}\,\BB_{[m_1m_2}\,\BB_{m_3m_4}\,\BC_{m_5m_6}\,\BC_{m_7m_8]}\,\BC_{n_1n_2} 
\nn\\
 &\quad
 -2310\,\BC_{[m_1m_2}\,\BC_{m_3m_4}\,\BC_{m_5m_6}\,\BB_{m_7|\overline{n}_1|}\,\BB_{m_8]\overline{n}_2}\,,
\\
 \TmE_{m_1\cdots m_9 , n_1n_2, p}
 &\simeq \BA_{m_1 \cdots m_9 , n_1n_2 , p}
 -126\,\BA_{[m_1\cdots m_5 |n_1n_2 , p|}\, \bigl(\BC_{m_6\cdots m_9]} -3\,\BB_{m_6m_7}\, \BC_{m_8m_9]} \bigr)
\nn\\
 &\quad
 +1260\,\BC_{[m_1\cdots m_4|n_1n_2|}\, \BB_{m_5m_6}\,\BC_{m_7m_8}\,\BC_{m_9]p}
 + 420\,\BC_{[m_1\cdots m_4|n_1n_2|}\, \BC_{m_5m_6m_7|p|}\,\BC_{m_8m_9]} 
\nn\\
 &\quad
 -420\,\BC_{[m_1\cdots m_4 |n_1n_2|}\, \BC_{m_5m_6m_7|p|}\,\BB_{m_8m_9]}\,\BC_0 
\nn\\
 &\quad
 +210\,\BC_{[m_1\cdots m_4}\,\BC_{m_5m_6m_7|p|}\,\bigl(\BB_{m_8m_9]}\,\BC_{n_1n_2} -3\, \BC_{m_8m_9]}\,\BB_{n_1n_2}\bigr)
\nn\\
 &\quad
 +315\,\BC_{[m_1\cdots m_4}\,\BC_{m_5m_6|n_1n_2|}\,\bigl(\BC_{m_7m_8}\,\BB_{m_9]p} - \BB_{m_7m_8}\,\BC_{m_9]p} \bigr)
\nn\\
 &\quad
 - 105\,\BC_{[m_1\cdots m_4}\,\BC_{m_5\cdots m_8}\,\BB_{m_9]p}\,\BB_{n_1n_2}\,\BC_0 
\nn\\
 &\quad
 +630\,\BC_{[m_1\cdots m_4}\,\BC_{m_5m_6}\,\BC_{m_7m_8}\,\BB_{m_9]p}\,\BB_{n_1n_2}
\nn\\
 &\quad
 +2520\,\BC_{[m_1m_2m_3|p|}\,\BC_{m_4m_5}\,\BC_{m_6m_7}\,\BB_{m_8|\overline{n}_1|}\,\BB_{m_9]\overline{n}_2}
\nn\\
 &\quad
 -1260\,\BC_{[m_1m_2m_3|p|}\,\BB_{m_4m_5}\,\BB_{m_6m_7}\,\BC_{m_8|\overline{n}_1|}\,\BC_{m_9]\overline{n}_2}
\nn\\
 &\quad
 -2520\,\BC_{[m_1m_2m_3|\overline{n}_1|}\,\BC_{m_4m_5}\,\BC_{m_6|\overline{n}_2|}\,\BB_{m_7m_8}\BB_{m_9]p} 
\nn\\
 &\quad
 - 630\,\BB_{[m_1m_2}\,\BB_{m_3m_4}\,\BB_{m_5m_6}\,\BC_{m_7m_8}\BC_{m_9]p}\, \BC_{n_1n_2} \,.
\end{align}

Similarly, even for the potentials $\AA_{10,1,1}$ and $\BF_{10}$\,, if we consider the redefinitions,
\begin{align}
 \TmF_{m_1\cdots m_{10}, b , c} 
 &\simeq \AA_{m_1\cdots m_{10}, b , c} 
 -120\,\AA_{[m_1\cdots m_7|b , c|}\, \AC_{m_8m_9m_{10}]}
\nn\\
 &\quad +1260\,\AB_{[m_1\cdots m_5 |b|}\,\AC_{m_6m_7m_8}\,\AC_{m_9m_{10}]c} 
 +42\,\AC_{[m_1\cdots m_6|b|}\,\AC_{m_7m_8m_9}\,\AC_{m_{10}]}\,\AC_c 
\nn\\
 &\quad 
 +63\,\AC_{[m_1\cdots m_5}\,\bigl(\AC_{m_6\cdots m_9|b|}\,\AC_{m_{10}]} - \AC_{m_6m_7m_8}\,\AC_{m_9m_{10}]b}\bigr)\,\AC_c
\nn\\
 &\quad -1575\,\AC_{[m_1\cdots m_4 |b|}\,\AC_{m_5m_6m_7}\,\AC_{m_8m_9|c|}\,\AC_{m_{10}]} \,,
\\
 \TmF_{m_1\cdots m_{10}}
 &= \BF_{m_1\cdots m_{10}}
 -45\,\BE_{[m_1\cdots m_8}\,\BC_{m_9m_{10}]} 
 +630\,\BB_{[m_1\cdots m_6}\,\BC_{m_7m_8}\,\BC_{m_9m_{10}]} 
\nn\\
 &\quad + \tfrac{21}{2}\, \BC_{[m_1\cdots m_6}\,\bigl(\BC_{m_7\cdots m_{10}]}\,\BC_0 -3\,\BC_{m_7m_8}\,\BC_{m_9m_{10}]}\bigr)\,\BC_0 
\nn\\
 &\quad - \tfrac{315}{2}\, \BC_{[m_1\cdots m_4}\,\bigl(\BC_{m_5\cdots m_8}\,\BC_0 +15\, \BC_{m_5m_6}\,\BC_{m_7m_8}\bigr)\,\BC_{m_9m_{10}]} 
\nn\\
 &\quad + 3780\,\BB_{[m_1m_2}\,\BC_{m_3m_4}\,\BC_{m_5m_6}\,\BC_{m_7m_8}\,\BC_{m_9m_{10}]} \,,
\end{align}
the $T$-duality rule \eqref{eq:F1011A-F10B} is simplified as
\begin{align}
 \TmF_{a_1\cdots a_9\Ay , \Ay , \Ay} \overset{\text{A--B}}{=} \TmF_{a_1\cdots a_9\By}\,.
\end{align}
As is studied in \cite{1201.5819,1210.1422}, potentials that couple to the $(9-p)^{(p,0)}_4$-branes ($p=1,3,5,7,9/0,2,4,6,8$ in type IIA/IIB theory) are packaged into the self-dual $\OO(10,10)$ tensor $\TmF^+_{M_1\cdots M_{10}}$\,.
Then, the above potentials, $\TmF_{10,1,1}$ and $\TmF_{10}$, will be identified as the particular components of $\TmF^+_{M_1\cdots M_{10}}$\,. 
The potential $\TmF^+_{M_1\cdots M_{10}}$ contains the following family of potentials:
\begin{align}
\vcenter{\xymatrix@C=0pt@R=18pt{
 \text{\underline{IIA}}\quad & \TmF_{10,9,9} \ar@{<.>}[rd] & & \TmF_{10,7,7} \ar@{<.>}[ld] \ar@{<.>}[rd] & & \TmF_{10,5,5} \ar@{<.>}[ld] \ar@{<.>}[rd] & & \TmF_{10,3,3} \ar@{<.>}[ld] \ar@{<.>}[rd] & & \TmF_{10,1,1} \ar@{<.>}[ld] \ar@{<->}[rd]^-{\tiny\textcircled{z}} & \\
 \text{\underline{IIB}}\quad & & \TmF_{10,8,8} & & \TmF_{10,6,6} & & \TmF_{10,4,4} & & \TmF_{10,2,2} & & \TmF_{10}
}}\,.
\end{align}
What we have explicitly confirmed is only the rightmost arrow \textcircled{z}, but the existence of the $\OO(10,10)$-covariant potential $\TmF^+_{M_1\cdots M_{10}}$ suggests the validity of other maps. 
Namely, we can define the family of potentials $\TmF_{10,p,p}$ through the simple $T$-duality rules,
\begin{align}
\begin{split}
 \TmF_{a_1\cdots a_9\Ay ,b_1\cdots b_{p-1} \Ay , b_1\cdots b_{p-1}\Ay} &\overset{\text{A--B}}{\simeq} \TmF_{a_1\cdots a_9\By , b_1\cdots b_{p-1}, b_1\cdots b_{p-1}} \,,
\\
 \TmF_{a_1\cdots a_9\Ay ,b_1\cdots b_p, b_1\cdots b_p} &\overset{\text{A--B}}{\simeq} \TmF_{a_1\cdots a_9\By , b_1\cdots b_p\By, b_1\cdots b_p\By}\,,
\end{split}
\end{align}
for $p=1,3,5,7,9$, without any non-linear correction. 
Of course, in order to discuss the M-theory uplifts or the $S$-duality rules for $\TmF_{10,p,p}$ ($p\geq 2$), we need to determine how they enter into the 1-form field, $\cA_\mu^I$ or $\bm{\cA}_\mu^\sfI$\,.

\subsection{$S$-duality rule}
\label{sec:S-duality}

In this paper, the type IIB fields are defined to be $S$-duality covariant, and under an $\SL(2)$ transformation $\Lambda^{\SLa}{}_{\SLb}$, the bosonic fields transform as
\begin{align}
\begin{split}
 &\Bg'_{mn} =\Bg_{mn}\,,\quad 
 \Bm'^{\SLa\SLb} = \Lambda^{\SLa}{}_{\SLc}\,\Lambda^{\SLb}{}_{\SLd}\,\Bm^{\SLc\SLd}\,, \quad
 \BA'^{\SLa}_2 = \Lambda^{\SLa}{}_{\SLb}\,\BA^{\SLb}_2\,, \quad 
 \BA'_4 = \BA_4\,, 
\\
 &\BA'^{\SLa}_6 = \Lambda^{\SLa}{}_{\SLb}\,\BA^{\SLb}_6\,,\quad
 \BA'_{7,1} = \BA_{7,1}\,,\quad 
 \BA'^{\SLa\SLb}_8 = \Lambda^{\SLa}{}_{\SLc}\,\Lambda^{\SLb}{}_{\SLd}\,\BA^{\SLc\SLd}_8 \,,\quad 
 \BA'^{\SLa}_{8,2} = \Lambda^{\SLa}{}_{\SLb}\,\BA^{\SLb}_{8,2} \,,
\\
 &\BA'^{\SLa_1\SLa_2\SLa_3}_{10} = \Lambda^{\SLa_1}{}_{\SLb_1}\,\Lambda^{\SLa_2}{}_{\SLb_2}\,\Lambda^{\SLa_3}{}_{\SLb_3}\,\BA^{\SLb_1\SLb_2\SLb_3}_{10}\,,\quad 
 \BA'_{9,2,1}=\BA_{9,2,1}\,.
\end{split}
\end{align}
In particular, under the $S$-duality, $\Lambda=\bigl(\begin{smallmatrix} 0 & 1 \\ -1 & 0 \end{smallmatrix}\bigr)$, the component fields are transformed as
\begin{align}
\begin{split}
 &\Bg'_{mn} =\Bg_{mn}\,,\quad 
 \BC'_0 = - \tfrac{\BC_0}{(\BC_0)^2+\Exp{-2\BPhi}}\,,\quad \Exp{-\BPhi'} = \tfrac{\Exp{-\BPhi'}}{(\BC_0)^2+\Exp{-2\BPhi}}\,,
\\
 &\BB'_2 =- \BC_2\,,\quad \BC'_2 = \BB_2\,,\quad \BC'_4 = \BC_4 - \BB_2\wedge \BC_2 \,, 
\\
 &\BC'_6 = - \bigl(\BB_6 - \tfrac{1}{2!}\, \BB_2\wedge \BC_2\wedge \BC_2\bigr) \,, \quad 
 \BB'_6 = \BC_6 - \tfrac{1}{2!}\, \BC_2\wedge \BB_2\wedge \BB_2\,, 
\\
 &\BC'_8 = \BE_8 - \tfrac{1}{3!}\, \BB_2\wedge \BC_2\wedge \BC_2\wedge \BC_2 \,, \quad 
 \BE'_8 = \BC_8 - \tfrac{1}{3!}\, \BC_2\wedge \BB_2\wedge \BB_2\wedge \BB_2\,, 
\\
 &\BC'_{10} = - \bigl(\BE_{10} - \tfrac{1}{4!}\, \BB_2\wedge \BC_2\wedge \BC_2\wedge \BC_2\wedge \BC_2\bigr) \,, \quad 
 \BF'_{10} = \BC_{10} - \tfrac{1}{4!}\, \BC_2\wedge \BB_2\wedge \BB_2\wedge \BB_2\wedge \BB_2\,, 
\\
 &\BA'_{7,1} = \BA_{7,1}\,,\quad \bBD'_{8,2} =- \bBE_{8,2}\,,\quad \bBE'_{8,2} = \bBD_{8,2}\,,\quad \BA'_{9,2,1} = \BA_{9,2,1} \,. 
\end{split}
\label{eq:S-duality}
\end{align}

It is sometimes useful to introduce the dual parameterization of $\Bm_{\SLa\SLb}$,
\begin{align}
 (\Bm_{\SLa\SLb}) = \Exp{\Phi} \begin{pmatrix}
 \Exp{-2\BPhi} + (\BC_0)^2 & \BC_0 \\
 \BC_0 & 1
 \end{pmatrix} \equiv \Exp{-\tilde{\phi}} \begin{pmatrix}
 1 & -\tilde{\gamma} \\
 -\tilde{\gamma} & \Exp{2\tilde{\phi}} + \tilde{\gamma}^2
 \end{pmatrix} ,
\end{align}
which is equivalent to
\begin{align}
 \tilde{\gamma} \equiv -\frac{\BC_0}{(\BC_0)^2+\Exp{-2\BPhi}}\,,\qquad 
 \Exp{\tilde{\phi}}\equiv \frac{\Exp{-\BPhi'}}{(\BC_0)^2+\Exp{-2\BPhi}}\,.
\end{align}
Then, the $S$-duality rule becomes
\begin{align}
 \BC'_0 = \tilde{\gamma}\,,\qquad \Exp{\BPhi'} = \Exp{-\tilde{\phi}} \,. 
\end{align}
The electric-magnetic duality for $\BH_9$ is also simplified as
\begin{align}
 \BH_9 = \Exp{-2\tilde{\phi}} *_{\rmE}\, \rmd\tilde{\gamma} \,. 
\label{eq:EM-7_3}
\end{align}

For the $T$-duality-covariant potentials, the $S$-duality transformation rules are complicated. 
For example, we find
\begin{align}
 \TmD'_6 &= \BC_6 -\tfrac{1}{2}\, \BC_4\wedge \BB_2 + \tfrac{\tilde{\gamma}}{2}\,\bigl(\TmD_6 + \tfrac{1}{2}\, \BC_6\,\BC_0 + \tfrac{1}{2}\, \BC_4\wedge \BC_2 -\tfrac{1}{2}\,\BB_2\wedge\BC_2^2 \bigr)\,,
\\
 \TmE'_8 &= \BC_8 - \BC_6\wedge\BB_2 + \tfrac{1}{3}\, \BC_4\wedge \BB_2^2 + \tfrac{\tilde{\gamma}}{3!}\,\bigl(\BC_4^2 -2\,\BC_4\wedge \BC_2\wedge \BB_2 + \BC_2^2 \wedge \BB_2^2 \bigr)\,,
\label{eq:E8-S-dual}
\\
 \TmF'_{10} &= \BC_{10} - \BC_8\wedge \BB_2 + \tfrac{1}{2}\,\BC_6\wedge \BB_2^2 - \tfrac{1}{8}\,\BC_4\wedge \BB_2^3 - \tfrac{1}{30}\,\BC_2\wedge \BB_2^4
\nn\\
 &\quad + \tfrac{\tilde{\gamma}}{40}\,\bigl(\BD_6\wedge \BB_2 + \tfrac{1}{2}\,\BC_6\wedge \BB_2\, \BC_0 - 2\,\BC_4^2 + \tfrac{9}{2}\,\BC_4\wedge \BC_2\wedge \BB_2 - \tfrac{5}{2}\,\BC^2_2\wedge\BB_2^2 \bigr) \wedge\BB_2 
\nn\\
 &\quad - \tfrac{\tilde{\gamma}^2}{40}\,\bigl[2\,\BD_6\wedge (\BC_4 - \BB_2\wedge \BC_2) + \BC_6\wedge (\BC_4 - \BB_2\wedge \BC_2)\,\BC_0
\nn\\
 &\qquad\qquad + (\BC_4^2 - 2\,\BC_4\wedge\BC_2 \wedge \BB_2 + \BC_2^2\wedge \BB_2^2)\wedge\BC_2 \bigr]\,.
\end{align}
The $S$-duality rules for other $T$-duality-covariant potentials also can be obtained from \eqref{eq:S-duality}.

\section{Field strengths and gauge transformations}
\label{sec:gauge}

In this section, we summarize the field strengths and gauge transformations studied in the literature in terms of our mixed-symmetry potentials, and make a small progress. 

\subsection{11D/Type IIA supergravity}
\label{sec:gauge-IIA}

In 11D supergravity, the field strengths $\MF_{\hat{4}}$ and $\MF_{\hat{7}}$ defined in section \ref{sec:11D-sugra} are invariant under
\begin{align}
 \delta \MA_{\hat{3}} = \rmd \Mv_{\hat{2}}\,, \qquad
 \delta \MA_{\hat{6}} = \rmd \Mv_{\hat{5}} - \tfrac{1}{2}\,\MA_{\hat{3}}\wedge \rmd\Mv_{\hat{2}} \,.
\end{align}
Here, we discuss the field strengths for the mixed-symmetry potentials $\MA_{\hat{8},\hat{1}}$ and $\MA_{\hat{10},\hat{1},\hat{1}}$\,. 

Since $\MA_{\hat{8},\hat{1}}$ and $\MA_{\hat{10},\hat{1},\hat{1}}$ are 11D uplifts of the R--R 7-form and 9-form, let us consider the 11D uplifts of the known R--R field strengths $\cG_8$ and $\cG_{10}$\,. 
The 10-form $\cG_{10}$ is the electric-magnetic dual to the Romans mass \cite{Romans:1985tz}, and in order to discuss the field strength of $\MA_{\hat{10},\hat{1},\hat{1}}$, we need to introduce the mass deformation. 
Thus, let us begin by summarizing the gauge transformations and field strengths in massive type IIA supergravity. 
In massive type IIA supergravity, the field strength in the $A$-basis has the 0-form field strength $F_0\equiv m$,
\begin{align}
 F \equiv F_0 + F_2 + \cdots + F_8 + F_{10}= \rmd \TmC + m \,,\qquad \TmC \equiv \TmC_1 + \TmC_3 + \TmC_5 + \TmC_7 + \TmC_9\,.
\end{align}
Accordingly, in the $C$-basis, the field strength is given by
\begin{align}
 \cG \equiv \cG_0 + \cG_2 + \cdots + \cG_8 + \cG_{10} = \Exp{\AB_2\wedge} \bigl[\rmd (\Exp{-\AB_2\wedge}\AC) + m\bigr] = \rmd \AC -\AH_3\wedge \AC + \Exp{\AB_2\wedge} m \,. 
\end{align}
The gauge transformation is given by
\begin{align}
 \delta\AB_2 = \rmd \chi_1\,,\qquad 
 \delta \AC = \Exp{\AB_2\wedge}\,\rmd \lambda - m\Exp{\AB_2\wedge}\chi_1 \,.
\end{align}

Now, let us review the uplifts of these relations to 11D. 
Since the R--R 1-form is contained in the 11D metric, under gauge transformations, the 11D metric is transformed as
\begin{align}
 \delta \Mg_{ij} = -m\, \bigl(\chi_i\, \Mg_{j\Az} + \chi_j\, \Mg_{i\Az}\bigr)\,,
\label{eq:g-massive}
\end{align}
where the coordinate $x^{\Az}$ is also transformed as $\delta x^{\Az} = - \lambda_0$\,. 
The gauge transformations for the R--R 3-form and the $B$-field are uplifted as
\begin{align}
 \delta \MA_{\hat{3}} &= \rmd \Mv_{\hat{2}} + m\,\iota_{\Az} \MA_{\hat{3}} \wedge \iota_{\Az} \Mv_{\hat{2}} \,,\qquad \Mv_{\hat{2}} \equiv \lambda_2 + \chi_1\wedge\rmd x^{\Az}\,,
\end{align}
Under these transformations, the field strength,
\begin{align}
 \MF_{\hat{4}} \equiv \rmd \MA_{\hat{3}} + \tfrac{m}{2}\,\iota_{\Az}\MA_{\hat{3}} \wedge\iota_{\Az}\MA_{\hat{3}} 
 \equiv \cG_4 + \AH_3 \wedge (\rmd x^{\Az} +\AC_1)\,,
\end{align}
transforms as
\begin{align}
 \delta \MF_{\hat{4}} = m\,\iota_{\Az} \Mv_2 \wedge \iota_{\Az} \MF_{\hat{4}} \,. 
\end{align}
The non-invariance is due to $\delta(\rmd x^{\Az} +\AC_1)=m\,\iota_{\Az}\Mv_2$\,, although $\cG_4$ and $\AH_3$ are invariant. 
From a similar consideration, the gauge transformations for the R--R 5-, 7-, and 9-forms are also uplifted as\footnote{In our convention, the gauge transformations of $\iota_{\Az} \MA_{\hat{8},\Az}$ and $\iota_{\Az} \MA_{\hat{10},\Az,\Az}$ does not include the mass deformation because the R--R potentials are included there such that the mass dependence is canceled out [see \eqref{eq:IIA-7-9}]. The R--R 6-form potential is also contained in $\MA_{\hat{6}}$ such that the mass dependence is canceled out, but the gauge transformation of the potential $\AB_6$ gives the mass dependence of $\delta \MA_{\hat{6}}$\,.}
\begin{align}
 \delta \MA_{\hat{6}} &= \rmd \Mv_{\hat{5}} - \tfrac{1}{2}\,\MA_{\hat{3}}\wedge \rmd\Mv_{\hat{2}} - m\,\bigl(\iota_{\Az}\Mv_{7,\Az} + \iota_{\Az} \MA_{\hat{6}} \wedge \iota_{\Az} \Mv_{\hat{2}} \bigr)\,,
\\
 \delta \bigl(\iota_{\Az} \MA_{\hat{8},\Az}\bigr) &= \iota_{\Az} \rmd \Mv_{\hat{7},\Az} + \iota_{\Az} \MA_{\hat{3}} \wedge \iota_{\Az}\rmd \Mv_{\hat{5}} - \tfrac{2}{3!}\, \iota_{\Az} \MA_{\hat{3}}\wedge\iota_{\Az} \bigl(\MA_{\hat{3}}\wedge \rmd\Mv_2\bigr)\,,
\label{eq:delta-M81}
\\
 \delta \bigl(\iota_{\Az} \MA_{\hat{10},\Az,\Az}\bigr) &= \iota_{\Az} \rmd \Mv_{\hat{9},\Az,\Az} + \iota_{\Az} \MA_{\hat{3}} \wedge \iota_{\Az}\rmd \Mv_{\hat{7},\Az}
\nn\\
 &\quad - \tfrac{1}{2}\, \iota_{\Az} \MA_{\hat{3}} \wedge \iota_{\Az} \MA_{\hat{3}} \wedge \iota_{\Az}\rmd \Mv_{\hat{5}} - \tfrac{3}{4!}\, \iota_{\Az} \MA_{\hat{3}}\wedge\iota_{\Az} \bigl(\MA_{\hat{3}}\wedge \rmd\Mv_2\bigr)\,,
\end{align}
and the associated field strengths are defined by
\begin{align}
 \MF_{\hat{7}} &\equiv \rmd \MA_{\hat{6}} + \tfrac{1}{2}\,\MA_{\hat{3}}\wedge \MF_{\hat{4}} -m\, \bigl(\iota_{\Az} \MA_{\hat{8},\Az} - \iota_{\Az} \MA_{\hat{3}}\wedge \iota_{\Az} \MA_{\hat{6}} + \tfrac{1}{2\cdot 3!}\,\iota_{\Az} \MA_{\hat{3}}\wedge\iota_{\Az} \MA_{\hat{3}}\wedge \MA_{\hat{3}}\bigr)\,,
\\
 \iota_{\Az} \MF_{\hat{9},\Az} &\equiv \iota_{\Az} \rmd \MA_{\hat{8},\Az} + \iota_{\Az} \MA_{\hat{6}}\wedge \iota_{\Az} \MF_{\hat{4}} + \tfrac{1}{3!}\,\iota_{\Az} \MA_{\hat{3}} \wedge \iota_{\Az}\bigl(\MF_{\hat{4}}\wedge \MA_{\hat{3}} \bigr) 
\nn\\
 &\quad - \tfrac{2m}{2\cdot 4!}\,\iota_{\Az} \MA_{\hat{3}} \wedge\iota_{\Az} \MA_{\hat{3}} \wedge\iota_{\Az} \MA_{\hat{3}} \wedge\iota_{\Az} \MA_{\hat{3}} \,,
\\
 \iota_{\Az} \MF_{\hat{11},\Az,\Az} &\equiv \iota_{\Az} \rmd \MA_{\hat{10},\Az,\Az} + \iota_{\Az} \MA_{\hat{8},\Az}\wedge \iota_{\Az} \MF_{\hat{4}} + \tfrac{1}{4!}\,\iota_{\Az} \MA_{\hat{3}} \wedge \iota_{\Az} \MA_{\hat{3}} \wedge \iota_{\Az}\bigl(\MF_{\hat{4}}\wedge \MA_{\hat{3}} \bigr)
\nn\\
 &\quad - \tfrac{3m}{2\cdot 5!}\,\iota_{\Az} \MA_{\hat{3}} \wedge\iota_{\Az} \MA_{\hat{3}} \wedge\iota_{\Az} \MA_{\hat{3}} \wedge\iota_{\Az} \MA_{\hat{3}} \wedge\iota_{\Az} \MA_{\hat{3}}\,.
\end{align}
We note that the 7-form field strength is not invariant similar to the 4-form,
\begin{align}
 \delta \MF_{\hat{7}} = m\,\iota_{\Az} \Mv_2\wedge \iota_{\Az} \MF_{\hat{7}} \,, 
\end{align}
while the projections of the 9-form and the 11-form are invariant as it is clear from
\begin{align}
 \iota_{\Az} \MF_{\hat{9},\Az} = \cG_8\,,\qquad 
 \iota_{\Az} \MF_{\hat{11},\Az,\Az} = \cG_{10} \,.
\label{eq:cG9-cG11}
\end{align}
The above gauge transformations have been discussed in \cite{hep-th/9712115,hep-th/9802199,hep-th/9806120,hep-th/9912030} and they are gauge symmetry of the ``massive 11D supergravity'' \cite{hep-th/9712115,hep-th/9806120,hep-th/9912030}, which reproduce the massive type IIA supergravity after the dimensional reduction. 

By using the above setup, we can easily consider the 11D uplift of the electric-magnetic duality $\cG_8 = * \cG_2$. 
For this purpose, we also introduce the 2-form field strength associated with a Killing vector $k\equiv \partial_{\Az}$ as \cite{hep-th/9912030} (see also \cite{0907.3614})
\begin{align}
 \MF_{\hat{2}} \equiv \rmd k_{\hat{1}} + m\,\abs{k}^2\,\iota_k\MA_{\hat{3}}\,,
\end{align}
where $k_{\hat{1}} \equiv k^i\, \Mg_{ij} \,\rmd x^j$\,. 
This transforms as
\begin{align}
 \delta \MF_{\hat{2}} = m\, \iota_k\Mv_{\hat{2}} \wedge \iota_k \rmd k_{\hat{1}} = m\, \iota_k\Mv_{\hat{2}} \wedge \iota_k \MF_{\hat{2}} \,. 
\end{align}
In terms of the type IIA fields, we have $k_{\hat{1}}=\Exp{\frac{4}{3}\APhi}(\rmd x^{\Az}+\AC_1)$, and this gives
\begin{align}
 \MF_{\hat{2}} \equiv \Exp{\frac{4}{3}\APhi} \cG_2 + \tfrac{4}{3}\Exp{\frac{4}{3}\APhi} \rmd\APhi \wedge \bigl(\rmd x^{\Az} + \AC_1\bigr)\,.
\end{align}
Then, the electric-magnetic duality $\cG_8 = * \cG_2$ becomes
\begin{align}
 \iota_{\Az} \MF_{\hat{9},\Az} = \iota_{\Az} \hat{*}\MF_2 \,,
\end{align}
which shows that the dual graviton is electric-magnetic dual to the Killing vector. 
In this paper, only the restricted component that couple to supersymmetric branes are considered, and the restriction corresponds to the projection $\iota_{\Az}$ in front of the field strength.\footnote{The full definition of the field strength $\MF_{\hat{9},\Az}$ without the projection has been proposed in \cite{0907.3614}. It can be found by regarding the dilaton equations of motion as the Bianchi identity, and by uplifting this to 11D.}
We can similarly consider the 11D uplift of the electric-magnetic duality $m = \cG_{0} = * \cG_{10}$ \cite{hep-th/9912030}
\begin{align}
 m\,\abs{k}^4 = \hat{*} \MF_{\hat{11},\Az,\Az}\,,
\end{align}
where
\begin{align}
 \MF_{\hat{11},\Az,\Az} \equiv \iota_{\Az} \MF_{\hat{11},\Az,\Az}\wedge \bigl(\rmd x^{\Az}+\AC_1\bigr)\,,
\end{align}
is the gauge-invariant field strength. 

Now, we consider the relation to the recent studies on mixed-symmetry potentials in DFT. 
If we decompose the 7-form field strength as
\begin{align}
 \MF_{\hat{7}} = \AH_7 + \cG_6 \wedge (\rmd x^{\Az} + \AC_1) \,,
\end{align}
the 7-form field strength $\AH_7$ becomes
\begin{align}
 \AH_7 = \rmd \TmD_6 - \tfrac{1}{2}\,\bigl(\cG_6 \wedge \AC_1 - \cG_4\wedge \AC_3 + \cG_2 \wedge \AC_5 - \cG_0\,\AC_7\bigr) + \tfrac{m}{2}\, \TmC_7\,,
\end{align}
which coincides with the expression given in \cite{1903.05601}, and is invariant under
\begin{align}
 \delta \TmD_6 =\rmd \chi_5 + \tfrac{1}{2}\,\bigl(\TmC_5\wedge \rmd\lambda_0 - \TmC_3\wedge \rmd\lambda_2 + \TmC_1\wedge \rmd\lambda_4 \bigr) - m\,\bigl(\lambda_6 +\tfrac{1}{2}\,\TmC_5\wedge \chi_1\bigr)\,.
\end{align}
Here, we have parameterized
\begin{align}
 \Mv_{\hat{5}} = \chi_5 + \lambda_4\wedge\rmd x^{\Az}\,,\qquad 
 \Mv_{\hat{7},\Az} = \overline{\lambda}_7 + \lambda_6 \wedge\rmd x^{\Az}\,.
\end{align}

Let us also clarify the relation between the field strength of the dual graviton and the field strength $\iota_n \AH_{8, n}$ defined in \eqref{eq:def-AH7}. 
To this end, we assume the existence of a Killing direction denoted by $n$ (i.e.~$\Lie_n \equiv \iota_n \rmd + \rmd \iota_n=0$) other than the M-theory circle. 
For simplicity, we turn off the mass parameter. 
Then, as before, we can easily show that the field strength
\begin{align}
 \iota_n \MF_{\hat{9},n} &\equiv \iota_n \rmd \MA_{\hat{8},n} + \iota_n \MA_{\hat{6}}\wedge \iota_n \MF_{\hat{4}} + \tfrac{1}{3!}\,\iota_n \MA_{\hat{3}} \wedge \iota_n\bigl(\MF_{\hat{4}}\wedge \MA_{\hat{3}} \bigr) \,,
\end{align}
is invariant under
\begin{align}
\begin{split}
 \delta \MA_{\hat{3}} &= \rmd \Mv_{\hat{2}}\,,\qquad
 \delta \MA_{\hat{6}} = \rmd \Mv_{\hat{5}} - \tfrac{1}{2}\,\MA_{\hat{3}}\wedge \rmd\Mv_{\hat{2}} \,,
\\
 \delta \bigl(\iota_n \MA_{\hat{8},n}\bigr) &= \iota_n \rmd \Mv_{\hat{7},n} + \iota_n \MA_{\hat{3}} \wedge \iota_n\rmd \Mv_{\hat{5}} - \tfrac{2}{3!}\, \iota_n \MA_{\hat{3}}\wedge\iota_n \bigl(\MA_{\hat{3}}\wedge \rmd\Mv_2\bigr)\,.
\end{split}
\end{align}
Now, we consider the reduction to type IIA theory. 
We define the field strength of the dual graviton in type IIA theory as
\begin{align}
 \iota_n \cG_{8,n} \equiv \iota_{\Az} \iota_n \bigl(\MF_{\hat{9},n} - \MF_{\hat{9},\Az} \AC_n\bigr) = \iota_{\Az} \iota_n \MF_{\hat{9},n} + \iota_n \cG_8 \,\iota_n \AC_1 \,,
\end{align}
which is also gauge invariant because $\iota_n \AC_1$ is gauge invariant due to the Killing equation, $\Lie_{n}\lambda_0= \iota_n\rmd\lambda_0=0$\,. 
Then, a straightforward but a bit long computation gives
\begin{align}
 \iota_n \cG_{8,n} &= \rmd \iota_n A_{7,n} + \iota_n \AC_5\wedge\iota_n \rmd\AC_3 -\iota_n \AB_6\wedge \iota_n \AH_3
 - \tfrac{1}{3}\,\iota_n\AB_2\wedge\AC_3\wedge\iota_n\rmd \AC_3 
 - \tfrac{2}{3}\,\AB_2\wedge\iota_n\AC_3\wedge\iota_n\rmd \AC_3 
\nn\\
 &\quad - \tfrac{1}{3}\,\iota_n\AB_2\wedge\rmd\AC_3\wedge\iota_n \AC_3 
 + \tfrac{1}{6}\,\iota_n\AH_3\wedge\iota_n\AC_3\wedge \AC_3 
 - \tfrac{1}{6}\,\AH_3\wedge\iota_n\AC_3\wedge \iota_n\AC_3 
 + \iota_n \cG_8 \,\iota_n \AC_1
\nn\\
 &= \iota_n \AH_{8, n} - \iota_n \AH_7\wedge \iota_n \AB_2 \,,
\end{align}
where we have used \eqref{eq:A-D71-form}. 
This shows that the invariant field strength $\iota_k \cG_{8,n}$ is the component of the untwisted tensor,
\begin{align}
 \hat{H}_{M_1M_2M_3} \equiv (\Exp{\bm{\AB}})_{M_1}{}^{N_1}\,(\Exp{\bm{\AB}})_{M_2}{}^{N_2}\,(\Exp{\bm{\AB}})_{M_3}{}^{N_3}\,\hat{H}_{N_1N_2N_3}\,,\qquad (\Exp{\bm{\AB}})_M{}^N \equiv \begin{pmatrix}
 \delta_m^n & \AB_{mn} \\ 0 & \delta^m_n \end{pmatrix}.
\end{align}
Indeed, we can easily check
\begin{align}
 \iota_n \cG_{8,n} &= \tfrac{1}{7!\,2!}\,\epsilon_{m_1\cdots m_7 a_1a_2 n}\, \hat{H}^{a_1a_2}{}_n \,\rmd x^{m_1}\wedge\cdots\wedge\rmd x^{m_7}
 = \iota_n \AH_{8,n} - \iota_n \AH_7\wedge \iota_n \AB_2 \,.
\end{align}
Namely, the field strength $H_{MNP}$ is similar to the R--R field strength $F$\,; $F$ is invariant under gauge transformations of the R--R potentials, but not under $B$-field gauge transformations, and it becomes invariant after untwisting the field strength as $\cG=\Exp{\AB_2}F$\,. 

For completeness, let us also show that the gauge transformation \eqref{eq:delta-M81} reproduces \cite{hep-th/9802199}
\begin{align}
 \delta(\iota_n\AA_{7,n}) &= \iota_n \rmd \lambda_{6,n} + \iota_n\AC_3\wedge\iota_n\rmd\lambda_4 
 - \iota_n\AB_2\wedge\iota_n\rmd \chi_5
 - \tfrac{1}{3}\,\iota_n\AC_3\wedge \bigl(\iota_n\AC_3\wedge \rmd \chi_1
 + 2\,\iota_n\AB_2\wedge\rmd \lambda_2\bigr)
\nn\\
 &\quad + \tfrac{1}{3}\,\bigl(\iota_n\AC_3\wedge \AB_2+\AC_3\wedge \iota_n\AB_2\bigr)\wedge\iota_n\rmd\lambda_2
 + \tfrac{1}{3}\,\AC_3\wedge\rmd\iota_n\AC_3\wedge \iota_n\rmd\chi_1\,,
\end{align}
where we have parameterized
\begin{align}
 \Mv_{\hat{7}, n} = \lambda_{7,n} + \lambda_{6,n} \wedge\rmd x^{\Az}\,.
\end{align}
In terms of the $T$-duality-covariant tensor, we have
\begin{align}
 \delta(\iota_n\TmD_{7,n}) &= \iota_n \rmd \lambda_{6,n} -\tfrac{1}{2}\,\bigl(\iota_n\TmC_5 \wedge\iota_n\rmd\lambda_2 - \iota_n\TmC_3 \wedge\iota_n\rmd\lambda_4 + \iota_n\TmC_1 \wedge\iota_n\rmd\lambda_6 \bigr)
\nn\\
 &\quad - \iota_n \TmD_6 \wedge\iota_n\rmd \chi_1\,.
\label{eq:D71-delta-A}
\end{align}

The field strength $\iota_n \MF_{\hat{9},n}$ also contains the field strength of the potential $\AA_{8,1}$. 
Since we have established the 11D--10D map, it is a straightforward task to compute the field strength. 
The relation between $\AA_{8,1}$ and the potential $\TmE_{8,1}$ is also given, and it will not be difficult to rewrite the field strength in a manifestly $T$-duality-covariant form. 
Similarly, we can also consider the reduction of the field strength $\iota_n \MF_{\hat{11},n,n}$, which gives the field strengths of $\AA_{9,1,1}$ and $\AA_{10,1,1}$\,. 
The former can be expressed in the $T$-duality-covariant form by rewriting $\AA_{9,1,1}$ into $\TmE_{9,1,1}$\,. 
Since this is a 9-form, we need to introduce another deformation parameter associated with the non-geometric $R$-flux $R^{1,1}$ (with non-vanishing component $R^{n,n}=m$), which is the magnetic flux of the exotic $7^{(1,0)}_3$-brane (see \cite{1805.12117}). 
On the other hand, the field strength of $\AA_{10,1,1}$ gives the field strength of the $T$-duality-covariant potential $\TmF_{10,1,1}$\,, although the field strength automatically vanishes in 10D. 

For the mixed-symmetry potential $\MA_{\hat{9},\hat{3}}$, it is not straightforward to define the field strength because it is not related to the standard fields. 
However, since this is the 11D uplift of the type IIA potential $\TmD_{8,2}$\,, it may be useful to define the type IIA field strength by using the component $\hat{H}^m{}_{n_1n_2}$ and uplift this to 11D. 

\subsection{Type IIB supergravity}
\label{sec:gauge-IIB}

In type IIB supergravity, the gauge transformations are given as follows \cite{hep-th/0506013,hep-th/0602280,hep-th/0611036,1004.1348}:
\begin{align}
 \delta \BA_2^{\SLa}&=\rmd \Lambda^{\SLa}_1\,,
\\
 \delta \BA_4&=\rmd \Lambda_3 - \tfrac{1}{2!}\,\epsilon_{\SLc\SLd}\, \BA^{\SLc}_2\wedge \rmd\Lambda^{\SLd}_1 \,,
\\
 \delta \BA_6^{\SLa}&=\rmd \Lambda^{\SLa}_5 + \BA_2^{\SLa} \wedge \rmd \Lambda_3 - \tfrac{2}{3!}\,\epsilon_{\SLc\SLd}\, \BA^{\SLa}_2\wedge \BA^{\SLc}_2 \wedge \rmd\Lambda^{\SLd}_1 \,,
\\
 \delta \BA_8^{\SLa\SLb}&=\rmd \Lambda^{\SLa\SLb}_7 + \BA_2^{(\SLa} \wedge \rmd \Lambda^{\SLb)}_5 + \tfrac{1}{2!}\,\BA_2^{\SLa} \wedge\BA_2^{\SLb} \wedge \rmd \Lambda_3 - \tfrac{3}{4!}\,\epsilon_{\SLc\SLd}\, \BA^{\SLa}_2\wedge \BA^{\SLb}_2\wedge \BA^{\SLc}_2 \wedge \rmd\Lambda^{\SLd}_1 \,,
\\
 \delta \BA_{10}^{\SLa\SLb\SLc}&=\rmd \Lambda^{\SLa\SLb\SLc}_9 + \BA_2^{(\SLa} \wedge \rmd \Lambda^{\SLb\SLc)}_7 + \tfrac{1}{2!}\,\BA_2^{(\SLa} \wedge\BA_2^{\SLb} \wedge \rmd \Lambda^{\SLc)}_5
\nn\\
 &\quad + \tfrac{1}{3!}\,\BA_2^{\SLa} \wedge\BA_2^{\SLb}\wedge\BA_2^{\SLc} \wedge \rmd \Lambda_3 - \tfrac{4}{5!}\,\epsilon_{\SLd\SLe}\, \BA^{\SLa}_2\wedge \BA^{\SLb}_2\wedge \BA^{\SLc}_2\wedge \BA^{\SLd}_2 \wedge \rmd\Lambda^{\SLe}_1 \,.
\end{align}
In terms of the component fields, we find
\begin{align}
 \delta \BB_2 &= \rmd \chi_1\,, \qquad 
 \delta \BC = \Exp{\BB_2\wedge}\rmd \lambda = \rmd \hat{\lambda} - \BH_3\wedge \hat{\lambda} \,, 
\\
 \delta \BB_6 &= \rmd \chi_5 + \BC_2\wedge \bigl(\rmd\lambda_3 + \BB_2\wedge \rmd\lambda_1\bigr) \,, 
\\
 \delta \BE_8 &= \rmd\zeta_7 +\BC_2\wedge \rmd \chi_5 + \tfrac{1}{2!}\,\BC_2\wedge\BC_2\wedge\bigl(\rmd\lambda_3 + \BB_2\wedge \rmd\lambda_1\bigr)\,, 
\\
 \delta \BF_{10} &= \rmd\eta_9 +\BC_2\wedge \rmd \zeta_7 + \tfrac{1}{2!}\,\BC_2\wedge\BC_2\wedge \rmd \chi_5 + \tfrac{1}{3!}\,\BC_2\wedge\BC_2\wedge\BC_2\wedge \bigl(\rmd\lambda_3 + \BB_2\wedge \rmd\lambda_1\bigr)\,, 
\end{align}
where the gauge parameters are parameterized as
\begin{align}
\begin{split}
 &(\Lambda^{\SLa}_1) = \begin{pmatrix} \chi_1 \\ - \lambda_1 \end{pmatrix},\quad 
 \Lambda_3 = \lambda_3\,,\quad 
 (\Lambda^{\SLa}_5) = \begin{pmatrix} \lambda_5 \\ - \chi_5 \end{pmatrix},
\\
 &\begin{pmatrix} \Lambda^{\SLE{1}\SLE{1}}_7 \\ \Lambda^{\SLE{2}\SLE{2}}_7 \end{pmatrix}= \begin{pmatrix} \lambda_7 \\ \zeta_7 \end{pmatrix},\quad 
 \begin{pmatrix} \Lambda^{\SLE{1}\SLE{1}\SLE{1}}_9 \\ \Lambda^{\SLE{2}\SLE{2}\SLE{2}}_9 \end{pmatrix}
 = \begin{pmatrix} \lambda_9 \\ -\eta_9 \end{pmatrix} ,
\end{split}\end{align}
and we have defined
\begin{align}
 \lambda \equiv \lambda_1+\lambda_3+\lambda_5+\lambda_7+\lambda_9\,,\qquad 
 \hat{\lambda} \equiv \hat{\lambda}_1+\hat{\lambda}_3+\hat{\lambda}_5+\hat{\lambda}_7+\hat{\lambda}_9 \equiv \Exp{\BB_2\wedge}\lambda \,.
\end{align}

Let us also consider the field strength for the dual graviton $\BA_{7,1}$. 
Similar to the case of type IIA supergravity, by introducing the untwisted tensor,
\begin{align}
 \hat{H}_{M_1M_2M_3} \equiv (\Exp{\bm{\BB}})_{M_1}{}^{N_1}\,(\Exp{\bm{\BB}})_{M_2}{}^{N_2}\,(\Exp{\bm{\BB}})_{M_3}{}^{N_3}\,\hat{H}_{N_1N_2N_3}\,,\qquad (\Exp{\bm{\BB}})_M{}^N \equiv \begin{pmatrix}
 \delta_m^n & \BB_{mn} \\ 0 & \delta^m_n \end{pmatrix}.
\end{align}
we define the field strength as
\begin{align}
 \iota_n \cG_{8,n} &\equiv \tfrac{1}{7!\,2!}\,\epsilon_{m_1\cdots m_7 a_1a_2 n}\, \hat{H}^{a_1a_2}{}_n \,\rmd x^{m_1}\wedge\cdots\wedge\rmd x^{m_7}
 = \iota_n \BH_{8, n} - \iota_n \BH_7\wedge \iota_n \BB_2 \,.
\end{align}
By assuming $\Lie_n=0$ for an arbitrary field, it is invariant under the gauge transformation
\begin{align}
 \delta(\iota_n\TmD_{7,n}) &= \iota_n \rmd \lambda_{6,n} +\tfrac{1}{2}\,\bigl(\iota_n\TmC_6 \wedge\iota_n\rmd\lambda_1 - \iota_n\TmC_4 \wedge\iota_n\rmd\lambda_3 + \iota_n\TmC_2 \wedge\iota_n\rmd\lambda_5 \bigr)
\nn\\
 &\quad - \iota_n \TmD_6 \wedge\iota_n\rmd \chi_1\,,
\label{eq:D71-delta-B-D}
\end{align}
which is the $T$-dual counterpart of \eqref{eq:D71-delta-A}. 
Since the dual graviton $\BA_{7,n}$ is $S$-duality invariant, it is natural to expect that this field strength is invariant under $S$-duality. 
Indeed, we can express the field strength in a manifestly $S$-duality-invariant form,
\begin{align}
 \iota_n \cG_{8,n}
 &= \rmd \iota_n \BA_{7,n} -\epsilon_{\SLc\SLd}\,\iota_n\rmd \BA_6^{\SLc}\wedge\iota_n\BA^{\SLd}_2
 - \tfrac{1}{2}\, \iota_n\BA_4\wedge\iota_n\rmd\BA_4
\nn\\
 &\quad +\tfrac{1}{2}\,\epsilon_{\SLc\SLd}\,\iota_n\bigl(\rmd\BA_4\wedge \BA^{\SLc}_2\bigr)\wedge\iota_n\BA^{\SLd}_2 
 + \epsilon_{\SLc\SLd}\,\iota_n \bigl(\BA_4\wedge \rmd\BA^{\SLc}_2\bigr)\wedge\iota_n\BA^{\SLd}_2 
\nn\\
 &\quad - \tfrac{1}{16}\,\epsilon_{\SLa\SLb}\,\epsilon_{\SLc\SLd}\,\BA^{\SLa}_2\wedge\rmd\BA^{\SLb}_2\wedge \iota_n\BA^{\SLc}_2\wedge\iota_n\BA^{\SLd}_2
 + \tfrac{1}{24}\,\epsilon_{\SLa\SLb}\,\epsilon_{\SLc\SLd}\,\BA^{\SLa}_2\wedge\iota_n\rmd\BA^{\SLb}_2\wedge \BA^{\SLc}_2\wedge\iota_n\BA^{\SLd}_2 \,.
\end{align}
The gauge transformation \eqref{eq:D71-delta-B-D} also can be expressed as
\begin{align}
 \delta(\iota_n \BA_{7,n}) &= \iota_n\rmd\lambda_{6,n}
 - \epsilon_{\SLc\SLd}\,\iota_n\BA^{\SLc}_6 \wedge\iota_n\rmd\Lambda^{\SLd}_1 
 -\tfrac{1}{2}\,\iota_n\BA_4 \wedge\iota_n\rmd\Lambda_3
\nn\\
 &\quad + \epsilon_{\SLc\SLd}\, \iota_n\bigl(\BA_4\wedge \BA^{\SLc}_2\bigr)\wedge\iota_n\rmd\Lambda^{\SLd}_1
 -\tfrac{1}{4}\,\epsilon_{\SLc\SLd}\,\iota_n\BA_4 \wedge \iota_n\bigl(\BA^{\SLc}_2\wedge\rmd\Lambda^{\SLd}_1\bigr)
\nn\\
 &\quad - \tfrac{1}{16}\,\epsilon_{\SLa\SLb}\,\epsilon_{\SLc\SLd}\,\BA^{\SLa}_2\wedge\rmd\Lambda^{\SLb}_1\wedge \iota_n\BA^{\SLc}_2\wedge\iota_n\BA^{\SLd}_2
 + \tfrac{1}{24}\,\epsilon_{\SLa\SLb}\,\epsilon_{\SLc\SLd}\,\BA^{\SLa}_2\wedge\iota_n \BA^{\SLb}_2\wedge \BA^{\SLc}_2\wedge\iota_n \rmd\Lambda^{\SLd}_1 \,.
\label{eq:D71-delta-B}
\end{align}

Here, we do not study additional potentials, but one can obtain their field strengths as follows. 
As mentioned in the type IIA case, the field strength of $\TmD_{8,2}$ will be obtained by computing $\hat{H}^m{}_{n_1n_2}$\,. 
Then, rewriting the field strength in terms of the $S$-duality covariant potentials, we can obtain the field strength of $\BA^{\SLa}_{8,2}$\,. 
In order to define the field strength of the potentials $\TmE_{8+n,p,n}$, it is useful to find the $T$-duality-covariant expression for $\ket{\TmE_{MN}}$ that reproduce the field strength $\BH_9$ as the particular component. 
This covariant field strength contains various field strengths, and by using the relation between $\TmE_{8+n,p,n}$ and our mixed-symmetry potentials, we can obtain the $S$-duality-covariant expressions for field strengths. 

\section{Conclusions}
\label{sec:conclusions}

In this paper, we have provided explicit definitions of mixed-symmetry potentials by finding their relation to the standard supergravity fields under $T$- and $S$-duality transformations. 
The obtained $T$-duality rules are generally very complicated, but by performing certain field redefinitions, they are considerably simplified. 
The redefined fields $\TmC_{p}$, $\TmD_{6+n,n}$, $\TmE_{8+n,p,n}$, and $\TmF_{10,p,p}$ are identified with certain components of the $\OO(10,10)$-covariant tensors $\TmC_{\dot{a}}$, $\TmD_{M_1\cdots M_4}$, $\TmE_{MN\dot{a}}$, and $\TmF^+_{M_1\cdots M_{10}}$\,. 
These $\OO(10,10)$-covariant tensors have been studied in the literature, but their relation to the standard supergravity fields have not discussed enough. 
For example, the potential $\TmE_8$ has been expected to the $S$-dual of the R--R 8-form $\BC_8$, but the $S$-duality rule \eqref{eq:E8-S-dual}, including the non-linear terms, is newly determined in this paper. 
The $S$-duality rule for $\TmF_{10}$ and more mixed-symmetry potentials are also newly determined. 
Additionally, we have also studied the field strengths of mixed-symmetry potentials. 
Most of the results has been known in the literature (where the mixed-symmetry potentials are treated as $p$-forms), but here we have clarified the relation to the field strength $H_{MNP}$ studied in DFT. 
We have also provided the $S$-duality-invariant expression for the field strength of the dual graviton. 

The linear map has been originally studied for the generalized metric \cite{hep-th/0402140,1701.07819,1909.01335}. 
By comparing two parameterizations, duality rules for potentials that appear in the $E_{n(n)}$ generalized metric ($n\leq 8$) are determined. 
The linear map for the 1-form $\cA_\mu^I$ studied here is more efficient to find the duality rules, but the parameterization is involved and the obtained $T$-duality rules are rather long. 
In the linear map for the generalized metric, the parameterization is more systematic, and by considering the linear map for the $E_{11}$ generalized metric, we may find a better definition of mixed-symmetry potentials which simplify the duality rules. 

As we have demonstrated, the linear map works well in finding the duality rules for mixed-symmetry potentials, and it is a straightforward to consider more mixed-symmetry potentials. 
In addition, having clarified the definitions of mixed-symmetry potentials, it is important to consider the application to the worldvolume theories of exotic branes. 
It is also interesting to study the supersymmetry transformations for mixed-symmetry potentials (where Killing vectors should be involved) by extending the series of works \cite{hep-th/0506013,hep-th/0602280,hep-th/0611036,1004.1348}.

Before closing this paper, let us comment on a relevant open issue, called the exotic duality \cite{1109.4484,1009.4657,1102.0934,1309.2653,1412.8769}, which is the electric-magnetic duality for exotic branes with co-dimension equal to (or higher than) two. 
As we have already mentioned, the mixed-symmetry potentials couple to various exotic branes electrically. 
On the other hand, the exotic branes (with co-dimension two) magnetically couple to certain dual fields \cite{1303.1413,1402.5972,1411.1043,1412.8769,1612.08738}, such as the $\beta$-fields\footnote{This is originally introduced for example in \cite{Shapere:1988zv,Giveon:1988tt,Duff:1989tf,Tseytlin:1990nb,hep-th/9401139} and utilized more recently in the $\beta$-supergravity \cite{1106.4015,1202.3060,1204.1979,1306.4381,1411.6640}.} or the $\gamma$-fields \cite{1007.5509,1411.6640,1412.0635,1412.8769}, which are associated with the non-geometric $Q$-flux \cite{hep-th/0508133} or the $P$-fluxes \cite{hep-th/0602089,0811.2900,1007.5509}.\footnote{See \cite{0705.3410,0709.2186,0711.2512,0907.2041,0907.5580,0911.2876,1212.4984,1301.7073,1302.0529,1304.0792,1405.2325,1508.01197,1510.01522,1603.01290,1603.08545,1712.07310,1805.05748,1909.07391,1909.08630,1909.10993} for an incomplete list of references utilizing non-geometric fluxes.} 
Thus, the exotic duality is the electric-magnetic duality between the mixed-symmetry potentials and the dual potentials. 
An example is given in \eqref{eq:EM-7_3},
\begin{align}
 \tfrac{1}{9!}\,\epsilon^{m n_1\cdots n_9}\, \BH_{n_1\cdots n_9} = - \Exp{-2\tilde{\phi}} \sqrt{\abs{\Bg}}\, \Bg^{mn}\, \partial_n \tilde{\gamma} \,,
\label{eq:EM-7_3-component}
\end{align}
where $\tilde{\gamma}$ roughly corresponds to the $\gamma$-field (as we explain below). 
Recently, the $T$-duality-covariant expression of the exotic duality has been investigated in \cite{1508.00780,1603.07380,1612.02691} but it has not yet fully succeeded. 
For the $T$-duality-covariant exotic duality, it will be important to establish the $T$-duality-covariant description of the dual potentials, and we make a small attempt below. 

First of all, let us explain the definition of the dual fields, such as the $\beta$- and $\gamma$-fields. 
In the $U$-duality formulations, the supergravity fields are embedded into the generalized metric $\cM_{IJ}$, which is defined as (see \cite{1111.0459} and references therein)
\begin{align}
 \cM_{IJ} \equiv (\cE^\rmT)_I{}^K\,\delta_{KL}\,\cE^L{}_J\,,
\end{align}
where the generalized vielbein $\cE^I{}_J$ is the matrix representation of an $E_{n(n)}$ element in the vector representation. 
The identity matrix $\delta_{IJ}$ is invariant under the maximal compact subgroup, and the generalized vielbein $\cE^I{}_J$ is generally parameterized by the Borel subalgebra. 
For example, in the type IIB theory, the Borel subalgebra is generated by
\begin{align}
 \{ \sfK^m{}_n\ (m\leq n),\ 2\,\sfR_{\SLE{1}\SLE{2}},\ \sfR_{\SLE{2}\SLE{2}},\ \sfR_{\SLa}^{m_1m_2},\ \sfR^{m_1\cdots m_4},\ \sfR^{m_1\cdots m_6}_{\SLa},\ \sfR^{m_1\cdots m_7,m},\ \cdots \}\,,
\end{align}
and the generalized vielbein can be parameterized as \cite{hep-th/0107181}
\begin{align*}
 \cE = \Exp{\sum h_m{}^n\,\sfK^m{}_n} \Exp{2\,\BPhi\,\sfR_{\SLE{1}\SLE{2}}} \Exp{-\BC_0\,\sfR_{\SLE{2}\SLE{2}}} \Exp{\frac{1}{2!}\,\bBA^{\SLa}_{m_1m_2}\,\sfR_{\SLa}^{m_1m_2}} \Exp{\frac{1}{4!}\,\bBA_{m_1\cdots m_4}\, \sfR^{m_1\cdots m_4}} \Exp{\frac{1}{6!}\,\bBA^{\SLa}_{m_1\cdots m_6}\,\sfR^{m_1\cdots m_6}_{\SLa}} \cdots \,,
\end{align*}
where the standard vielbein corresponds to $(e^a{}_m)=\Exp{-h^\rmT}$\,. 
On the other hand, we can also consider the negative Borel subalgebra, spanned by \cite{1612.08738}
\begin{align}
 \{ \sfK^m{}_n\ (m\geq n),\ 2\,\sfR_{\SLE{1}\SLE{2}},\ -\sfR_{\SLE{1}\SLE{1}},\ \sfR^{\SLa}_{m_1m_2},\ \sfR_{m_1\cdot m_4},\ \sfR_{m_1\cdots m_6}^{\SLa},\ \sfR_{m_1\cdots m_7,m},\ \cdots \}\,,
\end{align}
and introduce the dual parameterization as\footnote{Our dual fields have opposite sign compared to those introduced in \cite{1701.07819}.}
\begin{align*}
 \tilde{\cE} = \Exp{\sum \tilde{h}_m{}^n\,\sfK^m{}_n} \Exp{2\,\tilde{\BPhi}\,\sfR_{\SLE{1}\SLE{2}}} \Exp{-\gamma\,\sfR_{\SLE{1}\SLE{1}}} \Exp{-\frac{1}{2!}\,\tilde{\bBA}_{\SLa}^{m_1m_2}\,\sfR^{\SLa}_{m_1m_2}} \Exp{-\frac{1}{4!}\,\tilde{\bBA}^{m_1\cdots m_4}\, \sfR_{m_1\cdot m_4}} \Exp{-\frac{1}{6!}\,\tilde{\bBA}_{\SLa}^{m_1\cdots m_6}\,\sfR_{m_1\cdots m_6}^{\SLa}} \cdots \,.
\end{align*}
The dual vielbein is similarly defined by $(\tilde{e}^a{}_m)=\Exp{-\tilde{h}^\rmT}$ and the dual metric is $\tilde{\Bg}_{mn}\equiv (\tilde{e}^\rmT\,\tilde{e})_{mn}$\,. 
Then, by comparing the two parameterizations of the generalized metric,
\begin{align}
 (\cE^\rmT\,\cE)_{IJ} = \cM_{IJ} = (\tilde{\cE}^\rmT\,\tilde{\cE})_{IJ}\,,
\label{eq:standard-M-dual}
\end{align}
we can obtain the dual fields as a local redefinitions of the standard fields \cite{1612.08738}. 
For example, if we consider only the NS--NS fields, the relation \eqref{eq:standard-M-dual} is simplified as
\begin{align}
\begin{split}
 \begin{pmatrix} g_{mn} - B_{mp}\,g^{pq}\,B_{qn} & -B_{mp}\,g^{pn} \\ g^{mp}\,B_{pn} & g^{pq}
 \end{pmatrix} &= \begin{pmatrix} \tilde{g}_{mn} & \tilde{g}_{mp}\,\beta^{pn} \\ -\beta^{mp}\,\tilde{g}_{pn} & \tilde{g}^{mn} - \beta^{mp}\,\tilde{g}_{pq}\,\beta^{qn}
 \end{pmatrix} ,
\\
 \Exp{-2\BPhi}\sqrt{\abs{g}} &= \Exp{-2\tilde{\BPhi}}\sqrt{\abs{\tilde{g}}}\,,
\end{split}
\label{eq:DFT-non-geometric}
\end{align}
where $g_{mn}\equiv \Exp{\BPhi/2}\Bg_{mn}$ is the standard string-frame metric and $\tilde{g}_{mn}\equiv \Exp{\tilde{\BPhi}/2}\tilde{\Bg}_{mn}$ is the dual-string-frame metric. 
They are precisely the relations studied in the $\beta$-supergravity \cite{1106.4015,1202.3060,1204.1979,1306.4381,1411.6640}. 
On the other hand, if we only keep the metric $\Bg_{mn}$ and $\Bm_{\SLa\SLb}$, the relation \eqref{eq:standard-M-dual} reduces to
\begin{align}
 \Bg_{mn} = \tilde{\Bg}_{mn} \,,\qquad \tilde{\phi} = \tilde{\BPhi} \,,\qquad \tilde{\gamma} = \gamma \,,
\label{eq:S-dual-non-geometric}
\end{align}
and this shows that the $\gamma$-field is similar to the $\tilde{\gamma}$ appearing in \eqref{eq:EM-7_3}. 
In general, without any truncations, these relations receive non-linear corrections. 

Secondly, let us explain the $T$-duality rules for the dual fields. 
By following the discussion of \cite{1701.07819}, the $T$-duality rules for the dual fields are determined in the same manner as the standard potentials. 
To this end, we parameterize the dual fields in the same manner as \eqref{eq:IIB-param}--\eqref{eq:10-form-param}, for example,
\begin{align}
 \bigl(\tilde{\bBA}_{\SLa}^{m_1m_2}\bigr) \equiv \begin{pmatrix} \beta^{m_1m_2} \\\ -\gamma^{m_1m_2} \end{pmatrix} ,\qquad
 \tilde{\bBA}^{m_1\cdots m_4} \equiv \gamma^{m_1\cdots m_4} - 3\,\gamma^{[m_1m_2}\,\beta^{m_3m_4]} \,,\quad \cdots \,.
\end{align}
Then, we find that the gamma fields $\gamma^{m_1\cdots m_p}$ follow the same $T$-duality rules as those of the R--R field $\BC_{m_1\cdots m_p}$\,, although the position of the indices are opposite:\footnote{The 11D uplifts also have the same form as the standard potentials.}
\begin{align}
\begin{split}
 \gamma'^{a_1\cdots a_{n-1}\Ay}&= \gamma^{a_1\cdots a_{n-1}} - \tfrac{(n-1)\,\gamma^{[a_1\cdots a_{n-2}|y|}\,\tilde{\Bg}^{a_{n-1}]y}}{\tilde{\Bg}^{yy}}\,,
\\
 \gamma'^{a_1\cdots a_n} &= \gamma^{a_1\cdots a_ny} - n\, \gamma^{[a_1\cdots a_{n-1}}\, \beta^{a_n]y} - \tfrac{n\,(n-1)\,\gamma^{[a_1\cdots a_{n-2}|y|}\, \beta^{a_{n-1}|y|}\,\tilde{\Bg}^{a_n]y}}{\tilde{\Bg}^{yy}}\,.
\end{split}
\end{align} 
The dual fields in the NS--NS sector also transform as
\begin{align}
\begin{split}
 \tilde{g}'^{ab} &= \tilde{g}^{ab} - \tfrac{\tilde{g}^{a y}\,\tilde{g}^{b y}-\beta^{a y}\,\beta^{b y}}{\tilde{g}^{yy}}\,,\qquad 
 \tilde{g}'^{a \Ay}=-\tfrac{\beta^{a y}}{\tilde{g}^{yy}}\,,\qquad 
 \tilde{g}'^{\Ay\Ay}=\tfrac{1}{\tilde{g}^{yy}}\,,
\\
 \beta'^{ab} &= \beta^{ab} - \tfrac{\beta^{ay}\,\beta^{by}-\tilde{g}^{ay}\,\beta^{by}}{\tilde{g}^{yy}}\,,\qquad 
 \beta'^{ay} = -\tfrac{\tilde{g}^{ay}}{\tilde{g}^{yy}} \,, \qquad
 \Exp{2\tilde{\BPhi}'}= \tfrac{\Exp{2\tilde{\BPhi}}}{\tilde{g}^{yy}} \,.
\end{split}
\end{align}
Then, we find that
\begin{align}
 \tilde{\cH}_{MN} \equiv
 \begin{pmatrix} \tilde{g}_{mn} & \tilde{g}_{mp}\,\beta^{pn} \\ -\beta^{mp}\,\tilde{g}_{pn} & \tilde{g}^{mn} - \beta^{mp}\,\tilde{g}_{pq}\,\beta^{qn}
 \end{pmatrix} ,\qquad
 \Exp{-2\tilde{d}} \equiv \Exp{-2\tilde{\BPhi}}\sqrt{\abs{\tilde{g}}}\,,
\end{align}
transform covariantly under $T$-duality, which reduce to \eqref{eq:DFT-non-geometric} when only the NS--NS fields are present. 
Similarly, we can show that the $\gamma$-field also transform covariantly under $T$-duality. 
For this purpose, we define the dual field $\alpha$ associated with $\TmC = \Exp{-\AB_2\wedge}\AC$ (or $\TmC = \Exp{-\BB_2\wedge}\BC$) as
\begin{align}
 \alpha \equiv \Exp{-\beta \wedge} \gamma \qquad \bigl(\beta\equiv \tfrac{1}{2!}\,\beta^{mn}\,\partial_m\wedge\partial_n\bigr)\,,
\end{align}
where $\alpha\equiv \sum_p \frac{1}{p!}\,\alpha^{m_1\cdots m_p}\,\partial_{m_1}\wedge\cdots\wedge\partial_{m_p}$ and $\gamma\equiv \sum_p \frac{1}{p!}\,\gamma^{m_1\cdots m_p}\,\partial_{m_1}\wedge\cdots\wedge\partial_{m_p}$ are poly-vectors and $\wedge$ is the wedge product for poly-vectors. 
This $\alpha^{m_1\cdots m_p}$ transforms as
\begin{align}
 \alpha^{a_1\cdots a_p} = \alpha^{a_1\cdots a_py}\,,\qquad
 \alpha^{a_1\cdots a_{p-1}y} = \alpha^{a_1\cdots a_{p-1}}\,,
\end{align}
under the $T$-duality along the $x^y$-direction. 
In other words, 
\begin{align}
 \ket{\alpha} \equiv \sum_p \frac{1}{p!}\,\alpha^{m_1\dots m_p}\,\tilde{\Gamma}_{m_1\cdots m_p}\ket{\tilde{0}}\,, 
\end{align}
transforms as an $\OO(10,10)$ spinor,\footnote{To be more precise, it is a spinor density. The $\OO(10,10)$ spinor $\ket{\TmC}$ has weight $1/2$ and the weight can be removed by considering $\Exp{d}\ket{\TmC}$. On the other hand, $\ket{\alpha}$ has weight $-1/2$ and $\Exp{-\tilde{d}}\ket{\alpha}$ is weightless.} where we have defined a new vacuum annihilated by $\Gamma^m$\,,
\begin{align}
 \ket{\tilde{0}} \equiv C\,\ket{0} = \Gamma^{0\cdots 9}\,\ket{0} \,,\qquad \tilde{\Gamma}^M \equiv \Gamma^{11}\,\Gamma^M\,. 
\end{align}
Other dual fields, such as $\beta^{m_1\cdots m_6}$ \cite{1412.8769,1612.08738}, also can be embedded into $T$-duality tensors. 

Finally, let us consider the exotic duality, in particular \eqref{eq:EM-7_3-component}. 
The left-hand side is the field strength of the potential $\TmE_8$\,. 
Since $\TmE_8$ is a component of $\ket{\TmE_{MN}}$, the field strength will be also defined covariantly. 
The field strength has been discussed in \cite{1903.05601}, although the explicit form has not yet been determined,
\begin{align}
 \ket{K^M} \sim \partial_N \ket{\TmE^{MN}} + \cdots \,.
\end{align}
On the other hand, the right-hand side contains $\rmd\tilde{\gamma}$, which is roughly equal to the $P$-flux $P_1 \equiv \rmd \gamma$\,, as we have seen in \eqref{eq:S-dual-non-geometric}. 
The $P$-flux may be also defined $T$-duality covariantly,
\begin{align}
 \ket{P_M} \equiv \partial_M \ket{\gamma} + \cdots \,, 
\end{align}
and the exotic duality will be a covariant relation connecting $\ket{K^M}$ and $\ket{P_M}$\,. 
The non-trivial point is that although the field strength $\ket{K^M}$ is defined in the standard parameterization, the $P$-flux $\ket{P_M}$ is defined in the dual parameterization, and it is not easy to find the relation. 
In order to find the covariant expression for the exotic duality, it may be useful to consider the supergravity action for the dual fields. 
As it has been (partially) worked out in \cite{1612.08738}, by substituting the dual parameterization into the action of the $U$-duality-covariant supergravity, known as the exceptional field theory \cite{1308.1673,1312.0614,1312.4542,1406.3348} (which is based on DFT and earlier works \cite{hep-th/0104081,hep-th/0307098,0712.1795,0901.1581,1008.1763,1110.3930,1111.0459,1208.5884}), we obtain the action for the dual fields,
\begin{align}
 \cL &= \tilde{*} \tilde{\sfR} - \tfrac{1}{2}\,\rmd \tilde{\varphi}\wedge \tilde{*} \rmd \tilde{\varphi} - \tfrac{1}{2}\Exp{-2\tilde{\varphi}}\rmd \gamma\wedge \tilde{*} \rmd \gamma + \cdots \,.
\end{align}
Then, the equations of motion for the dual fields $\gamma^{m_1\cdots m_p}$ are precisely the exotic duality, as discussed in \cite{1412.8769}. 
Thus, in order to find the $T$-duality-covariant exotic duality, it will be useful to find the $T$-duality-covariant action for the dual fields, by defining the dual fields as $T$-duality-covariant tensors. 
From the covariant action, we obtain the $T$-duality-covariant equations of motion for the dual fields, and they will correspond to the exotic duality. 

\subsection*{Acknowledgments}

We would like to thank Jos\'e J.\ Fern\'andez-Melgarejo and Shozo Uehara for useful discussions. 
This work is supported by JSPS Grant-in-Aids for Scientific Research (C) 18K13540 and (B) 18H01214. 

\appendix

\section{Notations}
\label{app:notation}

\subsection{Differential forms and mixed-symmetry potentials}

We employ the following convention for differential forms:
\begin{align}
\begin{split}
 &(*\alpha_p)_{m_1\cdots m_{10-p}} =\tfrac{1}{(10-p)!}\,\varepsilon^{n_1\cdots n_p}{}_{m_1\cdots m_{10-p}}\,\alpha_{n_1\cdots n_p} \,, 
\\
 &*(\rmd x^{m_1}\wedge \cdots \wedge \rmd x^{m_p}) = \tfrac{1}{(10-p)!}\,\varepsilon^{m_1\cdots m_p}{}_{n_1\cdots n_{10-p}}\,\rmd x^{n_1}\wedge \cdots \wedge \rmd x^{n_{10-p}} \,,
\\
 &(\iota_v \alpha_p) = \tfrac{1}{(p-1)!}\,v^n\,\alpha_{n m_1\cdots m_{p-1}}\,\rmd x^{m_1}\wedge\cdots\wedge \rmd x^{m_{p-1}}\,,
\end{split}
\end{align}
where
\begin{align}
 \varepsilon_{m_1\cdots m_{10}}\equiv \sqrt{\abs{g}}\, \epsilon_{m_1\cdots m_{10}}\,,\quad 
 \varepsilon^{m_1\cdots m_{10}}\equiv \tfrac{1}{\sqrt{\abs{g}}}\,\epsilon^{m_1\cdots m_{10}}\,,\quad 
 \epsilon_{0\cdots 9}= 1 \,,\quad 
 \epsilon^{0\cdots 9}= -1\,.
\end{align}
The symmetrization and antisymmetrization are normalized as
\begin{align}
 A_{(m_1\cdots m_n)} \equiv \tfrac{1}{n!}\,\bigl(A_{m_1\cdots m_n} + \cdots \bigr) \,,\qquad
 A_{[m_1\cdots m_n]} \equiv \tfrac{1}{n!}\,\bigl(A_{m_1\cdots m_n} \pm \cdots \bigr) \,.
\end{align}
Indices separated by ``$|$'' are not (anti-)symmetrized. 
For example,
\begin{align}
 3\,\MA_{[i_1|k_1k_2|}\,\MA_{i_2i_3]j} = \MA_{i_1 k_1k_2}\,\MA_{i_2i_3j} + \MA_{i_2 k_1k_2}\,\MA_{i_3i_1j} + \MA_{i_3 k_1k_2}\,\MA_{i_1i_2j} \,.
\end{align}
When two groups of indices are antisymmetrized, we have used overlines. 
For example,
\begin{align}
 3\,\MA_{[i_1|\bar{k}_1\bar{k}_2|}\,\MA_{i_2\cdots i_6]\bar{k}_3} 
 = \MA_{[i_1|k_1k_2|}\,\MA_{i_2\cdots i_6]k_3} 
 + \MA_{[i_1|k_2k_3|}\,\MA_{i_2\cdots i_6]k_1} 
 + \MA_{[i_1|k_3k_1|}\,\MA_{i_2\cdots i_6]k_2} \,.
\end{align}

In this paper, we consider only the components of the mixed-symmetry potentials that satisfy the restriction rule \eqref{eq:M-restriction} or \eqref{eq:B-restriction}. 
For convenience, by using the equality $\simeq$, we have expressed various equations without making the restriction rule manifest. 
However, we can always convert the equality $\simeq$ into the exact equality $=$ by making the restriction rule manifest.
For example, let us consider the equation \eqref{eq:D71A},
\begin{align}
 \TmD_{m_1\cdots m_7, n} 
 &\simeq \AA_{m_1\cdots m_7, n}
 + 7\, \TmD_{[m_1\cdots m_6} \,\AB_{m_7]n}
 - \tfrac{1}{2}\, \AC_{m_1\cdots m_7} \,\AC_{n} 
 - \tfrac{21}{2}\, \AC_{[m_1\cdots m_5}\,\AC_{m_6m_7]n}
\nn\\
 &\quad + 70\, \AC_{[m_1m_2m_3}\, \AC_{m_4m_5 |n|}\, \AB_{m_6m_7]} \,.
\end{align}
In this example, the restriction rule is $\{m_1,\dotsc, m_7\}\ni n$, and this is automatically satisfied by choosing $m_7=n$\,. 
We then obtain
\begin{align}
 \TmD_{m_1\cdots m_6 n, n} 
 &= \AA_{m_1\cdots m_6 n, n}
 - 6\, \TmD_{[m_1\cdots m_5 |n|} \,\AB_{m_6]n}
 - \tfrac{1}{2}\, \AC_{m_1\cdots m_6 n} \,\AC_{n} 
 - \tfrac{15}{2}\, \AC_{[m_1\cdots m_4|n|}\,\AC_{m_5m_6]n}
\nn\\
 &\quad
 + 20\, \AC_{[m_1m_2m_3}\, \AC_{m_4m_5 |n|}\, \AB_{m_6]n} 
 + 30\, \AC_{[m_1m_2|n|}\, \AC_{m_3m_4 |n|}\, \AB_{m_5m_6]} \,.
\label{eq:D6-A-explicit}
\end{align}
In general, this makes the expressions longer, and that is the reason why we are using $\simeq$\,. 

In order to simplify expressions, it is also useful to use the notation of the differential form. 
For example, several relations for the $T$-duality-covariant potentials become
\begin{align}
 \TmD_6 &= \AB_6 - \tfrac{1}{2} \,\AC_5\wedge \AC_1\,,
\\
 \iota_n \TmD_{7 , n} 
 &= \iota_n \AA_{7, n}
 - \iota_n \TmD_6 \wedge \iota_n \AB_2
 - \tfrac{1}{2}\, \bigl(\iota_n \AC_7 \, \iota_n \AC_1 + \iota_n \AC_5 \wedge \iota_n \AC_3\bigr)
\nn\\
 &\quad
 + \tfrac{1}{3}\, \bigl(\AB_{2}\wedge \iota_n \AC_3 - \AC_{3} \wedge \iota_n \AB_2 \bigr) \wedge \iota_n \AC_3 \,,
\label{eq:A-D71-form}
\\
 \iota_n \TmE_{8 , n} &= \iota_n \AA_{8, n} + \iota_n \AB_6 \wedge \iota_n \AC_3 
 - \tfrac{1}{3} \, \iota_n \AC_5 \wedge \bigl(\AC_3\, \iota_n \AC_1 - \iota_n \AC_3 \wedge \AC_1\bigr)\,,
\\
 \iota_n\TmF_{10, n , n} 
 &= \iota_n\AA_{10, n , n} - \iota_n \AA_{8, n}\wedge \iota_n\AC_3 
 + \tfrac{1}{2}\,\iota_n \AB_6 \wedge \iota_n \AC_3\wedge \iota_n \AC_3
\nn\\
 &\quad
 +\tfrac{1}{20}\, \iota_n\AC_7 \wedge\bigl(\iota_n\AC_3 \wedge \AC_1 - \AC_3 \, \iota_n\AC_1 \bigr)\,\iota_n\AC_1
\nn\\
 &\quad 
 +\tfrac{1}{20}\, \iota_n\AC_5\wedge \bigl(\iota_n\AC_5\wedge \AC_1 +2\,\AC_3\wedge \iota_n \AC_3\bigr)\,\iota_n\AC_1
\nn\\
 &\quad 
 -\tfrac{1}{20}\, \AC_5\wedge \bigl(\iota_n\AC_5\wedge \iota_n\AC_1 -\tfrac{1}{2}\,\iota_n\AC_3\wedge \iota_n \AC_3\bigr)\,\iota_n\AC_1
\nn\\
 &\quad
 -\tfrac{1}{8}\,\iota_n\AC_5\wedge \iota_n\AC_3\wedge \iota_n\AC_3\wedge\AC_1 \,,
\\
 \TmD_6 &= \BB_6 - \tfrac{1}{2}\,\bigl(\BC_6\,\BC_0 + \BC_4 \wedge \BC_2 \bigr) \,,
\\
 \iota_n\TmD_{7, n} 
 &= \iota_n\BA_{7, n}
 - \tfrac{1}{2}\, \iota_n\BC_6\wedge \iota_n \BC_2
 + \tfrac{1}{2}\, \BC_0\, \iota_n\BC_6\wedge\iota_n \BB_2 
 + \tfrac{1}{2}\, \BC_4\wedge\iota_n\BB_2\wedge \iota_n \BC_2
\nn\\
 &\quad
 + \tfrac{1}{4}\, \iota_n\BC_4\wedge\bigl(\BB_2\wedge \iota_n \BC_2 - \BC_2\wedge \iota_n \BB_2\bigr)
 - \tfrac{1}{4}\, \BB_2\wedge \BC_2\wedge \iota_n\BB_2\wedge\iota_n \BC_2\,,
\\
 \TmE_8 &= \BE_8 - \BB_6\wedge \BC_2 +\tfrac{1}{3}\,\BC_4 \wedge \BC_2\wedge \BC_2 + \tfrac{1}{6}\,\BC_4 \wedge \BC_4 \,\BC_0 \,,
\\
 \TmF_{10}
 &= \BF_{10} - \BE_8\wedge \BC_2
 +\tfrac{1}{2}\,\BB_6 \wedge \BC_2\wedge \BC_2
 + \tfrac{1}{20}\, \BC_6\wedge \BC_4\,\BC_0^2 
\nn\\
 &\quad 
 - \tfrac{1}{40}\, \BC_6\wedge \BC_2\wedge \BC_2 \,\BC_0 
 - \tfrac{1}{20}\, \BC_4\wedge \BC_4\wedge\BC_2\,\BC_0 
 - \tfrac{1}{8}\, \BC_4\wedge \BC_2\wedge \BC_2 \wedge \BC_2 
\nn\\
 &\quad 
 + \tfrac{1}{30}\, \BB_2\wedge \BC_2\wedge \BC_2 \wedge \BC_2\wedge \BC_2 \,.
\end{align}

\subsection{Supergravity fields}
\label{app:sugra}

Our gauge potentials are related to those used in \cite{hep-th/9802199,hep-th/9812188,hep-th/9806169,hep-th/9908094} as follows. 
In type IIA theory, their fields (left) and our fields (right) are related as
\begin{align}
 B = \AB_2 \,,\quad 
 C^{(p)} = \AC_p\,,\quad
 B^{(6)} = - \AB_6\,, \quad
 N^{(7)} = \AA_{7,n} \,,\quad 
 N^{(8)} = -\AA_{8,n}\,,
\end{align}
where $n$ represents a Killing direction. 
In type IIB theory, the relation is summarized as
\begin{align}
 \cB =& \BB_2\,,\quad 
 C^{(0)} = - \BC_0\,,\quad
 C^{(2)} = - \BC_2\,,\quad
 C^{(4)} = - \BA_4\,,\quad
 C^{(6)} = -\bigl( \BC_6-\tfrac{1}{4}\, \BB_2\wedge \BB_2\wedge \BC_2\bigr)\,,\quad 
\nn\\
 C^{(8)} &= -\bigl(\BC_8 - \tfrac{1}{3!}\,\BC_2\wedge \BB_2\wedge \BB_2\bigr)\,,\quad 
 B^{(6)} = -\bigl( \BB_6-\tfrac{1}{4}\, \BC_2\wedge \BC_2\wedge \BB_2\bigr) \,,\quad
 \widetilde{C}^{(8)} = \BE_8\,.
\end{align}
Although the full $T$-duality rule for the dual graviton $N^{(7)}$ has not been obtained there, by comparing the gauge transformation \eqref{eq:D71-delta-B} with Eq.~(B.4) of \cite{hep-th/9806169}, we find
\begin{align}
 N^{(7)} = \iota_n \BA_{7,n} - \tfrac{1}{4}\,\epsilon_{\SLc\SLd}\,\BA^{\SLc}_2\wedge\iota_n \BA^{\SLd}_2 \,.
\end{align}
In addition, $N^{(8)}$ and $\cN^{(8)}$ correspond to our $\bBD_{8,2}$ and $\bBE_{8,2}$ at least under $\BB_2=0$ and $\BC_2=0$. 
We have not identified the precise relation between their $N^{(9)}$ and our $\bBA_{9,2,1}$\,. 

Our 11D fields $\MA_{\hat{3}}$, $\MA_{\hat{6}}$, and $\MA_{\hat{8},\hat{1}}$ are the same as those used in \cite{hep-th/9802199,hep-th/9912030,hep-th/0003240}, where $\MA_{\hat{8},n}$ is denoted as $\hat{N}^{(8)}$\,. 
The 9-form $\iota_n\hat{C}^{(10)}$ used in \cite{hep-th/9912030,hep-th/0003240} can be defined as
\begin{align}
 \iota_n \hat{C}^{(10)} = \iota_n \MA_{\hat{10},n,n} + \tfrac{1}{4!}\,\MA_{\hat{3}}\wedge \iota_n \MA_{\hat{3}}\wedge \iota_n \MA_{\hat{3}}\wedge \iota_n \MA_{\hat{3}}\,.
\end{align}

Let us also identify the relation between our type IIB fields and those used in \cite{hep-th/0506013,hep-th/0602280,hep-th/0611036,1004.1348}. 
For this purpose, it is useful to perform a redefinition,
\begin{align}
\begin{split}
\begin{alignedat}{2}
 \tilde{\BA}^{\SLa}_2 &\equiv \BA^{\SLa}_2\,,\qquad\quad 
 \tilde{\BA}_4 \equiv \BA_4\,,\qquad &
 \tilde{\BA}^{\SLa}_6 &\equiv \BA^{\SLa}_6 - \tfrac{1}{3}\,\BA_4\wedge \BA_2^{\SLa} \,, 
\\
 \tilde{\BA}^{\SLa\SLb}_8 &\equiv \BA^{\SLa\SLb}_8 - \tfrac{1}{4}\,\BA^{(\SLa}_6\wedge \BA^{\SLb)}_2 \,,\qquad&
 \tilde{\BA}^{\SLa\SLb\SLc}_{10} &\equiv \BA^{\SLa\SLb\SLc}_{10} - \tfrac{1}{5}\,\BA^{(\SLa\SLb}_8\wedge \BA^{\SLc)}_2\,,
\end{alignedat}
\end{split}
\end{align}
which makes the field strengths to have the schematic form $\BF \sim \rmd\tilde{\BA} + \sum \BF\wedge \tilde{\BA}$\,,
\begin{align}
\begin{split}
 \BF^{\SLa}_3 &= \rmd \tilde{\BA}^{\SLa}_2 \,,\qquad
 \BF_5 = \rmd \tilde{\BA}_4 + \tfrac{1}{2}\, \epsilon_{\SLa\SLb}\, \BF^{\SLa}_3\wedge \tilde{\BA}^{\SLb}_2\,,
\\
 \BF^{\SLa}_7 &= \rmd \tilde{\BA}^{\SLa}_6 + \tfrac{1}{3}\,\BF_5 \wedge \tilde{\BA}^{\SLa}_2 - \tfrac{2}{3}\,\BF^{\SLa}_3\wedge \tilde{\BA}_4\,,
\\
 \BF^{\SLa\SLb}_9 &= \rmd \tilde{\BA}^{\SLa\SLb}_8 + \tfrac{1}{4}\,\BF^{(\SLa}_7 \wedge \tilde{\BA}^{\SLb)}_2 - \tfrac{3}{4}\,\BF^{(\SLa}_3 \wedge \tilde{\BA}^{\SLb)}_6\,,
\\
 \BF^{\SLa\SLb\SLc}_{11} &= \rmd \tilde{\BA}^{\SLa\SLb\SLc}_{10} + \tfrac{1}{5}\,\BF^{(\SLa\SLb}_9 \wedge \tilde{\BA}^{\SLc)}_2 - \tfrac{4}{5}\,\BF^{(\SLa}_3 \wedge \tilde{\BA}^{\SLb\SLc)}_8 = 0\,. 
\end{split}
\end{align}
The gauge transformation also can be expressed as $\delta \tilde{\BA} \sim \rmd\tilde{\Lambda} + \sum \BF\wedge \tilde{\Lambda}$,
\begin{align}
\begin{split}
 \delta \tilde{\BA}_2^{\SLa}&=\rmd \tilde{\Lambda}^{\SLa}_1\,, \qquad
 \delta \tilde{\BA}_4 =\rmd\tilde{\Lambda}_3 + \tfrac{1}{2}\,\epsilon_{\SLa\SLb}\, \BF^{\SLa}_3\wedge \tilde{\Lambda}^{\SLb}_1 \,,
\\
 \delta \tilde{\BA}_6^{\SLa} &=\rmd \tilde{\Lambda}_5^{\SLa} + \tfrac{1}{3}\, \BF_5\wedge \tilde{\Lambda}^{\SLa}_1 - \tfrac{2}{3}\, \BF^{\SLa}_3 \wedge \tilde{\Lambda}_3\,,
\\
 \delta \tilde{\BA}^{\SLa\SLb}_8 &= \rmd \tilde{\Lambda}^{\SLa\SLb}_7 - \tfrac{3}{4}\,\BF_3^{(\SLa}\wedge\tilde{\Lambda}^{\SLb)}_5 + \tfrac{1}{4}\,\BF^{(\SLa}_7\wedge \tilde{\Lambda}^{\SLb)}_1\,,
\\
 \delta \tilde{\BA}^{\SLa\SLb\SLc}_{10} &= \rmd \tilde{\Lambda}^{\SLa\SLb\SLc}_9 + \tfrac{1}{5}\, \BF^{(\SLa\SLb}_9\wedge \tilde{\Lambda}^{\SLc)}_1 - \tfrac{4}{5}\,\BF^{(\SLa}_3\wedge \tilde{\Lambda}_7^{\SLb\SLc)} \,.
\end{split}
\end{align}
by considering a field-dependent redefinitions of gauge parameters:
\begin{align}
\begin{split}
 \tilde{\Lambda}^{\SLa}_1 &\equiv \Lambda^{\SLa}_1\,,\qquad 
 \tilde{\Lambda}_3 \equiv \Lambda_3 -\tfrac{1}{2!}\,\epsilon_{\SLc\SLd}\,\tilde{\BA}_2^{\SLc}\wedge \Lambda^{\SLd}_1\,,
\\
 \tilde{\Lambda}^{\SLa}_5 &\equiv \Lambda^{\SLa}_5 + \tfrac{2}{3}\, \tilde{\BA}^{\SLa}_2 \wedge \Lambda_3 - \tfrac{1}{3}\, \tilde{\BA}_4\wedge \Lambda^{\SLa}_1 - \tfrac{1}{3!}\, \epsilon_{\SLc\SLd}\, \tilde{\BA}^{\SLa}_2\wedge \tilde{\BA}^{\SLc}_2 \wedge \Lambda^{\SLd}_1 \,,
\\
 \tilde{\Lambda}^{\SLa\SLb}_7 &\equiv \Lambda^{\SLa\SLb}_7 + \tfrac{3}{4}\, \tilde{\BA}^{(\SLa}_2\wedge \Lambda^{\SLb)}_5 + \tfrac{1}{4}\, \tilde{\BA}^{\SLa}_2\wedge \tilde{\BA}^{\SLb}_2\wedge \Lambda_3 
\\
 &\quad - \tfrac{1}{4}\, \tilde{\BA}^{(\SLa}_2\wedge \Lambda^{\SLb)}_1 - \tfrac{1}{4!}\, \epsilon_{\SLc\SLd}\, \tilde{\BA}^{\SLa}_2\wedge \tilde{\BA}^{\SLb}_2 \wedge \tilde{\BA}^{\SLc}_2 \wedge \Lambda^{\SLd}_1 \,,
\\
 \tilde{\Lambda}^{\SLa\SLb\SLc}_9 &\equiv \Lambda^{\SLa\SLb\SLc}_9 + \tfrac{4}{5}\, \tilde{\BA}^{(\SLa}_2\wedge \Lambda^{\SLb\SLc)}_7 + \tfrac{3}{10}\, \tilde{\BA}^{(\SLa}_2\wedge \tilde{\BA}^{\SLb}_2\wedge \Lambda^{\SLc)}_5 + \tfrac{1}{15}\,\tilde{\BA}^{\SLa}_2\wedge \tilde{\BA}^{\SLb}_2\wedge \tilde{\BA}^{\SLc}_2\wedge \Lambda_3
\\
 &\quad -\tfrac{1}{5}\,\tilde{\BA}^{(\SLa\SLb}_8\wedge\Lambda^{\SLc)}_1 + \tfrac{1}{5}\,\tilde{\BA}^{(\SLa}_2\wedge \tilde{\BA}^{\SLb}_6\wedge\Lambda^{\SLc)}_1 - \tfrac{1}{5!}\, \epsilon_{\SLd\SLe}\, \tilde{\BA}^{\SLa}_2\wedge \tilde{\BA}^{\SLb}_2 \wedge \tilde{\BA}^{\SLc}_2 \wedge \tilde{\BA}^{\SLd}_2 \wedge \Lambda^{\SLe}_1 \,.
\end{split}
\end{align}
Then, the tilde potentials and the field strengths are related to the fields used in \cite{hep-th/0506013,hep-th/0602280,hep-th/0611036,1004.1348} (appearing on the right-hand sides) as follows:
\begin{align}
\begin{split}
\begin{alignedat}{5}
 \tilde{\BA}_2 &= A_2\,,\quad&
 \tilde{\BA}_4 &= -4 A_4 \,,\quad&
 \tilde{\BA}^{\SLa}_6 &= -A^{\SLa}_6\,,\quad&
 \tilde{\BA}^{\SLa\SLb}_8 &= -4 A^{\SLa\SLb}_8 \,,\quad&
 \tilde{\BA}^{\SLa\SLb\SLc}_{10} &=12 A^{\SLa\SLb\SLc}_{10} \,,
\\
 \BF^{\SLa}_3 &= F^{\SLa}_3\,,\quad&
 \BF_5 &= - 4 F_5\,,\quad&
 \BF^{\SLa}_7 &= -F^{\SLa}_7\,,\quad&
 \BF^{\SLa\SLb}_9 &= - 4 F^{\SLa\SLb}_9 \,,\quad&
 \BF^{\SLa\SLb\SLc}_{11} &= 12 F^{\SLa\SLb\SLc}_{11}\,,
\\
 \tilde{\Lambda}^{\SLa}_1 &= \Lambda^{\SLa}_1\,,\quad&
 \tilde{\Lambda}_3 &= - 4 \Lambda_3\,,\quad&
 \tilde{\Lambda}^{\SLa}_5 &= -\Lambda^{\SLa}_5\,,\quad&
 \tilde{\Lambda}^{\SLa\SLb}_7 &= -4 \Lambda^{\SLa\SLb}_7 \,,\quad&
 \tilde{\Lambda}^{\SLa\SLb\SLc}_9 &=12 \Lambda^{\SLa\SLb\SLc}_9 \,.
\end{alignedat}
\end{split}
\end{align}

\end{document}